\pdfoutput=1

\documentclass{article}
\usepackage{arxiv}

\usepackage{xcolor}
\usepackage{graphicx}
\usepackage{hyperref}
\usepackage{natbib}
\usepackage{amsfonts}
\usepackage{amsmath}

\AtBeginDocument{%
\providecommand\BibTeX{{%
\normalfont B\kern-0.5em{\scshape i\kern-0.25em b}\kern-0.8em\TeX}}}

\usepackage{booktabs} 

\def\real{\mathbb{R}}
\usepackage{xspace}
\usepackage{multirow}
\usepackage{fbox}
\usepackage[normalem]{ulem}

\newcommand{\revision}[1]{{#1}\xspace}

\newcommand{\nico}[1]{{#1}\xspace}
\newcommand{\lo}[1]{{#1}\xspace}
\newcommand{\chems}[1]{{#1}\xspace}
\newcommand{\thib}[1]{{#1}\xspace}

\definecolor{colorMinutes}{rgb}{0.27, 0.35, 0.27}
\definecolor{colorHours}{rgb}{0.0, 0.18, 0.39}
\definecolor{colorDays}{rgb}{0.5, 0.09, 0.09}

\usepackage{siunitx,xinttools,xintfrac}
\usepackage{xfp}

\newcommand\duration[1]{%
\xintifLt{#1}{60}{
\num[round-mode=places,round-precision=2]{#1} s%
}{
\xintAssign\xintiiDivision{\xintNum{#1}}{86400}\to\days\remain%
\xintAssign\xintiiDivision{\xintNum{#1}}{3600}\to\hours\minutes%
\xintAssign\xintiiDivision{\minutes}{60}\to\minutes\seconds%
\xintiiifGt{\days}{0}{
{\color{colorDays}{\days} d}%
}{
\xintiiifGt{\hours}{0}{
{\color{colorHours}\hours:\num[minimum-integer-digits=2]{\minutes}:\num[minimum-integer-digits=2]{\seconds} h}%
}{
{\color{colorMinutes}\minutes:\num[minimum-integer-digits=2]{\seconds} m}%
}%
}
}%
}
\newcommand\durationT[2]{%
\xintifLt{#2}{60}{
\num[round-mode=places,round-precision=2]{#1} (\num[round-mode=places,round-precision=2]{#2}) s%
}{
\xintAssign\xintiiDivision{\xintNum{#1}}{86400}\to\days\remain%
\xintAssign\xintiiDivision{\xintNum{#1}}{3600}\to\hours\minutes%
\xintAssign\xintiiDivision{\minutes}{60}\to\minutes\seconds%
\xintAssign\xintiiDivision{\xintNum{#2}}{86400}\to\daysT\remainT%
\xintAssign\xintiiDivision{\xintNum{#2}}{3600}\to\hoursT\minutesT%
\xintAssign\xintiiDivision{\minutesT}{60}\to\minutesT\secondsT%
\xintiiifGt{\daysT}{0}{
{\color{colorDays}\days (\daysT) d}%
}{
\xintiiifGt{\hoursT}{0}{
{\color{colorHours}\hours:\num[minimum-integer-digits=2]{\minutes}:\num[minimum-integer-digits=2]{\seconds} (\hoursT:\num[minimum-integer-digits=2]{\minutesT}:\num[minimum-integer-digits=2]{\secondsT}) h}%
}{
{\color{colorMinutes}\minutes:\num[minimum-integer-digits=2]{\seconds} (\minutesT:\num[minimum-integer-digits=2]{\secondsT}) m}%
}%
}
}%
}

\newcommand{\Sum}[2]{\fpeval{ #1 + #2}}
\newcommand\twodurations[2]{\durationT{#1}{\Sum{#1}{#2}}}

\def\dtsDFLT{\texttt{Default}\xspace}
\def\dtsSHREC{\texttt{SHREC}\xspace}
\def\dtsABC{\texttt{ABC}\xspace}
\def\dtsEC{\texttt{EC}\xspace}

\def\pced{\texttt{PCEDNet}\xspace}
\def\pcedtc{\texttt{PCEDNet-2C}\xspace}
\def\fc{\texttt{FC}\xspace}
\def\fctc{\texttt{FC-2C}\xspace}
\def\cnn{\texttt{CNN}\xspace}
\def\cnntc{\texttt{CNN-2C}\xspace}
\def\ca{\texttt{CA}\xspace}
\def\fee{\texttt{FEE}\xspace}
\def\pcp{\texttt{PCPNet}\xspace}
\def\pcptc{\texttt{PCPNet-2C}\xspace}
\def\ecnet{\texttt{ECNet}\xspace}
\def\pienet{\texttt{PIENet}\xspace}

\newcommand{\dataStr}[1]{{\texttt{#1}}}

\def\dataEmpire{\dataStr{Empire}\xspace}
\def\dataLans{\dataStr{Lans}\xspace}
\def\dataChurch{\dataStr{Church}\xspace}
\def\dataPisa{\dataStr{Pisa Cathedral}\xspace}
\def\dataEuler{\dataStr{Euler}\xspace}
\def\dataMunich{\dataStr{Munich}\xspace}
\def\dataTrainSt{\dataStr{Train~St.}\xspace}
\def\dataTrainStation{\dataStr{Train Station}\xspace}
\def\dataLoudun{\dataStr{Loudun}\xspace}
\def\dataLoudunSmall{\dataStr{Loudun~1}\xspace}
\def\dataLoudunBig{\dataStr{Loudun~35}\xspace}
\def\dataRueMadame{\dataStr{Paris rue Madame}\xspace}

\def\dataCube{\dataStr{Cube}\xspace}
\def\dataHexagon{\dataStr{Icosahedron}\xspace}
\def\dataFandisk{\dataStr{Fandisk}\xspace}
\def\dataTwoCubes{\dataStr{2-cube}\xspace}
\def\dataFourCubes{\dataStr{4-cube}\xspace}
\def\dataAngle{\dataStr{Angle}\xspace}
\def\dataOctogon{\dataStr{Dodecahedron}\xspace}
\def\dataCone{\dataStr{Cone}\xspace}

\makeatletter
\def\input@path{{./images}{./}{../}}
\makeatother

\usepackage{import}

\begin{document}

\title{PCEDNet : A Lightweight Neural Network for Fast and Interactive Edge Detection in 3D Point Clouds}
\date{}

\author{ Chems-Eddine Himeur \\
CNRS, IRIT, Universit\'{e} de Toulouse, France\\
\texttt{chems-eddine.himeur@irit.fr} \\
\And
Thibault Lejemble \\
CNRS, IRIT, Universit\'{e} de Toulouse, France\\
\texttt{thibault.lejemble@irit.fr}
\And
\href{https://orcid.org/0000-0001-8984-1399}{\includegraphics[scale=0.06]{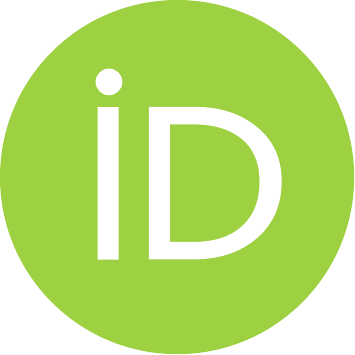}\hspace{1mm}Thomas Pellegrini} \\
CNRS, IRIT, Universit\'{e} de Toulouse, France\\
\texttt{thomas.pellegrini@irit.fr} \\
\And
\href{https://orcid.org/0000-0001-5606-9654}{\includegraphics[scale=0.06]{orcid.pdf}\hspace{1mm}Mathias Paulin} \\
CNRS, IRIT, Universit\'{e} de Toulouse, France\\
\texttt{mathias.paulin@irit.fr} \\
\And
\href{https://orcid.org/0000-0001-9908-3640}{\includegraphics[scale=0.06]{orcid.pdf}\hspace{1mm}Lo\"{i}c Barthe} \\
CNRS, IRIT, Universit\'{e} de Toulouse, France\\
\texttt{loic.barthe@irit.fr} \\
\And
\href{https://orcid.org/0000-0003-2180-4318}{\includegraphics[scale=0.06]{orcid.pdf}\hspace{1mm}Nicolas Mellado} \\
\texttt{nicolas.mellado@irit.fr} \\
}
\renewcommand{\undertitle}{}
\renewcommand{\shorttitle}{PCEDNet : A Lightweight Neural Network for Fast and Interactive Edge Detection in 3D Point Clouds}
\renewcommand{\headeright}{}

\maketitle

\begin{abstract}
In recent years, Convolutional Neural Networks (CNN) have proven to be efficient analysis tools for processing point clouds, e.g., for reconstruction, segmentation and classification.
In this paper, we focus on the classification of edges in point clouds, where both edges and their surrounding are described.
We propose a new parameterization adding to each point a set of differential information on its surrounding shape reconstructed at different scales.
These parameters, stored in a \emph{Scale-Space Matrix (SSM)}, provide a well suited information from which an adequate neural network can learn the description of edges and use it to efficiently detect them in acquired point clouds.
After successfully applying a multi-scale CNN on SSMs for the efficient classification of edges and their neighborhood, we propose a new \nico{lightweight} neural network architecture outperforming the CNN in learning time, processing time and classification capabilities.
Our architecture is compact, requires small learning sets, is very fast to train and classifies millions of points in seconds.
\end{abstract}

%
%
%


\begin{figure*}
\def\svgwidth{\linewidth}
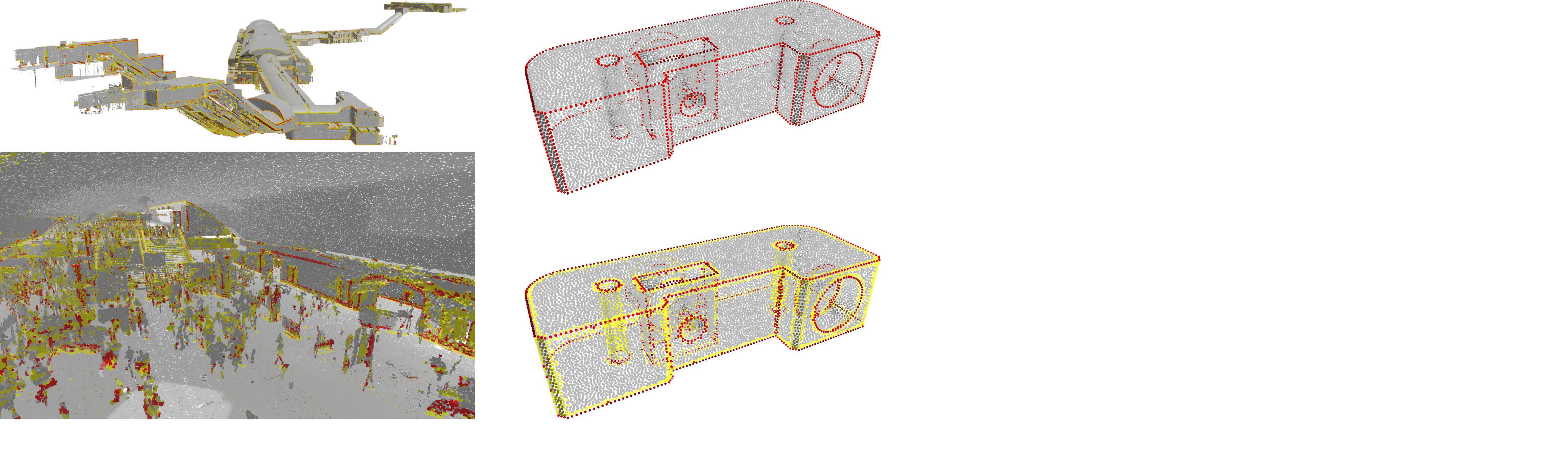
\caption{
Three examples of edge detection in point clouds by our \pced neural network.
It handles both: (a) the imperfect edges of large scale scans (here $12$ million vertices) subject to irregular sampling and noise while detecting both sharp (in red) and smoother (in yellow) edges in few minutes (here less than $6$) - and - (b) accurate CAD data
on which it can focus on sharp edges if desired, in a few seconds for this model.
(c) Our network can also be trained in a few seconds to detect edges following the edge definition provided by a user in an interactive model annotation.
We show two annotations corresponding to different user expectations.
Most of the processing is precomputed and at runtime edges of this model are classified in less than a second.
\label{fig:teaser}
}
\end{figure*}

\section{Introduction}

Nowadays, acquired point clouds \nico{is} a very common and widespread representation.
The large range of acquisition devices and Computer Vision techniques massively generate clouds with tens or hundreds of millions of points.
The shapes sampled by these large unstructured data are arbitrarily complex, and their processing remains extremely tedious.
Edges are fundamental features for processing point clouds and their automatic detection is thus useful in a wide range of applications in the fields of computer vision (e.g. feature extraction), computer graphics (e.g. contour line reconstruction) and others. 
Despite regular advances over the years, it remains 
an open, very challenging problem. 

In general, edges are defined strictly as sharp edges, e.g. for manufactured objects~\cite{Koch2019ABCAB}, or as feature lines \nico{for} objects.
When asked to draw feature lines, users tend to follow more complex rules that may vary from people to people~\cite{Cole:2008}.
This leads to a lack of clear theoretical definition of an edge, especially on 3D surfaces acquired from models including features of different scales and more or less damaged/clean edges (e.g. stone or plastered buildings, progressively smoothed edges, polished mechanical parts, etc). 
In addition, an edge can be considered as sharp or smooth depending on the observation scale.
This generates contextualized and potentially ambiguous interpretation of what edges are. 
For instance, different persons produced different annotations on the models provided for the recent feature curve detection contest~\cite{Thompson:2019}. 
These models include a large variety of different edges, from sharp to smooth and rounded. 
This underlines that the frontier between an edge and a smooth surface (i.e. a feature boundary and a feature continuity) remains subjective, especially in the case of real world models. 

Machine Learning (ML) approaches have gain a lot of interest for their ability to efficiently reproduce theoretically ``fuzzy'' or complicated classifications for which we are able to provide a sufficiently large set of annotated data. 
At first glance, they thus seem particularly attractive to create efficient edge detectors. 
However, in the context of point cloud processing, ML remains very challenging to use as there is neither natural ordering~\cite{NIPS2017_6931} nor intrinsic parameterization of the input data. 
In the past years, patches of points have been used by several approaches to apply Convolution Neural Networks (CNN) for classification~\cite{qi2016pointnet}{, local shape property estimation~\cite{GuerreroEtAl:PCPNet:EG:2018}, and continuous sharp edges \lo{classification/}reconstruction~\cite{PieNet2020}.}

For edge detection, geometric approaches avoid using patches by considering geometric descriptors parameterizing each individual point~\cite{demarsin2007detection,Weinmann2013FeatureRA}.
Following this idea,~\citet{Hackel_2016_CVPR} propose to train a random forest classifier taking as input geometric quantities obtained at multiple scales by linear regression in order to classify points belonging to edges.

In this paper, we introduce both a new way to individually parameterize points, together with a dedicated edge detection \lo{lightweight} neural network classifier called \emph{Point Cloud Edge Detection Network} (\pced).
Points are parameterized with a so-called \emph{Scale-Space Matrix} (SSM) (Section~\ref{sec:glsdescriptor}) encoding extrinsic geometric properties of a locally reconstructed surface surrounding each point of the input point cloud at multiple scales. 
The use of this multi-scale parameterization allows us to propose a compact architecture (Section~\ref{subsec_treeArchitecture}) based on a simple neural network enabling both its training in seconds or few minutes on small data sets, and the processing of 500 thousands of points per second on average with our current CPU implementation (Table~\ref{tab:visualTime}). 
This fast training and the limited requirement on annotated data allow to easily specialize our network to specific edge definitions, as illustrated in Figures~\ref{fig:teaser}-a and~\ref{fig:teaser}-b.
It is fast enough to allow a user to interactively label small sub-parts of a point cloud and 
let our network annotate millions of points in seconds (Figure~\ref{fig:teaser}-c, Section~\ref{subsec:interactive}).

The choice of our SSM values is validated by an ablation study (Section~\ref{subsec:ablation}), and we evaluate the effectiveness of our network by comparing its speed and accuracy with existing geometric edge detection approaches, point-based processing networks, and two baselines: a Convolutional Neural Network (\cnn) and a Fully Connected neural network (\fc) (Section~\ref{subsec_CNN}).
Our results (Section~\ref{sec_results}) demonstrate the superiority of our \pced over previous works, and illustrate its efficiency on a large variety of data, from CAD models to real world examples with tens of millions of points (Figures~\ref{fig:teaser}-a and~\ref{fig:loudoncomp}).

Current networks processing point clouds rely on very deep architectures in an end-to-end strategy.
While the way our point parameterization is computed is not novel in geometry processing, the use of these parameters structured in our SSM as input of a network is new. 
Our proposition illustrates how a simple network exploiting these multi-scale geometric surface descriptors overpasses current approaches. 
We believe that its superiority in training and evaluation speed, scalability and resource requirements (computation, memory) makes it an example for the creation of lightweight point {cloud} processing solutions opening new perspectives regarding the system interactivity and adaptation to user's wishes. 
This is also a step forward to reach real-time point cloud processing with limited resources \lo{and low energy consumption (Section~\ref{subsec:power}), which are both becoming} 
 very important challenges \lo{when also considering embedded systems and environmental impact}.

\section{Related Work}

In this section, we first present the way unorganized point clouds are parameterized before being processed for geometric learning (Section~\ref{subsec_pointParam}). 
We introduce the different existing architectures for point-based machine learning and we discuss methods for edge detection from point clouds (Section~\ref{subsec_networkStructure}). We then review geometric approaches (Section~\ref{subsec_edgeDetection}) dedicated to this topic.

\subsection{Point cloud parameterization}
\label{subsec_pointParam}
Point clouds are most of the time defined as unordered and unstructured set of points sampling an unknown surface.
When using neural networks on point clouds, a first challenge is to define 
a regular structure of the cloud that fits a network architecture, i.e. to \emph{parameterize} the point cloud according to the network architecture. 
Several approaches have been proposed to tackle this challenge.
A first class of approaches parameterizes unstructured point clouds in regular grids, e.g. using series of images taken from different viewpoints~\cite{su15mvcnn, 3dor.20171047,DBLP:journals/corr/KalogerakisAMC16,BOULCH2018189}, or using voxel grids~\cite{DBLP:journals/corr/abs-1711-06396,DBLP:journals/corr/abs-1711-08488,DBLP:journals/corr/QiSNDYG16,wu20153d,Maturana-2015-6018}. 
The main limitation of these approaches is the cells memory requirement that limits their scalability to very large models %
especially when detecting thin structures and details, {as} edges.

A second class of approaches processes each point and its neighborhood, so that the point coordinates are directly processed by the network. 
PointNet and its variants~\cite{qi2016pointnet,Qi:2017,GuerreroEtAl:PCPNet:EG:2018} use multi-layer perceptron (MLP) to consecutively sample and group point coordinates and build geometric features around a point.
Several approaches have been proposed to extend convolution to 3D point clouds, using local spectral convolution~\cite{DBLP:journals/corr/abs-1803-05827}, parameterized convolutional filters~\cite{xu2018spidercnn}, transformed points~\cite{Li:2018}, sparse lattices~\cite{Su:2018}, adaptive kernels~\cite{gdcfupc2019} and kernel point convolution~\cite{Thomas2019}. 

The strong benefit of point-based convolutions is to allow the design of networks whose first layers learn features at multiple scales directly from the point locations.
On the other hand, the characterization of geometric structures relies on the way convolution layers capture those features, only based on the local spatial organization of points. 
On real data, point sets also include sampling variation, noise, outliers and missing data that also have to be learned.

%

\subsection{Network architectures}
\label{subsec_networkStructure}
Most of the point-based geometry analysis neural networks are tailored for point cloud segmentation~\cite{Dai:2017} and classification~\cite{Wu:2015,hackel2017isprs}.
They follow the rational introduced by PointNet, where point coordinates are abstracted by successive layers, then reduced using max pooling for feature extraction and finally processed with MLP for the final decision.
Such architectures aim at abstracting the input point cloud and estimating high level or semantic properties.

Some approaches also aim at processing point clouds according to their local geometric properties, e.g. normal estimation~\cite{Boulch:2016}.
PCPNet~\cite{GuerreroEtAl:PCPNet:EG:2018} learns local shape properties (e.g. normal vectors) directly from raw point clouds.
Interestingly, this work suggests a multi-scale architecture where several networks process the surrounding of the analyzed point at increasing neighborhood size (one network per size). 
The features learned by the different networks are stacked to form a feature vector, and processed by a fully connected network to produce the final decision.
Recently, the deep Learning Point Network architecture~\cite{le2020deeperleanpn} improves the implementation of convolution-based point cloud processing networks such as PointNet, DGCNN~\cite{dgcnn} and SpiderNet~\cite{SpiderNet}.

Overall, all these networks have been designed to extract semantic information from point clouds, which justify their both deep and large architectures. 
In our more geometric and specialized context, these architectures become unnecessarily complex while requiring long training and processing times.


%


\subsection{Edge detection}
\label{subsec_edgeDetection}
Detecting edges in unstructured point clouds is usually cast as a sharp feature, a feature contour or a curve detection problem.
It is often the first step of constrained surface reconstruction algorithms and over the years, many approaches have been proposed in this context. 
We refer the interested reader to the survey by~\citet{Berger:2017}.
A standard approach is to compute at each point a geometric descriptor using the eigenstructure of the covariance matrix~\cite{Gum2001}. It can be a ratio between the eigenvalues, taking into account their evaluation at different scales~\cite{Pauly2003} or not~\cite{Xia2017}, or directly a curvature estimation~\cite{Lin2015,Wei2018}.
The ratio between eigenvalues is considered as a more reliable parameter and it is, for instance, used in the \nico{CGAL Library~\cite{cgal:ass-psp-21a}} with a Delaunay-based feature estimation~\cite{Merigot2011}.
While well established, all these approaches suffer from a sensibility to noise and they perform at a given scale with a strong dependence to a decision threshold. The methods introduced by~\citet{Pauly2003} and~\citet{Bazazian15} consider curvature ratio at different scales that reduces the dependence to the scale of analysis and the sensibility to noise, but they remain subject to a decision threshold.

Another family of methods relies on Moving Least Squares surface reconstruction~\cite{demarsin2007detection,weber2012sharp,ni2016edge}. Using this reconstruction, edge detection can be performed on a Gaussian map clustering computed in a local neighborhood~\cite{SFD}. 
Adaptive reconstruction kernels~\cite{Fleish2005} can also be combined with polynomial fit~\cite{Daniels2008}. 
Other approaches rely  either on subspace detection and feature intersection computation~\cite{FERNANDES2012}, on the mean-shift algorithm to select the farthest points from the centroid of their neighborhood~\cite{Ahmed2018}, on the average of neighbors altitude over a local tangent plane~\cite{Li2017}, or on the intersection of planes detected using RANSAC~\cite{Mitropoulou:2019}.
In other contexts, edges are defined as a specific type of lines on the surface.
\citet{Lin2015} use RANSAC to spatially regularize the response of a sharp feature detector.
Recently, \citet{Hackel_2016_CVPR} proposed a data-driven {method} where the points are classified and then structured using a graph-based approach. Extending local sharp feature detectors by a global analysis improves the robustness of the detection but
limits the scalability, a critical aspect when processing large sets of {point} clouds defined by tens of millions of points or more.

With an architecture {extending PU-Net~\cite{PUNET18}}, 
\ecnet~\cite{ecnet2018} is a network consolidating edges after up-sampling the point cloud and detecting its edges. 
\ecnet is thus a very deep network requiring important resources for training and processing points at inference time, while being 
limited in scalability by its up-sampling.
\lo{Finally, the recent \pienet~\cite{PieNet2020} first 
trains 
two networks based on
PointNet++~\cite{Qi:2017} 
%
for respectively classifying edge points and corner points.
The classified points are filtered with a non-maximal suppression, and clustered by feature using 
the third 
variant of PointNet++. 
A resulting set of curves is then generated by a two-headed PointNet network~\cite{qi2016pointnet}.
Even if it is closely
related to our proposal, the Pie-Net edge classification is based on a 
concatenation of several networks with
deep architecture, and it is thus also prone to high resource requirements.}

In this work, as suggested by~\citet{Hackel_2016_CVPR}, we propose to take advantage of the discrimination offered by geometric descriptors in a machine learning approach. We increase the robustness to noise and the adaptation to the different feature size \nico{and} shape by the use of stable multi-scale descriptors including derivatives over scales. 
We then use these parameters together with a specifically designed neural network. 
By relying on this set of descriptive parameters, our network avoids the deep design of existing approaches by only requiring a very compact architecture that recursively combines the features learned at different analysis scales. 

\section{Method}

\subsection{Problem statement}
\label{subsec_problemstatement}
Given a surface sampled by a point cloud, a sharp edge is commonly defined as a tangential discontinuity.
\citet{SFD} define edges as sharp features (crests and valleys) between two meeting planes, as well as corners at the point of intersection between three or more planes.
\citet{Hackel_2016_CVPR} call them "wire-frame contours", and define edges as linear features along which the orientation (normals) of the underlying surface exhibits an unusual discontinuity. 

\begin{figure}[t]
     \centering
     \def\svgwidth{0.5\linewidth}
    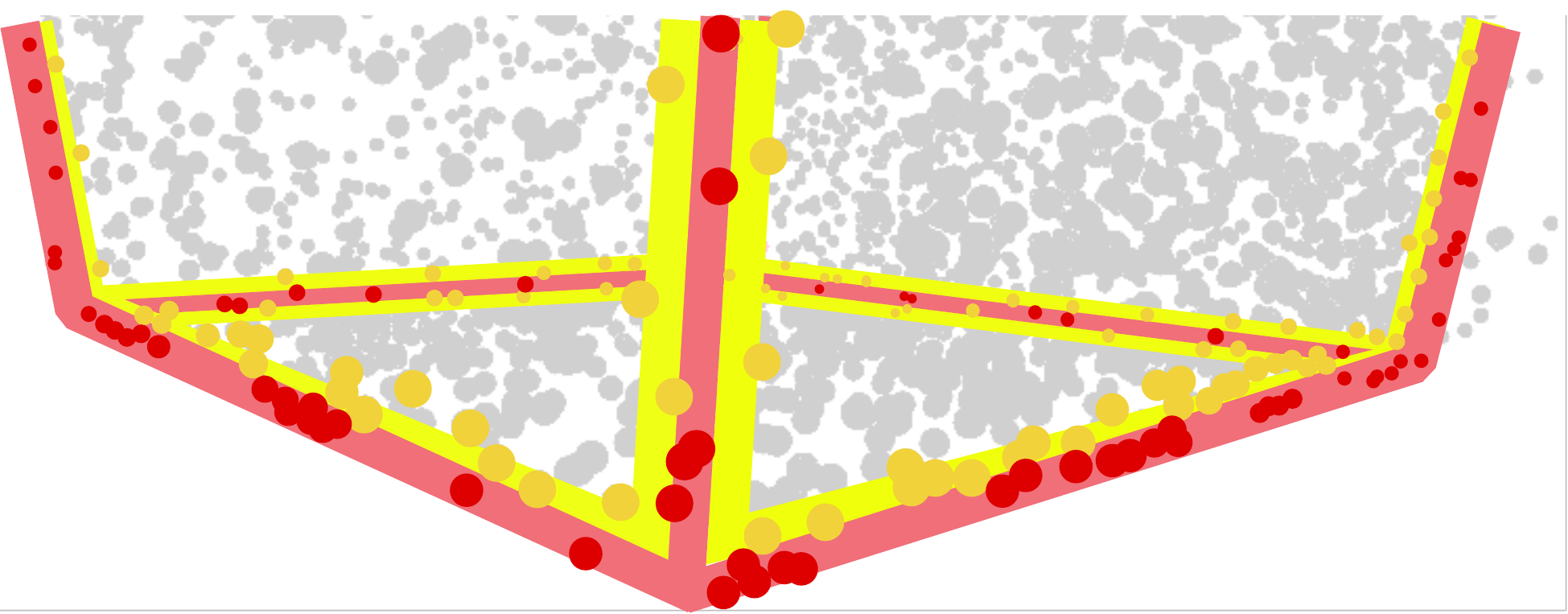
    \caption{Synthetic point cloud sampling a cube with sharp edges (red lines).
    Due to the sparse sampling and noise, only a few points lie on edges (red points). Detecting the edge surrounding (yellow points) improves the robustness to noise and sampling variation.}
    \label{fig:edgedef}
\end{figure}

There are at least two main reasons why these definitions are restrictive and not really practical for the analysis of point clouds:
\begin{enumerate}
    \item \textbf{Sparsity}: Points of acquired point clouds are unequally distributed over the object surfaces. Sharp edges are by nature very sparse, and it is very unlikely that the points of a point cloud actually sample the exact edge of the underlying surface as illustrated by the red points in Figure~\ref{fig:edgedef}.
    
    \item \textbf{Rounding}: Acquired point clouds are composed by points sampling real-world objects, on which edges are always more or less {rounded (Figure~\ref{fig:shrec-comp})/damaged (Figure~\ref{fig:loudoncomp})}. 
    For instance, two facades of a building might be locally connected by a continuous curved surface considered as an edge at the scale of the building, which may be unlikely to be detected at 
    {finer} scales such as the one of the bricks. 
\end{enumerate}

In order to handle these situations, we propose to classify points as \textbf{sharp-edge} when they lie on a sharp edge, and as \textbf{smooth-edge} when they are (1) near a sharp edge or (2) on a rounded edge.
Other points are then labeled as \textbf{non-edge}.
Depending on the application context and the artifacts found in the data, one might consider either one or the two edge classes in the results (we experimentally validate this proposition of two edge classes in Section~\ref{subsec:ablation}).
{In this paper, we denote our network as \pced when it is used with three classes and as \pcedtc when it is used with two classes.} 
Considering that we do not have a universal and practical definition of an edge, we propose to learn the point-based classification of this feature from examples.
For instance, when edges
\lo{have similar geometries and discretization}
over all models, \pcedtc may be preferred while \pced is suggested when the edge definition is more fuzzy. 
In that case, the \textbf{sharp-edge} class is to be used to annotate what is considered as an edge (whatever sharp or smooth/rounded) and the \textbf{smooth-edge} class is rather to annotate points at proximity of the points tagged as sharp-edges or on features that are close to what is considered to be an edge.
We do not suggest any automatic or threshold-based annotation approach as, so far, they are not robust enough and they do not enable the flexibility of letting the user defining his definition of edges through annotations.

In order to be robust to acquisition artifacts, sampling issues and edge roundness, we first reconstruct the surface described by the input points using a robust reconstruction algorithm
at multiple scales.
We then compute geometric descriptors of the reconstructed surfaces, which are parameters to be processed by a machine learning algorithm (see Section~\ref{sec:glsdescriptor}).
We show how this parameterization can be used with a common CNN and a fully connected neural network (Section~\ref{subsec_CNN}), and we propose a dedicated architecture (Section~\ref{subsec_treeArchitecture}) that outperforms these networks as well as previous edge classification approaches.

\subsection{Scale-Space Matrix}
\label{sec:glsdescriptor}

Our approach is inspired by the Growing Least Squares (GLS) approach~\cite{Mellado:2012:GLS}, where the geometry surrounding a point is described by the differential properties of the surfaces reconstructed from neighborhoods of increasing size. 
We first present the basics of GLS, and then introduce the Scale-Space Matrix (SSM), which wraps in a regular structure the differential properties of the surfaces reconstructed at different scales.

The GLS extends the concept of Scale-Space analysis~\cite{witkin1987scale,Lindeberg:1993} to point-based shape analysis.
The key idea is to detect pertinent geometric structures and scales as \emph{stabilities} in scale-space.
Stabilities are found when the magnitude of the derivatives of the surface is minimized when the scale varies.
Due to its multi-scale nature, this approach disambiguates between noise and features, and detects geometric features defined at arbitrary scales.
As such, points on a rounded and a sharp edge have \lo{discriminative} descriptors and can be disambiguated during classification.

We denote $\mathcal{S}^t$ the \emph{continuous surface} reconstructed at scale $t$, defined as the 0-isosurface of a scalar field $S^t(\mathbf{x}) : \real^3\rightarrow \real$.
We use the Algebraic Point Set Surfaces (APSS)~\cite{Guennebaud:2007:APS} to reconstruct continuous surfaces from raw point clouds.
This approach has been proven to be fast and stable at large scales~\cite{Guennebaud:2008:APS}.
As many other previous work, the scale is controlled by varying the size of a neighborhood ball centered around the evaluation point.

In its original formulation, the pertinent scale extraction introduced by \citet{Mellado:2012:GLS} combines several descriptors measuring the variation of local relief {(i.e. the algebraic distance from the point to the locally reconstructed surface)} $\frac{\delta\tau(\mathbf{x},t)}{\delta t}$, normal orientation $\frac{\delta\eta(\mathbf{x},t)}{\delta t}$ and mean curvature $\frac{\delta\kappa(\mathbf{x},t)}{\delta t}$, where $\tau(\mathbf{x},t):\real^4\rightarrow\real$, $\eta(\mathbf{x},t):\real^4\rightarrow\real^3$ and $\kappa(\mathbf{x},t):\real^4\rightarrow\real$ are scale-invariant properties of the surfaces $\mathcal{S}^t(\mathbf{x})$ (mathematical formulations are presented in Appendix~\ref{app:gls}).
They also use $\phi(\mathcal{S}^t(\mathbf{x}))$, the residuals of the fitting process, as confidence value.

In this work we propose to extend this idea further by measuring the surface variation in scale and space around each sample $\mathbf{p}_i$ of the input point cloud. 
We do not seek at defining an hand-crafted descriptor, but rather providing differential properties of the surface that are \nico{discriminative} for edge detection.
According to the GLS formalism, the parameters $\tau$, $\eta$ and $\kappa$ are differentiable both in scale and space, which leads to a Jacobian matrix of $5\times 4$ entries (see the description in Appendix~\ref{app:gls}).
By keeping only the quantities that are invariant by rigid transformations and not linearly dependent, we obtain the following feature vector $X^t_i$:
\begin{equation}
X^t_i =
    \begin{bmatrix}
    \tau_i^t & 
    \kappa_i^t & 
    k1_i^t & 
    \frac{\delta \tau_i^t}{\delta t} & 
    \frac{\delta \kappa_i^t}{\delta t} & 
    \phi(S^t_{\mathbf{p}_i})
    \end{bmatrix}
\end{equation}
where $\tau_i^t = \tau(\mathbf{p}_i,t)$ and $\kappa_i^t = \kappa(\mathbf{p}_i,t)$. 
$k1_i^t$ measures the magnitude of the first principal curvature of the surface, and is computed from $\frac{\delta \eta_i^t}{\delta \mathbf{x}}$.
According to our experiments, high values of $\tau$, which measure the local relief, helps to disambiguate between rounded and sharp edges.
The scale derivatives $\frac{\delta \tau_i^t}{\delta t}$ and $\frac{\delta \kappa_i^t}{\delta t}$ denote the stability of the reconstruction scale $s_j$.

The Scale-Space Matrix (SSM) defines a structured parameterization of the surfaces $\mathcal{S}^t$ described by the feature vectors $X^t_i$ computed at $N$ scales on each of the $M$ points of the cloud.
It is thus defined as follows:
\begin{equation}
SSM = 
    \begin{bmatrix}
     X_{1}^{1} &   X_{2}^{1} &   X_{3}^{1} & \dots  &   X_{M}^{1} \\
     X_{1}^{2} &  X_{2}^{2} &  X_{3}^{2} & \dots  &  X_{M}^{2} \\
    \vdots & \vdots & \vdots & \ddots & \vdots \\
  X_{1}^{N} & X_{2}^{N} &  X_{3}^{N} & \dots  &  X_{M}^{N}.
\end{bmatrix}.
\end{equation}
The minimum scale $s_\text{min}$ is defined 
{as the mean distance between points computed using the k nearest neighbors of each point (k=$10$), }
and the maximum scale $s_\text{max}$ is 10\% of the 
{diagonal} of the point cloud axis aligned bounding box (the effect of the variation of these values is presented in Section~\ref{sec:compexp}).
According to previous work~\cite{5539838,mellado2015relative}, we use a logarithmic scale sampling to obtain scale-invariant feature vectors, such that: 
 \begin{equation}
 \label{eq:scales}
     s_i = {\left(\frac{s_\text{max}}{s_\text{min}}\right)^\frac{i-1}{N-1}}*s_\text{min} {{~,~~~~~i = 1..N}}.
 \end{equation}
For all our experiments we use 16 scales distributed according to Equation~\ref{eq:scales} {(the use of different number of input scales is discussed in Section~\ref{subsec:numberOfScales})}.
Even though we carefully selected the set of descriptors defining our SSM for their geometric meaning with respect to edge detection, we validate the optimality of this choice when parameterizing our neural network by an ablation study presented in Section~\ref{subsec:ablation}.

\begin{figure}[t]
\centering
{
   \scriptsize
   \def\svgwidth{0.4\linewidth}
   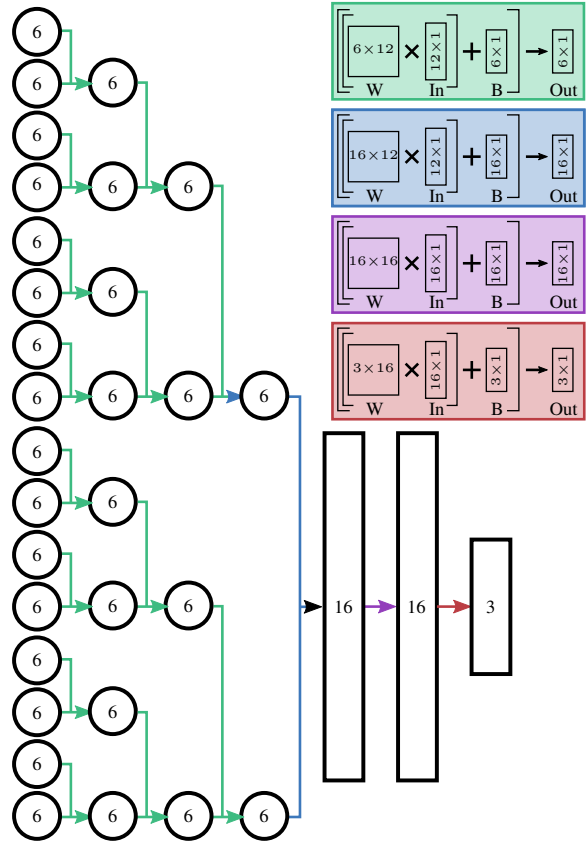
}
   \caption{Our \pced architecture. {The left column of circles represents the input sixteen $6$-dimensional vectors $X_i^t$. They are successively pairwise concatenated and processed by dense layers coupled with $6$ neurons as depicted in the green box (W: weights, In: input and B: bias). The output vector of size $12$ is then processed by a multi-layer perceptron (boxes blue, purple and red) for providing the output classification.}}
    \label{fig:arch8}
\end{figure}

\subsection{Our network: \pced}
\label{subsec_treeArchitecture}


The architecture of our Point Cloud Edge Detection Network, denoted \pced, is depicted in Figure~\ref{fig:arch8}. 
The input data is provided to the network as sixteen 6-dimensional vectors $X_i^t$ per point, each vector corresponding to a scale {(Figure~\ref{fig:arch8}-left column)}.  
The sixteen scale vectors are concatenated by groups of two in order to form eight 12-dimensional vectors. 
More precisely, the first scale is grouped with the second one, the third one with the fourth one, and so on. 
The idea is to halve the number of scales iteratively until obtaining a single 12-dimensional vector as a final feature representation {(four left columns of black circles in Figure~\ref{fig:arch8})}.

All the layers are fully-connected layers, also called dense layers (linear layer with biases{, depicted in Figure~\ref{fig:arch8}-green box}), followed by a sigmoid activation function. 
For the first layer, each vector of size 12 is given to a dense layer comprised of 6 neurons. There are eight input vectors and each vector is fed to its own 6-neuron layer. 
The weights are not shared between these small layers, allowing to process the scales differently, as would do a convolution layer, but here, we combine scales by groups of two. 
The subsequent three layers perform similarly, but with 48, 24, 12 neurons, respectively. 

The final 12-dimensional feature vector is given to a multi-layer perceptron (MLP) responsible for the classification {(\nico{MLP}
architecture is presented in Figure~\ref{fig:arch8}-blue/purple/red boxes}). 
This MLP is composed of two 16-neuron\nico{s} dense+sigmoid layers followed by the output layer with 3 neurons and a softmax activation function. 
The total number of weights is about 2.1k, which makes our architecture very compact.

The goal of grouping different scales together is to observe the input shape at different scales in a simultaneous and more intricate way than without grouping them.
We expect that this eases the simultaneous detection of higher-scale and sharp geometric properties of the input point clouds. 
Another reason is that such a representation helps the model to cope with noise in the point clouds.

The "tree structure" merging scales two by two iteratively can be seen as a variant of a grouped convolution operation. 
It preserves the high computing speed of keeping the scales separate while also mixing the scales layer after layer. 
It is expected to reduce the chance of being biased by a specific scale and the sensitivity to scale-dependent noise. 


\begin{figure}[h]
    \centering
    \includegraphics[height=.5\linewidth, angle=90 ]{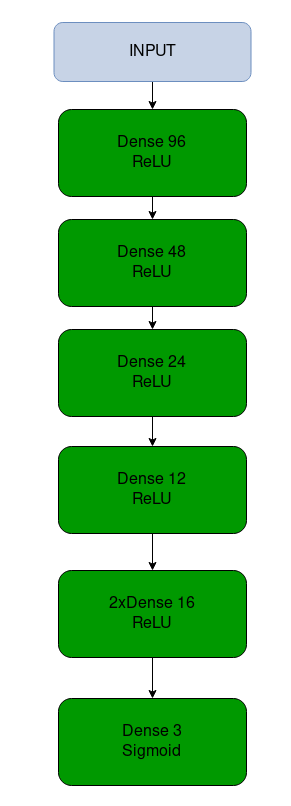}
    \caption{Architecture of the \fc baseline\label{fig:arch_fc}}
\end{figure}

\begin{figure}[h]
    \centering
    \includegraphics[height=.5\linewidth , angle =90]{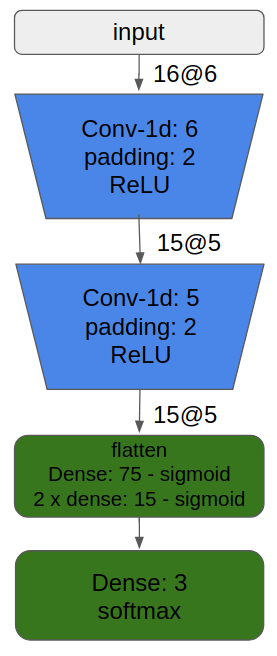}
    \caption{Architecture of the \cnn baseline\label{fig:arcch_cnn}}
\end{figure}

\subsection{Baseline models}
\label{subsec_CNN}
The use of this new architecture is justified only if it performs better than more common choice of networks. We thus incorporate as baselines two networks in our evaluation:
\begin{itemize}
    \item \fc: A \emph{Fully Connected} variant of \pced where all scales are connected (Figure~\ref{fig:arch_fc}).
    \item \cnn: A 1-d \emph{Convolutional Neural Network} with a number of layers similar to our \pced architecture (Figure~\ref{fig:arcch_cnn}).
\end{itemize}  

\paragraph{\fc}
This architecture presented in Figure~\ref{fig:arch_fc} allows us to measure the impact of the pairwise scale connection introduced in \pced.
\fc and \pced architectures differ by the four first layers: instead of consecutively combining scales by pairs, the \fc baseline combines all the scales in unique dense layers while keeping the same reduction rate, i.e. 16-8-4-2 scales.
We use the same number of neurons than \pced to get comparable architectures.
In total, the \fc baseline is comprised of about 6663 weights (three times more than \pced).
The input data is provided to the network by concatenating the sixteen 6-dimensional vectors used by \pced in a single 96-dimensional vector.

\paragraph{\cnn}
We chose a standard convolutional neural network architecture presented in Figure~\ref{fig:arcch_cnn}, composed of two blocks: a first block for representation learning with two 1D convolution layers, each followed by a ReLU activation function, and a second block dedicated to classification with three fully-connected layers and one 3D output layer. 
We tried a large number of variants and the one we present is the best we obtained by trials and errors.
The two convolution layers are composed of 15 filters, with a kernel size of $6$ and $5$, respectively.
Zero padding of size $2$ was used to get outputs of shape $15\times 5$. 
The three dense layers are composed of 75, 15 and 15 neurons, respectively, followed by the output layer with $3$ neurons. 
A sigmoid activation function is used after all the dense layers, except the output layer that uses a softmax non-linearity function. 
In total, \cnn is composed of about 8.7k weights.
The input data is also provided to the network as sixteen 6-dimensional vectors. 
Using the different scales as channels
enables a potentially larger expressivity power to the model since each convolution filter uses different weights to combine the scales.

\section{Experimental setup}
\label{sec:exp_set}

\begin{figure}[hb]
\centering
    \includegraphics[width=0.4\linewidth]{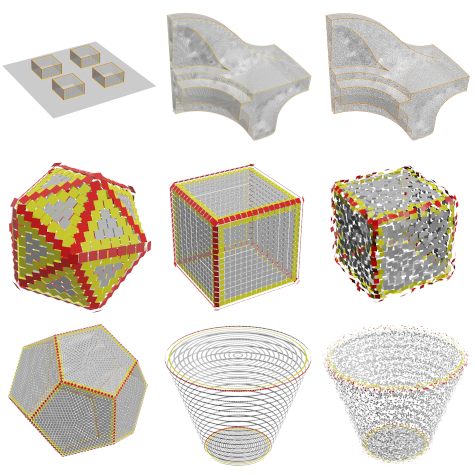}
    \caption{Training set of the dataset \dtsDFLT, with sharp edges points in red, and smooth edges points in yellow.}
    \label{fig:Tres}
\end{figure}

\subsection{Point cloud dataset}
\label{sec:training_set}

We measure the efficiency and adaptability of our network on three different datasets with ground truth for learning and evaluation: edges on point clouds with acquisition artifacts (\dtsDFLT), sharp edges on CAD models (\dtsABC) and annotated curves on challenging shapes (\dtsSHREC).
Each dataset is split in three sets: training set (denoted T, used for training), validation set (denoted V, used for learning to monitor accuracy) and evaluation set (denoted E).
Visual evaluations are also conducted on a set of acquired models.
Point clouds with classification results are shown in the website accompanying the submission.

In addition to 
annotated learning sets, we also show in Section~\ref{subsec:interactive} how \pced can learn from small set of data annotated interactively by a user and classify point clouds without requiring additional predefined training data.

\begin{figure}[tbh]
    \centering
    \includegraphics[width=0.7\linewidth]{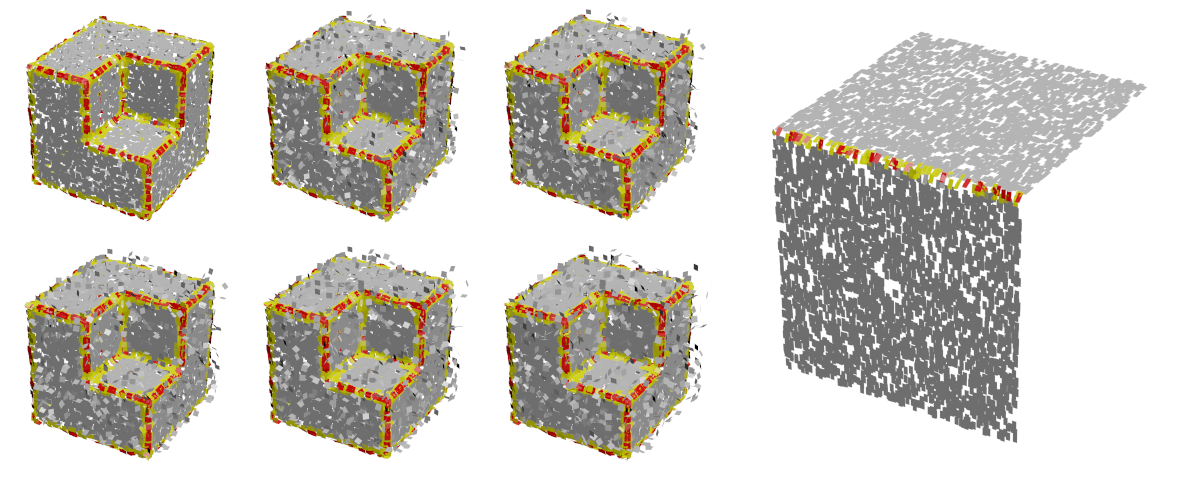}
    \caption{Evaluation set of the dataset \dtsDFLT. Left: \dataTwoCubes, with increasing Gaussian noise ($\sigma$ ranging between 0 and 0.14 units, cube edge length 1). Right: \dataAngle (90 degrees). \label{fig:testset1} }
\end{figure}

\paragraph{\dtsDFLT dataset}
We design this dataset to emulate geometric structures regularly found in acquired point clouds.
It is as small as possible, in order to demonstrate that \nico{a} few annotated data \nico{used for} training our network in a short time 
are enough for an effective detection \nico{(see timings in Table~\ref{table:trainingtimeours})}.
%
The \dtsDFLT{} dataset is composed of {$8$} synthetic 
{models} containing edges of different shapes illustrated in {Figure~\ref{fig:Tres} and~\ref{fig:testset1}, and detailed in Table~\ref{tab:training_stats}}.
{We also included some of our models characterized by varying densities and shapes (i.e. \dataCube, \dataCone, \dataFandisk and \dataTwoCubes), with different levels of noise generated by a random motion on the points following a Gaussian distribution with standard deviation denoted $\sigma$ (the different values of $\sigma$ per model are given in Table~\ref{tab:training_stats}-Noise).}
We defined the validation set by selecting a random subset within the training set with an equal number of points across the three classes  (1k per class\nico{,} 3k in total).
We also created an evaluation set with specific features to better evaluate the networks against noise and sharp angles. 
These point clouds, not used for training, are taken from the testing set introduced by \citet{Bazazian15} to evaluate the algorithm Covariance Analysis (denoted \ca in the following). 
{The models used for training, validation and evaluation are detailed in Table~\ref{tab:training_stats}-Usage.}
Table~\ref{tab:training_stats} presents quantitative details (including the values used for $\sigma$), and the point clouds are shown in Figures~\ref{fig:Tres},~\ref{fig:testset1} and in the joined website.

\begin{table}[]
\centering
{
\footnotesize
\setlength\tabcolsep{3.4pt} 
\begin{tabular}{|c|c|c|c|c|c|c|c|}
\hline
\multirow{2}{*}{Dataset}   & \multirow{2}{*}{Model}   & \multirow{2}{*}{Usage} & \multirow{2}{*}{\begin{tabular}[c]{@{}c@{}}\#k \\ points\end{tabular}} & \multicolumn{3}{c|}{\% points/class}                                  & \multirow{2}{*}{\begin{tabular}[c]{@{}c@{}}Noise\\ ($\sigma$)\end{tabular}} \\ \cline{5-7}
  &                         &                        &                                                                        & Edge                  & S-Edge                & Other                  &             \\ \hline
\multirow{9}{*}{\dtsDFLT{}}  & \dataCube                     & T, V                   & 1.5                                                                    & 12.2                  & 21.8                  & 65.9                  & 0-0.1       \\ \cline{2-8}
 & \dataHexagon                  & T, V                   & 0.6                                                                    & 34.6                  & 46.7                  & 18.6                  & -           \\ \cline{2-8}
 & \dataCone                     & T, V                   & 7.7                                                                    & 6.1                   & 11.7                  & 82.2                  & 0-0.1       \\ \cline{2-8}
 & \dataOctogon                  & T, V                   & 12                                                                     & 3.5                   & 7.7                   & 88.7                  & -           \\ \cline{2-8}
 & \dataFandisk                  & T, V                   & 106.5                                                                  & 5.5                   & 5                     & 89.4                  & 0-0.1       \\ \cline{2-8}
 & \dataFourCubes                  & T,V                    & 43.9                                                                   & 6.8                   & 6.6                   & 86.5                  & -           \\ \cline{2-8}
 & \multirow{3}{*}{\dataTwoCubes} & \multirow{3}{*}{E}     & \multirow{3}{*}{7.2}                                                   & \multirow{3}{*}{6}    & \multirow{3}{*}{17.3} & \multirow{3}{*}{76.6} & 0 - 0.01    \\  
   &                      &                        &                                                                        &                       &                       &                       & 0.02 - 0.03         \\
   &                      &                        &                                                                        &                       &                       &                       & 0.12 - 0.14         \\ \cline{2-8}
 & \dataAngle                & E                      & 6                                                                      & 0.6                   & 1.6                   & 97.8                  & -           \\ \hline \hline
 \multicolumn{2}{|c|}{}                  & T  &   287.9 &  9    & 11   &  80 & -           \\ \cline{3-8}
 \multicolumn{2}{|c|}{\dtsDFLT (total)}  & V  &  3    &  33.3    &  33.3   &  33.3   & -           \\ \cline{3-8}
 \multicolumn{2}{|c|}{}                 & E  & 55  & 4.9   & 14.54  & 80.56  & -           \\ \hline
 \multicolumn{2}{|c|}{}         & T  & 89.4   & 13.98   &  -  & 86.02  & -           \\ \cline{3-8}
 \multicolumn{2}{|c|}{\dtsSHREC}  & V  & 94.4   & 8.68   & -  & 91.32  & -           \\ \cline{3-8}
 \multicolumn{2}{|c|}{}         & E  & 654.7   & 10.57  & -  & 89.43  & -           \\ \hline
 \multicolumn{2}{|c|}{}         & T  &  3212.4  & 4.66   & -  & 95.34  & -           \\ \cline{3-8}
 \multicolumn{2}{|c|}{\dtsABC}  & V  & 690   & 5.79   & -  & 94.21  & -           \\ \cline{3-8}
 \multicolumn{2}{|c|}{}         & E  & 312314   & 5.52   & -  & 94.48  & \nico{0 - 0.04}           \\ \hline
\end{tabular}
}
\caption{
Statistics of the datasets used for training (T), validation (V) and evaluation (E). The Edge column gives the percentage of points of the class sharp-edge, the S-Edge, the one of the class smooth-edge and Other is for the non-edge class. 
\label{tab:training_stats}
}
\end{table}

\paragraph{\dtsABC dataset}
We use the \dtsABC dataset~\cite{Koch2019ABCAB} to evaluate the performance of our network on the detection of sharp edges in CAD models.
We use the chunk \texttt{000}, which contains 7167 models represented as OBJ files. 
We generate the point clouds by exporting the vertices and the normal vectors of the meshes.
The ground truth classification is produced using the vertices associated to a sharp feature in the feature files.
A notable difference with the \dtsDFLT dataset is that no ground truth information is provided for the 
smooth-edge
label.
Only sharp edges are considered and other smoother features are thus labeled as 
non-edge.
We define the training and evaluation sets by randomly selecting 200 and 50 models respectively (see the model list in the joined website).

\paragraph{\dtsSHREC dataset}
We use the dataset produced for feature curve extraction by~\citet{Thompson:2019} in order to evaluate the performance of our network on challenging data annotated by humans.
As for \dtsABC, we consider the vertices and normal vectors of the given meshes, and mark the annotated vertices as edge  vertices.
Similarly, we define the training and evaluation sets by randomly selecting models, e.g., (M5, M6 and M14) and (M2, M11) respectively.
This procedure is not ideal as these models exhibit very heterogeneous edge shapes, but we found it fairer than randomly sampling the point clouds.

\paragraph{Acquired point clouds}
We also perform visual evaluations on $9$ acquired point clouds whose number of vertices are reported in Table~\ref{tab:visualTime}.
\dataLoudunSmall and \dataLoudunBig are down-sampled versions (of respectively 1 and 35 million vertices) of a photogrammetric acquisition of the Square Tower of Loudun (France).
We used \dataEuler, \dataEmpire and \dataLans as provided by \citet{MonszpartEtAl:RAPter:2015}, 
\dataPisa as provided by \citet{mellado2015relative}, 
\dataMunich as provided by~\citet{Hackel_2016_CVPR}, 
\revision{\dataRueMadame as provided by~\citet{serna:hal-00963812}, }
and we downloaded \dataChurch and \dataTrainStation from Sketchfab\footnote{\dataChurch: \url{https://sketchfab.com/3d-models/christ-church-and-dublin-city-council-b5f6bcce8ebc44a3b4bbb6b0fef067b3}. \dataTrainStation: \url{ https://sketchfab.com/3d-models/station-rer-6c636ca4793345e8ae12beb97b7d6359}}.

\paragraph{Normal estimation}
The normals used for the GLS computation are either directly those estimated with the acquisition technique (e.g. photogrammetry, laser scanner), which are actually provided for most current acquisition devices, or those computed automatically from the point samples (we used Meshlab~\cite{Meshlab} to estimate normals on \dtsDFLT dataset and on \dataEuler, \dataLans, \dataChurch and \dataTrainStation).
We did not encounter any difficulty related to normal estimation. 
Eventually, oriented normals may be avoided and replaced by unoriented normals~\cite{Jiazhou:2013}, which could be easily estimated using local fitting.



\subsection{Networks training}

\paragraph{Implementation details}
We implemented \pced in C++ and ran our experiments on a \nico{10-cores Intel Xeon(R) CPU E5-2640 v4 (20 threads)}, with 128GB RAM. 
We use the Ponca library~\cite{Ponca} for surface fitting and derivative computation. 
\lo{These surface fitting computations (GLS) are performed on two 10-cores Intel Xeon(R) CPU E5}. 
The network modeling and evaluation for \pced and \fc is implemented in our own prototype using Eigen~\cite{eigen} for linear algebra. 
The baseline \cnn is implemented, trained and evaluated using Pytorch.

Categorical cross-entropy is used as objective function for both the baseline architectures and \pced. 
On each dataset, \cnn, \fc and \pced are trained for 200, 200 and 40 epochs respectively, all reaching 98\% accuracy on the validation sets. 
Weights are initialized randomly using the~\citet{glorot2010understanding} method.
For the three architectures, learning rate is set to $0.01$, momentum to $0.9$, with batch size of $100$ points.

\begin{figure}[t]
{
    \centering
    \footnotesize
    \def\svgwidth{0.25\linewidth}
    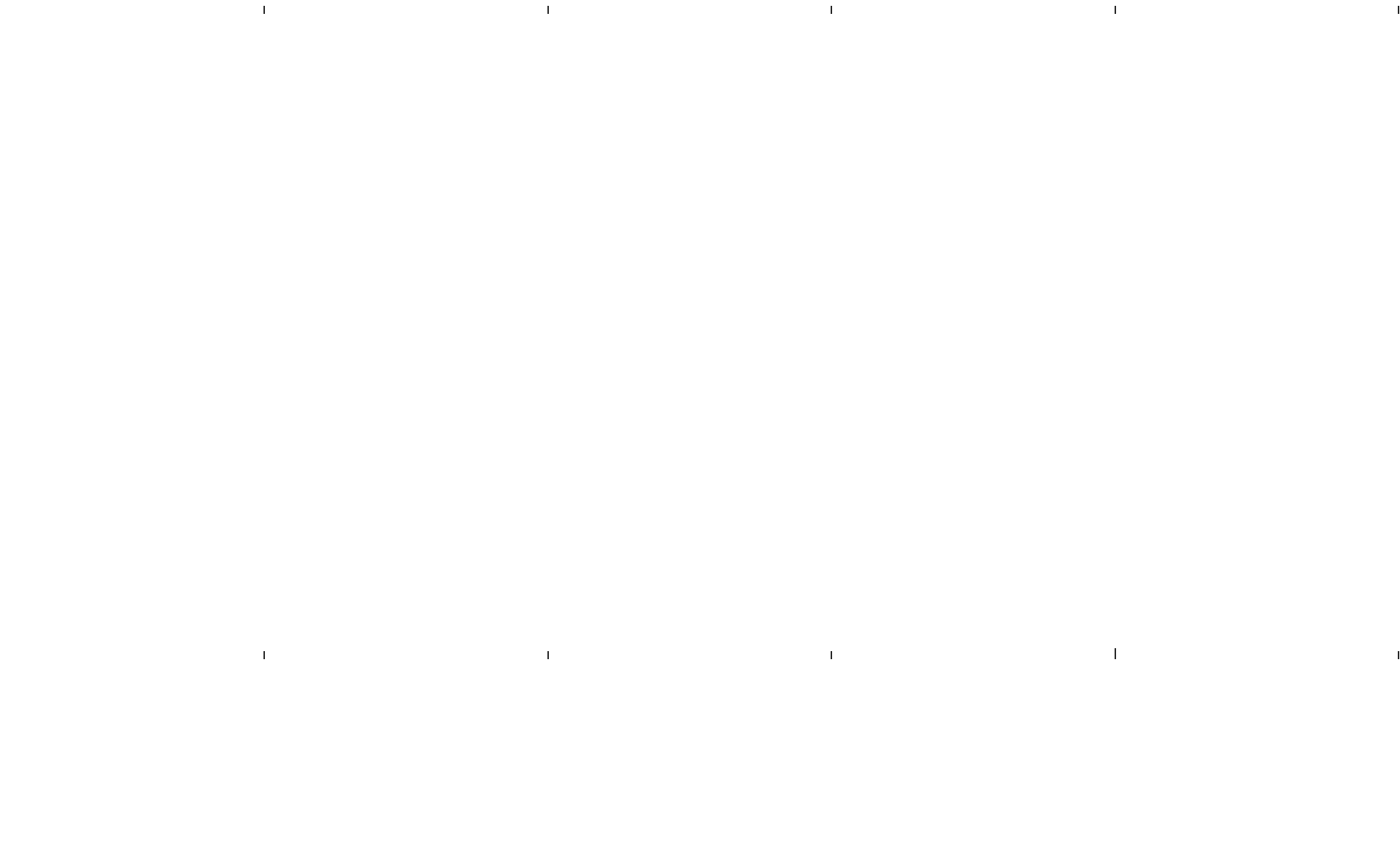 
    \hspace{5mm}
    \def\svgwidth{0.25\linewidth}
    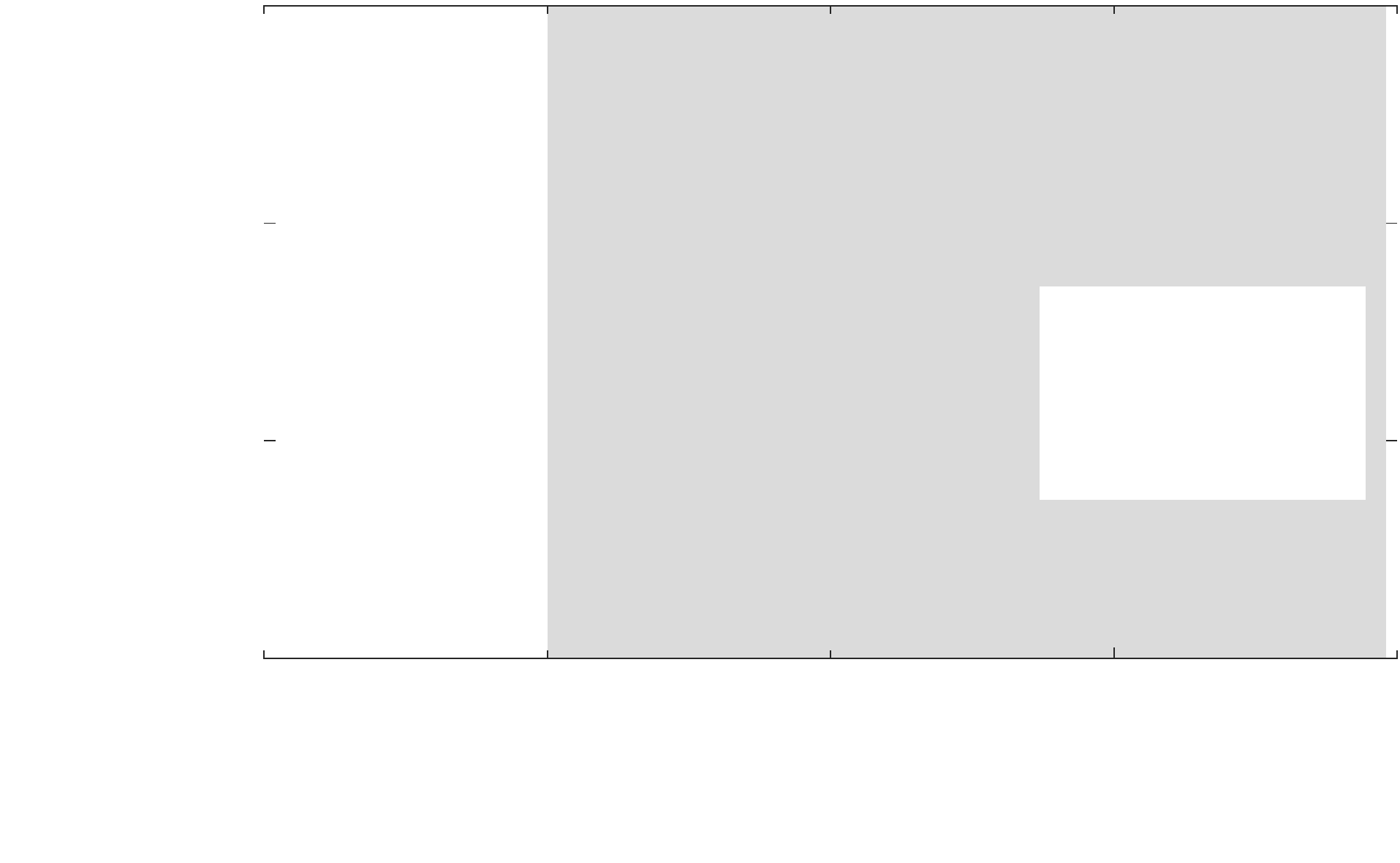 
    \\
    \def\svgwidth{0.25\linewidth}
    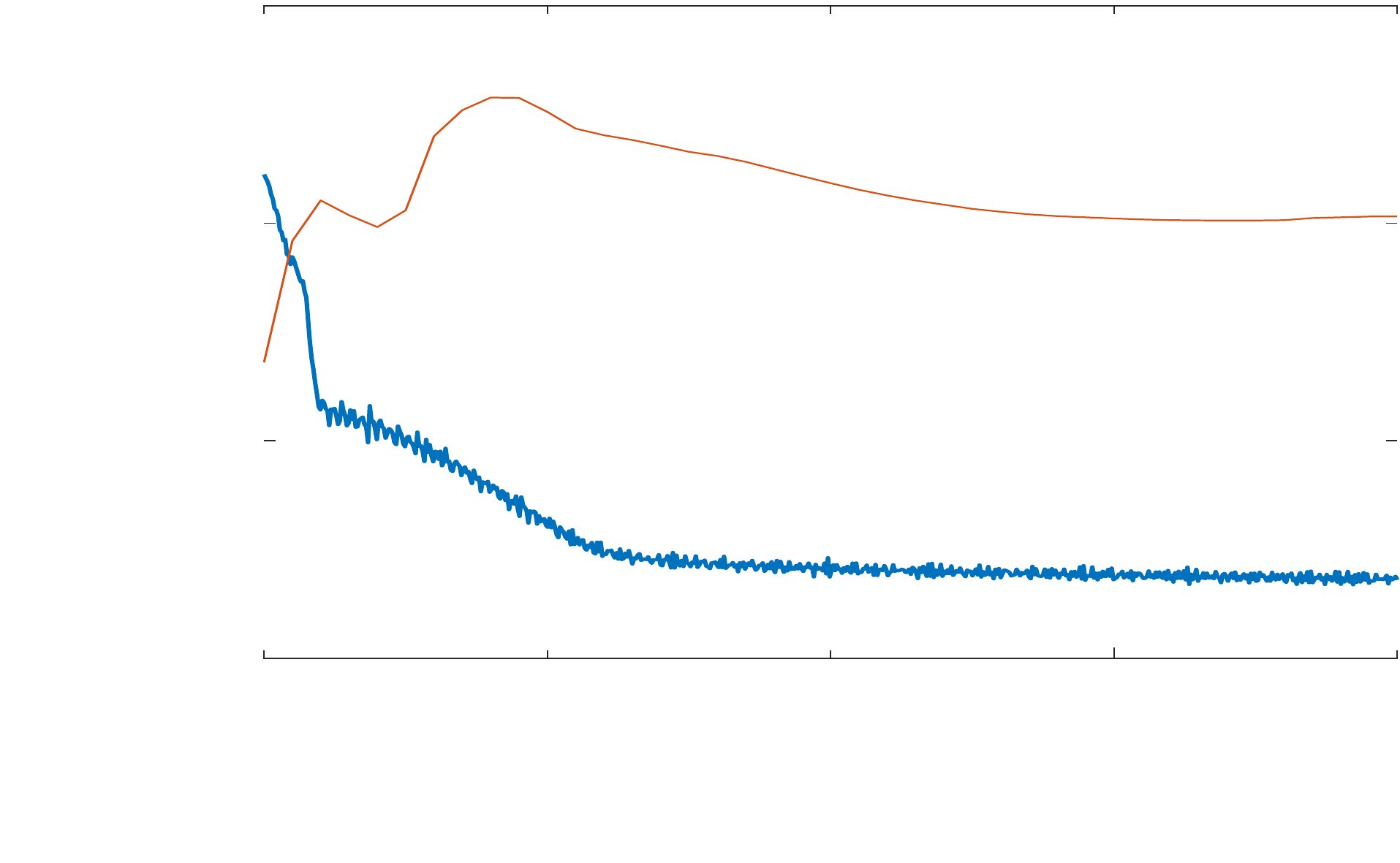 
    \hspace{5mm}
    \def\svgwidth{0.25\linewidth}
    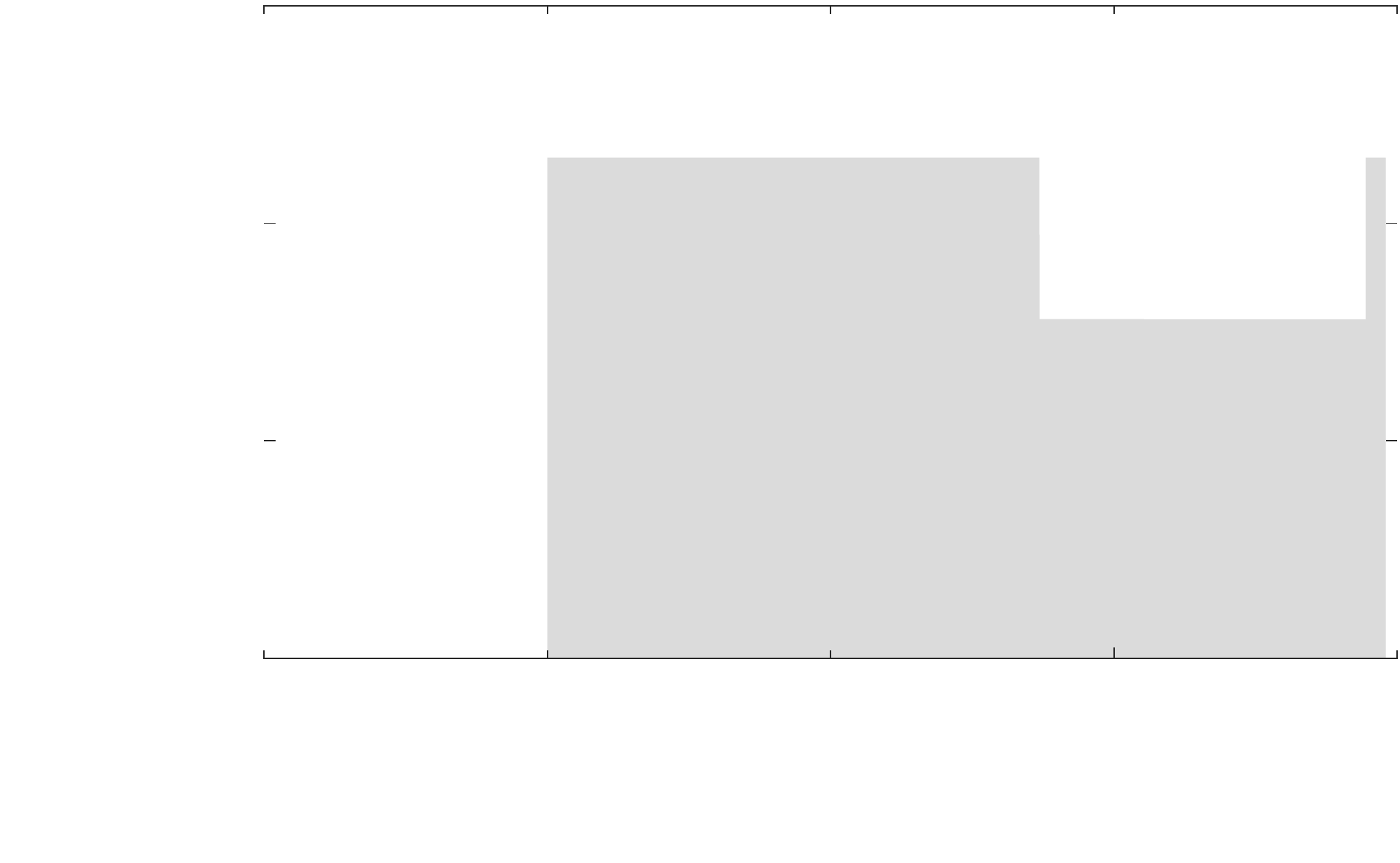  
    
}
    \caption{\pced: learning curves measured on Training (T) and Validation (V) sets on the (left) \dtsDFLT and (right) \dtsABC. For the latter, learning stops automatically at epoch 10, when loss and accuracy stop to be improved.
    \label{fig:lc}
    }
\end{figure}

\paragraph{Training}
As illustrated in Table~\ref{tab:training_stats}, the three training sets contain mostly non-edge surfaces.
The small number of sharp and smooth edge samples (4-14\% of the points) implies that we are in presence of a highly unbalanced training set biased towards the non-edge class. 
We handle this issue by generating balanced batches of points during training \lo{and validation}.
Batches are balanced by replicating sharp and smooth edge points until the number of points in each class is equal.
We recorded training curves for \pced on \dtsDFLT{} and \dtsABC datasets (see Figure~\ref{fig:lc} and Table~\ref{table:trainingtime}). 
As we can see in \lo{Figure~\ref{fig:lc}-bottom-left, the behavior of the validation loss may appear unusual, even though in practice, it is sometimes observed in machine learning. During the first epochs, starting from a random initialization, the network does not generalize yet on the validation set, which may result in an increase of the loss. Then, during the following epochs, even though decreasing, the loss may remain higher than its value at the random initialization, which may be due to the strong impact of wrong predictions on the loss value. The validation accuracy keeps increasing during training, probably because it is less sensitive to wrong predictions than the loss.}
Overall, training usually takes a couple of minutes for \pced, a dozen for the baseline \fc, and more than one hour for \cnn.

\begin{table}[h]
\centering
\setlength\tabcolsep{4pt} 
\begin{tabular}{|l|c|c|c|}
\cline{2-4}
 \multicolumn{1}{c|}{}        
& GLS              & Training          & Total  \\ 
\cline{2-4}
\hline
\pced(\dtsDFLT)    & \duration{19.32} & \duration{172.4}  & \duration{\Sum{19.32}{172.4}}    \\ \hline
\fc(\dtsDFLT)     & \duration{19.32} & \duration{838.06} & \duration{\Sum{19.32}{838.06}}   \\ \hline
\cnn(\dtsDFLT)    & \duration{19.32} & \duration{4748.28}& \duration{\Sum{19.32}{4748.28}}  \\ \hline \hline
{\pcedtc(\dtsABC)}    & \duration{131}   & \duration{1200}   & \duration{\Sum{131}{1200}}       \\ \hline
{\fctc(\dtsABC) }     & \duration{131}   & \duration{7265.35} & \duration{\Sum{131}{7265.35}}   \\ \hline
{\cnntc(\dtsABC)   }  & \duration{131}   & \duration{32400}  & \duration{\Sum{131}{32400}}      \\ \hline \hline
{\pcedtc(\dtsSHREC)}  & \duration{4}     & \duration{28.09}  & \duration{\Sum{4}{28.09}}        \\ \hline
{\fctc(\dtsSHREC) }   & \duration{4}     & \duration{603.26} & \duration{\Sum{4}{603.26}}       \\ \hline
{\cnntc(\dtsSHREC) }  & \duration{4}     & \duration{1561.71}& \duration{\Sum{4}{1561.71}}    \\   \hline
\end{tabular}
\caption{
Training times of our networks on the different datasets.
\label{table:trainingtimeours}
}
\end{table}


\section{Results}
\label{sec_results}

In this Section, we first validate our choice of input parameters with an ablation study (Section~\ref{subsec:ablation}) {and discuss the number of input scales (Section~\ref{subsec:numberOfScales}). 
We then compare our network with our baseline (i.e. \cnn and \fc), with \ecnet~\cite{ecnet2018} and \pcp~\cite{GuerreroEtAl:PCPNet:EG:2018} and two geometric feature detection methods, i.e. the covariance analysis method proposed by~\citet{Bazazian15} (\ca) and the Feature Edge Estimation (\fee) implemented in the CGAL Library~\cite{cgal:ass-psp-21a,Merigot2011}.
\lo{We also report the evaluation of \pienet and \ecnet performed on the \dtsABC dataset by~\citet{PieNet2020}.}
For clarity, we denote \texttt{A}(\texttt{D}) the approach \texttt{A} trained on dataset \texttt{D}, e.g., \pced trained on \dtsDFLT is denoted \pced(\dtsDFLT).
We present training and classification times (Section~\ref{subsec:timings}), \lo{quantitative}
comparisons (Section~\ref{subsec:comparisons}), visual comparisons (Section~\ref{subsec:visualEvaluation}), networks behavior on noisy data (Section~\ref{subsec:noise}), the way our \pced can be used for interactive learning (\lo{Section}~\ref{subsec:interactive}) and some complementary experiments (sensibility to variation in sampling, choice of the maximum scale and choice of the surface reconstruction method - Section~\ref{sec:compexp}).}
All of the Figures of the next sections are available in full resolution on the joined website homepage, with left/right interactive comparisons between the methods.
Results of the quantitative experiments on each dataset are documented in dedicated webpages presenting interactive distribution plots, histograms, tables, and a 3D point cloud viewer.

\paragraph{Comparison metrics}
We compare positive and negative matches of the classifications w.r.t. ground truth by measuring True Positives (TP), True Negatives (TN), False Positives (FP) and False Negatives (FN).
To do so, we consider the following metrics: Precision, Recall, F1-Score, \lo{Intersection over Union (IoU),} Matthews Correlation Coefficient (MCC), and Accuracy, whose formulations are given in Appendix~\ref{app:scores}.
Note that F1 and \lo{IoU scores ignore} the True Negatives and thus 
{may be} misleading for unbalanced classes, which is our case in general, i.e. the number of TP edge points is in general very small in comparison to the TN non-edge points.
It is however reported for comparison as F1 and \lo{IoU are} commonly used in similar contexts {and it still provides a score aggregating Precision and Recall}. \lo{Also note that the IoU metric is known to be very correlated to the F1 score (see their Equations in Appendix~\ref{app:scores}) and they differ as the IoU is more sensitive to high errors (i.e. it is more affected by worst cases) while the F1 score tends to measure an average performance.}
In our case (i.e. with \lo{a} low number of highly unbalanced classes), \lo{the number of TP significantly influences the result evaluation and}~\citet{Chicco2020TheAO}  
\lo{explain in details how a measure of the correlation between observation and prediction relying on all TP, TN, FP, FN, as the MCC, better represent the quality of the network classification.}

\paragraph{Comparison of classifications with two and three classes}
As previously stated, \pced outputs three classes corresponding to sharp-edges, smooth-edges and non-edges.  
However if edges are defined as sharp features only (\lo{e.g.}
the \dtsABC dataset), results from the smooth-edge class does not bring significant quantitative information and only the sharp-edge class is of interest.
In that case, we consider the sharp-edge as the unique positive class, and merge the smooth-edge and non-edge classes as the negative class.
When a fuzzier definition is used for edges, as for the \dtsSHREC dataset where both rounded and sharp edges co-exists, we present additional results where the positive class is defined either as sharp edge only, or, 
as the union of sharp-edges and smooth-edges classes.
Accordingly, three-classes approaches are trained on two-classes datasets as if no vertex is labelled as smooth-edge, but only has sharp-edge or non-edge.


\subsection{Ablation study}
\label{subsec:ablation}
In this Section, we validate both the choice of input parameters and the number of output classes (i.e two or three) of \pced.
Quantitative results are given respectively in Tables~\ref{tab:tablecompexp1} and~\ref{tab:tablecompexp2}.

\begin{table}[t]
\centering
{
\footnotesize
\setlength\tabcolsep{3.3pt} 
    \begin{tabular}{|c|c|c|c|c|c|c|c|c|}
            \cline{3-9}
            \multicolumn{2}{c|}{}  &\multicolumn{6}{c|}{\dataTwoCubes ($\sigma$)}  &\multirow{2}{*}{\dataAngle} \\
            \cline{3-8}
            \multicolumn{2}{c|}{}  & 0 & 0.01 & 0.02 & 0.03 & 0.12 & 0.14 & \\
            \hline
            All GLS            & P  & 0.529          & 0.305          & 0.348          & 0.413          & 0.310          & 0.340          & 0.327          \\
            derivatives        & R  & 0.415          & 0.424          & 0.454          & 0.481          & 0.427          & 0.45           & 0.5            \\
            (28 parameters)    & F1 & 0.465          & 0.355          & 0.394          & 0.445          & 0.360          & 0.387          & 0.396          \\
                               & MCC& 0.435          & 0.310          & 0.353          & 0.407          & 0.315          & 0.345          & 0.400          \\
            \hline
            Invariant to rigid & P  & \textbf{0.983} & \textbf{0.839} & \textbf{0.866} & \textbf{0.882} & \textbf{0.845} & \textbf{0.853} & \textbf{0.871} \\
             transformations   & R  & 0.823          & 0.904          & 0.921          & 0.912          & 0.904          & 0.919          & \textbf{0.984} \\
            (7 parameters)     & F1 & 0.897          & 0.871          & 0.893          & 0.897          & 0.874          & 0.885          & \textbf{0.925} \\
                               & MCC& 0.870          & 0.828          & 0.858          & 0.864          & 0.832          & 0.848          & \textbf{0.924} \\
            \hline
            Our                & P  & 0.966          & 0.807          & 0.832          & 0.868          & 0.817          & 0.826          & 0.724          \\
            configuration      & R  & \textbf{0.897} & \textbf{0.960} & \textbf{0.970} & \textbf{0.934} & \textbf{0.961} & \textbf{0.964} & \textbf{0.984} \\
            (6 parameters)     & F1 & \textbf{0.931} & \textbf{0.877} & \textbf{0.896} & \textbf{0.901} & \textbf{0.884} & \textbf{0.890} & 0.835          \\
                               & MCC& \textbf{0.908} & \textbf{0.839} & \textbf{0.864} & \textbf{0.869} & \textbf{0.847} & \textbf{0.856} & 0.840          \\
            \hline
            Remove             & P  & 0.360          & 0.217          & 0.250          & 0.265          & 0.217          & 0.233          & 0.428          \\
            scale              & R  & 0.340          & 0.489          & 0.513          & 0.696          & 0.475          & 0.495          & 0.416          \\
            (4 parameters)     & F1 & 0.350          & 0.301          & 0.337          & 0.385          & 0.298          & 0.317          & 0.423          \\
                               & MCC& 0.305          & 0.260          & 0.299          & 0.373          & 0.256          & 0.277          & 0.419          \\
            \hline
            Remove             & P  & 0.173          & 0.103          & 0.115          & 0.355          & 0.105          & 0.109          & 0.242          \\
            curvature          & R  & 0.292          & 0.290          & 0.277          & 0.783          & 0.290          & 0.286          & 0.222          \\
            (3 parameters)     & F1 & 0.217          & 0.152          & 0.163          & 0.489          & 0.155          & 0.158          & 0.232          \\
                               & MCC& 0.152          & 0.079          & 0.094          & 0.484          & 0.084          & 0.087          & 0.227          \\
             \hline
    \end{tabular}
}
    \caption{
    Precision, recall, F1 and MCC computed on the \dataTwoCubes (with a Gaussian noise generated with different value of $\sigma$) and \dataAngle models of the \dtsDFLT evaluation set (see Table~\ref{tab:training_stats}) for the different parameterization of \pced.
    \label{tab:tablecompexp1}
    }
\end{table}

\paragraph{SSM}
Our Scale-Space Matrix is composed of $6$ parameters computed by differentiating the GLS implicit surface both in scale and space.
In order to measure the relevance of our parameterization, we tested four alternative sets of parameters, ranging from $28$ to $3$ parameters per scale.
For each scenario, the \pced architecture is modified as follows: the number of weights of the four first layers is set according to the number of input parameters, as well as the connection between the fourth and fifth layers.
We define the different parameter sets as follows: 
\begin{itemize}
    \item $28$ parameters: taking all the derivatives of the GLS descriptor in scale and space,
    \item $7$ parameters: keeping only the derivatives that are invariant to rigid transformations (rotation, translation and scale), leading to $\begin{bmatrix}
    \tau_i^t & 
    \kappa_i^t & 
    k1_i^t & 
    k2_i^t & 
    \frac{\delta \tau_i^t}{\delta t} & 
    \frac{\delta \kappa_i^t}{\delta t} & 
    \phi(S^t_{\mathbf{p}_i})
    \end{bmatrix}$,
    \item $6$ parameters (selected \pced input): removing $k2$, which is linearly dependent to $k1$ and $\kappa$,
    \item $4$ parameters: removing scale information from \pced input, leading to $\begin{bmatrix}
    \tau_i^t & 
    \kappa_i^t & 
    k1_i^t & 
    \phi(S^t_{\mathbf{p}_i})
    \end{bmatrix}$
    \item $3$ parameters: removing curvature information from \pced input, leading to $\begin{bmatrix}
    \tau_i^t & 
    \frac{\delta \tau_i^t}{\delta t} & 
    \phi(S^t_{\mathbf{p}_i})
    \end{bmatrix}$.
\end{itemize}
We trained these five parameter configurations on \dtsDFLT and reported in Table~\ref{tab:tablecompexp1} the Precision, Recall and F1 scores when applied on the evaluation set.
As we compare 3-classes approaches on a 3-classes dataset in this experiment, we sum-up TP and FP for both the sharp-edge and smooth-edge labels, and use non-edge for negatives.
Our choice of $6$ parameters is validated as it always has the higher Recall, F1 and MCC, while providing a good precision in comparison with the other sets of parameters.

\paragraph{Number of output classes}
We measure the impact of the smooth-edge class on the quality of the sharp-edge classification
by implementing a version, denoted \pcedtc, in which the decision layers return scores for $2$ classes instead of $3$.
We trained \pcedtc on the \dtsDFLT dataset by considering points labelled as sharp-edge for one class and
as the union 
of non-edge and smooth-edge for the other class.
We report in Table~\ref{tab:tablecompexp2} the scores obtained by \pced and \pcedtc on the evaluation set.
We can see that the addition of the third class allows \pced to obtain higher scores than \pcedtc, for all the metrics.
Note that one class could have been the union of sharp and smooth edges and the other the non-edges. This increases the false positive rate and leads to the same best parameterization, with overall lower performance.
{This gain in performance can be due to the unequal sampling of point clouds and the lack of robustness of geometric thresholding, that makes the accurate labeling of edges very tedious. 
The addition of the smooth-edge class enables a more conservative labeling of points.}

\begin{table}[t]
\centering
{
\footnotesize
\setlength\tabcolsep{4.1pt} 
    \begin{tabular}{|c|c|c|c|c|c|c|c|c|}
    \cline{3-9}
            \multicolumn{2}{c|}{}  &\multicolumn{6}{c|}{\dataTwoCubes ($\sigma$)}  &\multirow{2}{*}{\dataAngle} \\
            \cline{3-8}
            \multicolumn{2}{c|}{}  & 0 & 0.01 & 0.02 & 0.03 & 0.12 & 0.14 & \\
            \hline
            \pcedtc     & P  & 0.576          & 0.298          & 0.366          & 0.364          & 0.308          & 0.330          & 0.435         \\
            (\dtsDFLT)  & R  & 0.564          & 0.610          & 0.652          & 0.689          & 0.611          & 0.614          & 0.278         \\
                        & F1 & 0.570          & 0.400          & 0.469          & 0.477          & 0.410          & 0.430          & 0.339         \\
                        & MCC& 0.539          & 0.374          & 0.445          & 0.458          & 0.383          & 0.402          & 0.344         \\
            \hline
            \pced      & P  & \textbf{0.966} & \textbf{0.807} & \textbf{0.832} & \textbf{0.868} & \textbf{0.817} & \textbf{0.826} & \textbf{0.724} \\
            (\dtsDFLT) & R  & \textbf{0.897} & \textbf{0.960} & \textbf{0.970} & \textbf{0.934} & \textbf{0.961} & \textbf{0.964} & \textbf{0.984} \\
            (Ours)     & F1 & \textbf{0.931} & \textbf{0.877} & \textbf{0.896} & \textbf{0.901} & \textbf{0.884} & \textbf{0.890} & \textbf{0.835} \\
                       & MCC& \textbf{0.908} & \textbf{0.839} & \textbf{0.864} & \textbf{0.869} & \textbf{0.847} & \textbf{0.856} & \textbf{0.840} \\
            \hline

    \end{tabular}
}
    \caption{Precision, Recall, F1 and MCC computed on the \dataTwoCubes (with different Gaussian noise) and \dataAngle models of the \dtsDFLT evaluation set (presented in Table~\ref{tab:training_stats}) for \pced and its two-classes implementation \pcedtc. As we can see, the addition of the third class improves the classification quality of the sharp-edge class.
    \label{tab:tablecompexp2} 
    }
\end{table}

{
\subsection{Different number of input scales}
\label{subsec:numberOfScales}
\begin{figure*}[htb]
\includegraphics[width = \linewidth]{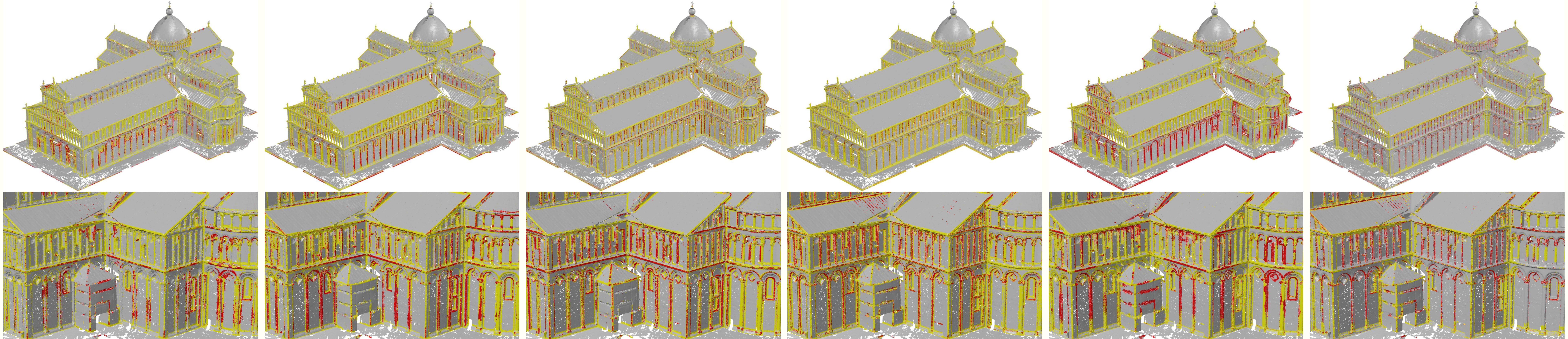}\\
\includegraphics[width = \linewidth]{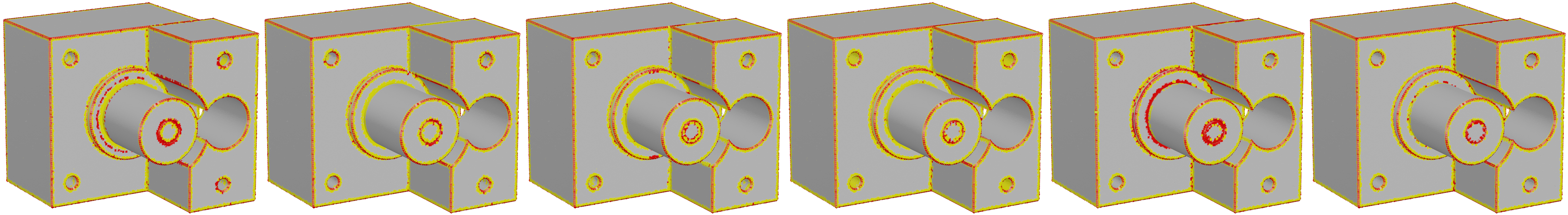}
\caption{{Different classifications produced by \pced with, from left to right, $4, 8, 16, 32, 64$ and $128$ input scales, on \lo{the} \dataPisa model (two top rows) and model $7024$ of the \dtsABC dataset (bottom row).}
}
\label{fig:differentScales}
\end{figure*}

\begin{figure*}
    \centering
    \scriptsize
    \def\svgwidth{0.158\linewidth} 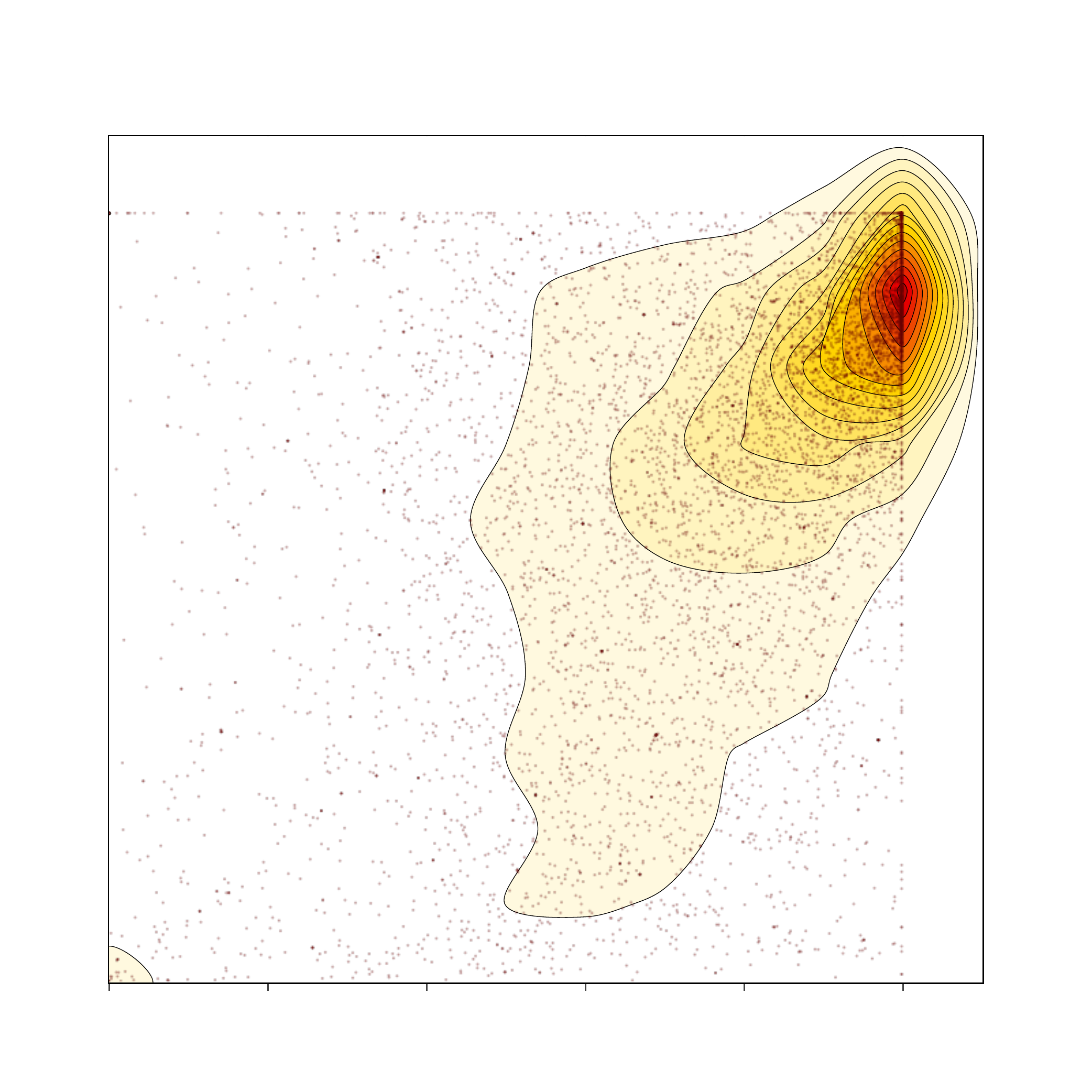
    \def\svgwidth{0.158\linewidth} 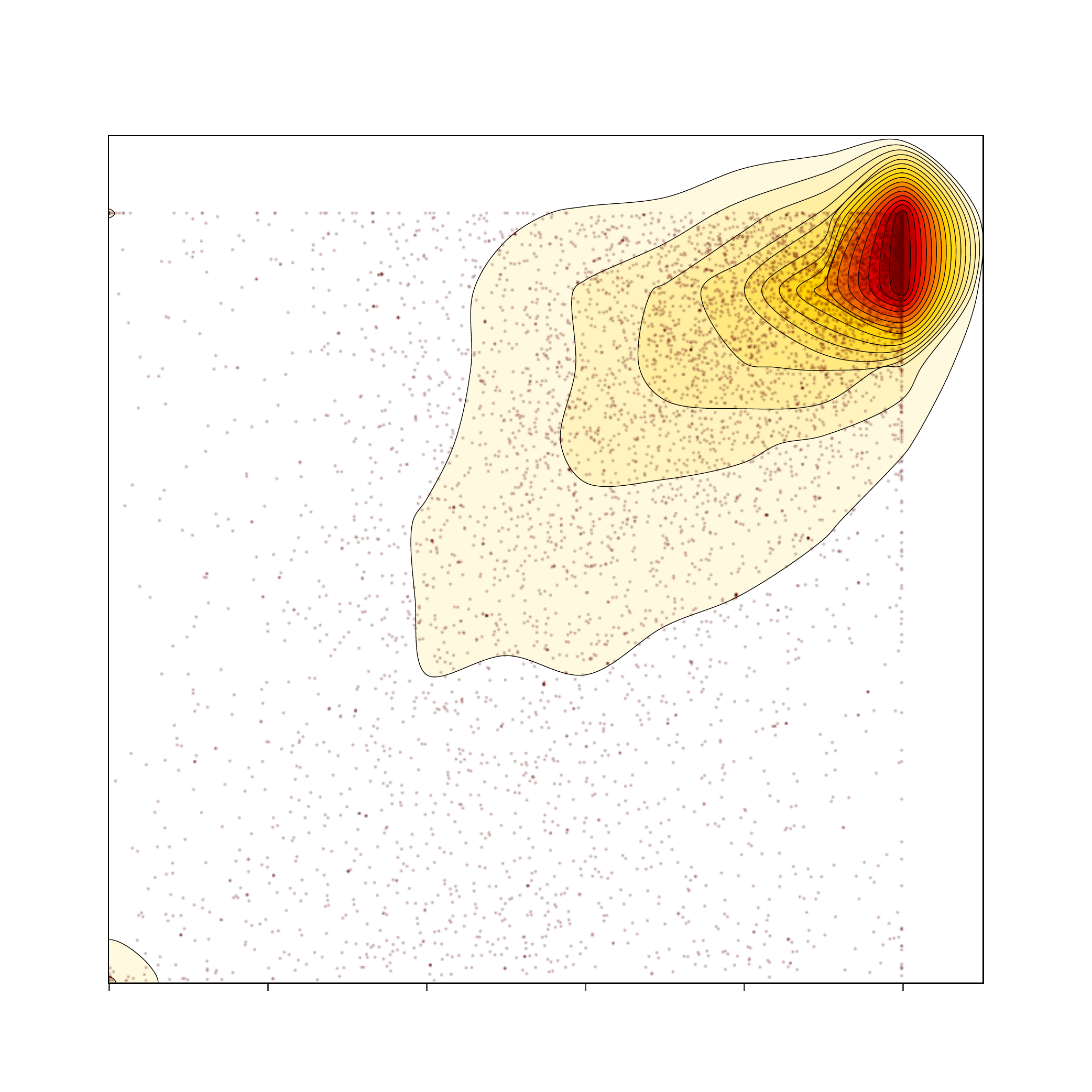
    \fbox[trlb]{
    \def\svgwidth{0.158\linewidth} 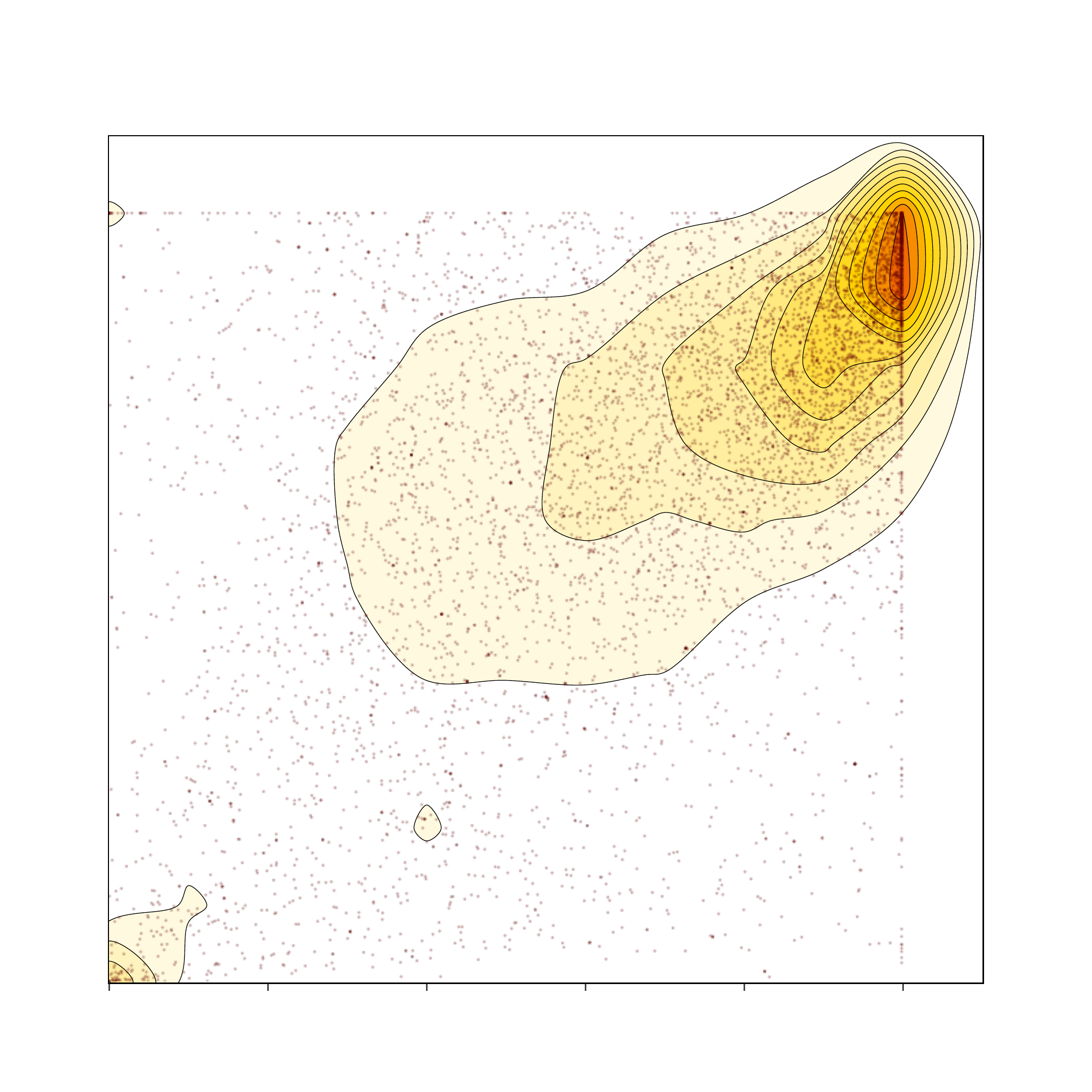
    }
    \def\svgwidth{0.158\linewidth} 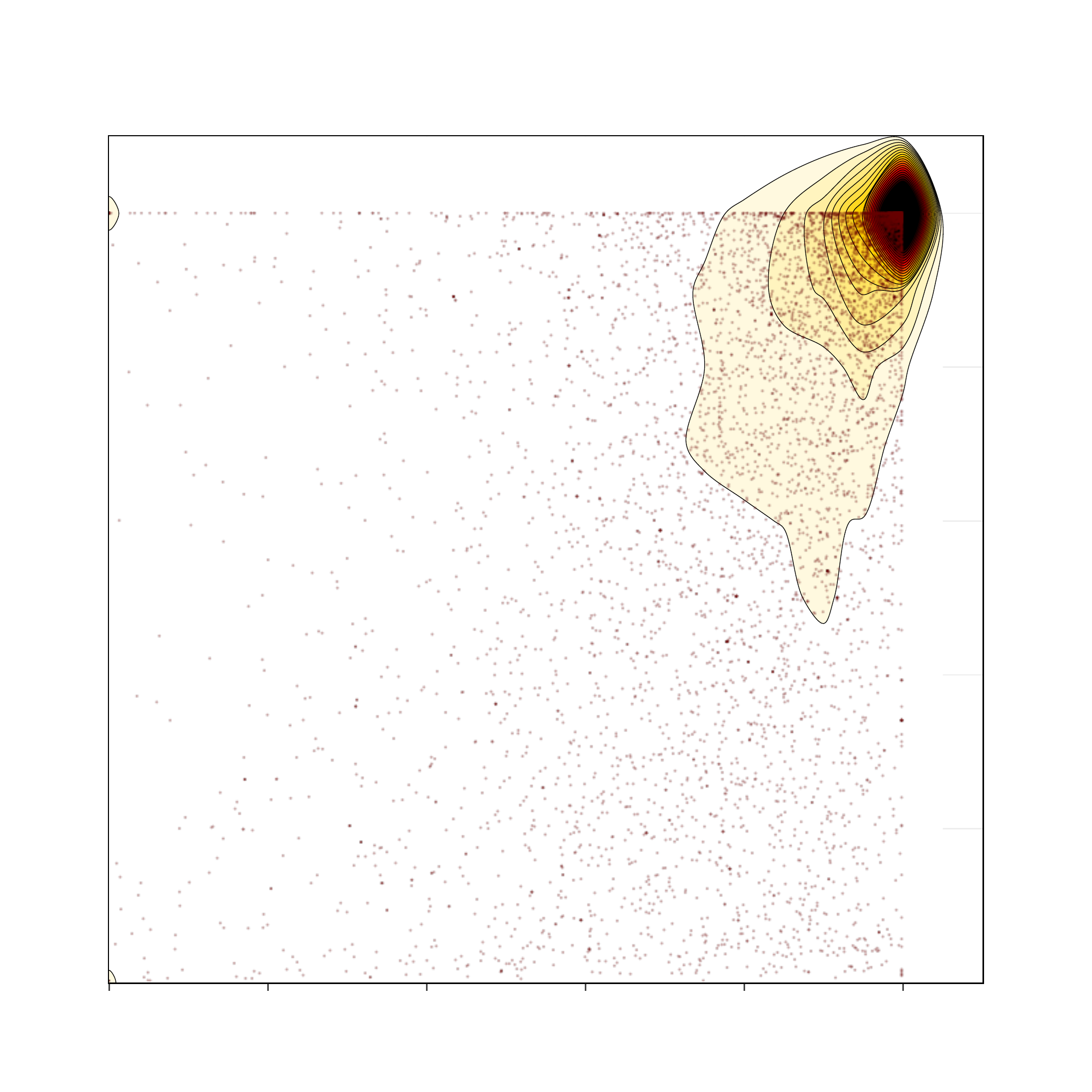
    \def\svgwidth{0.158\linewidth} 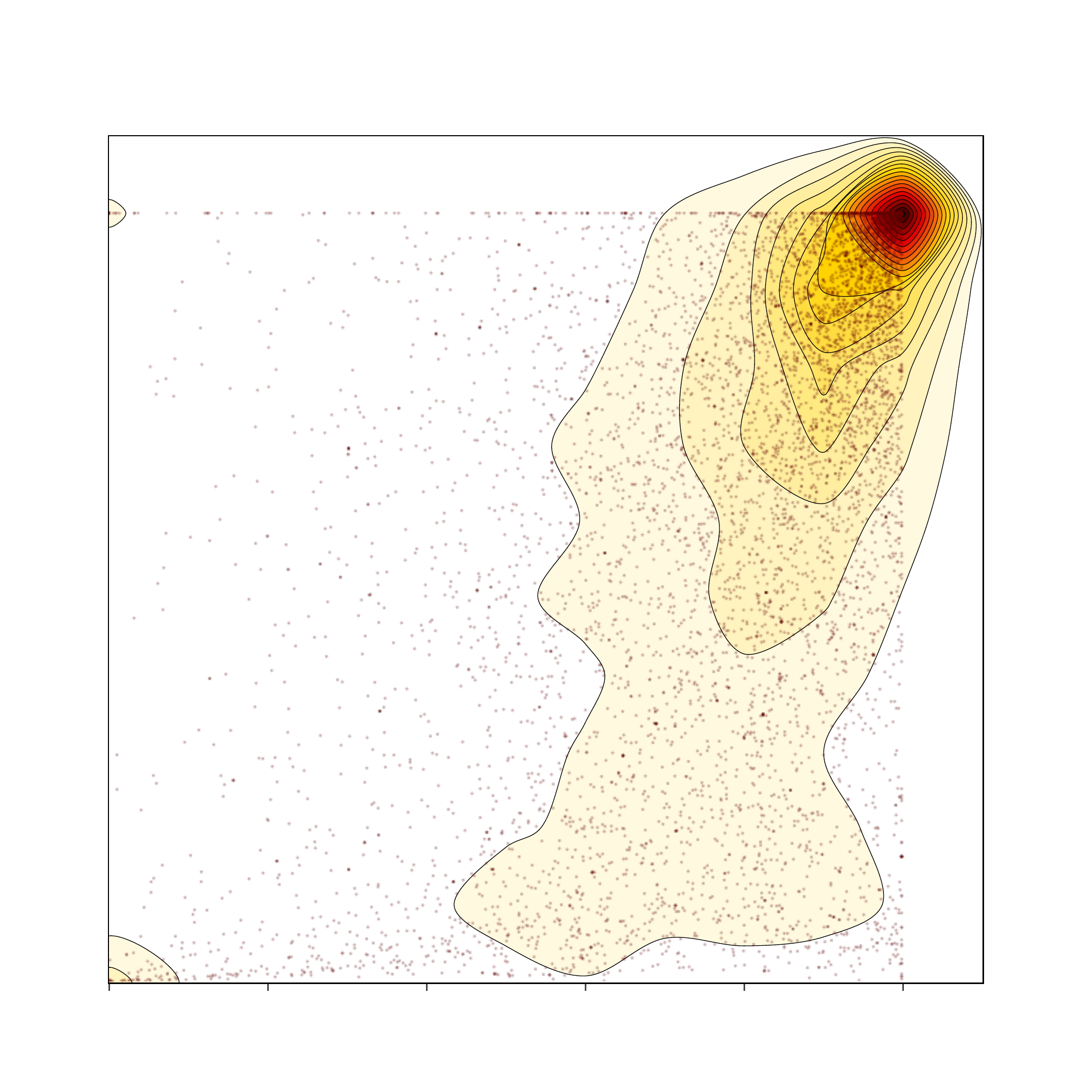
    \def\svgwidth{0.158\linewidth} 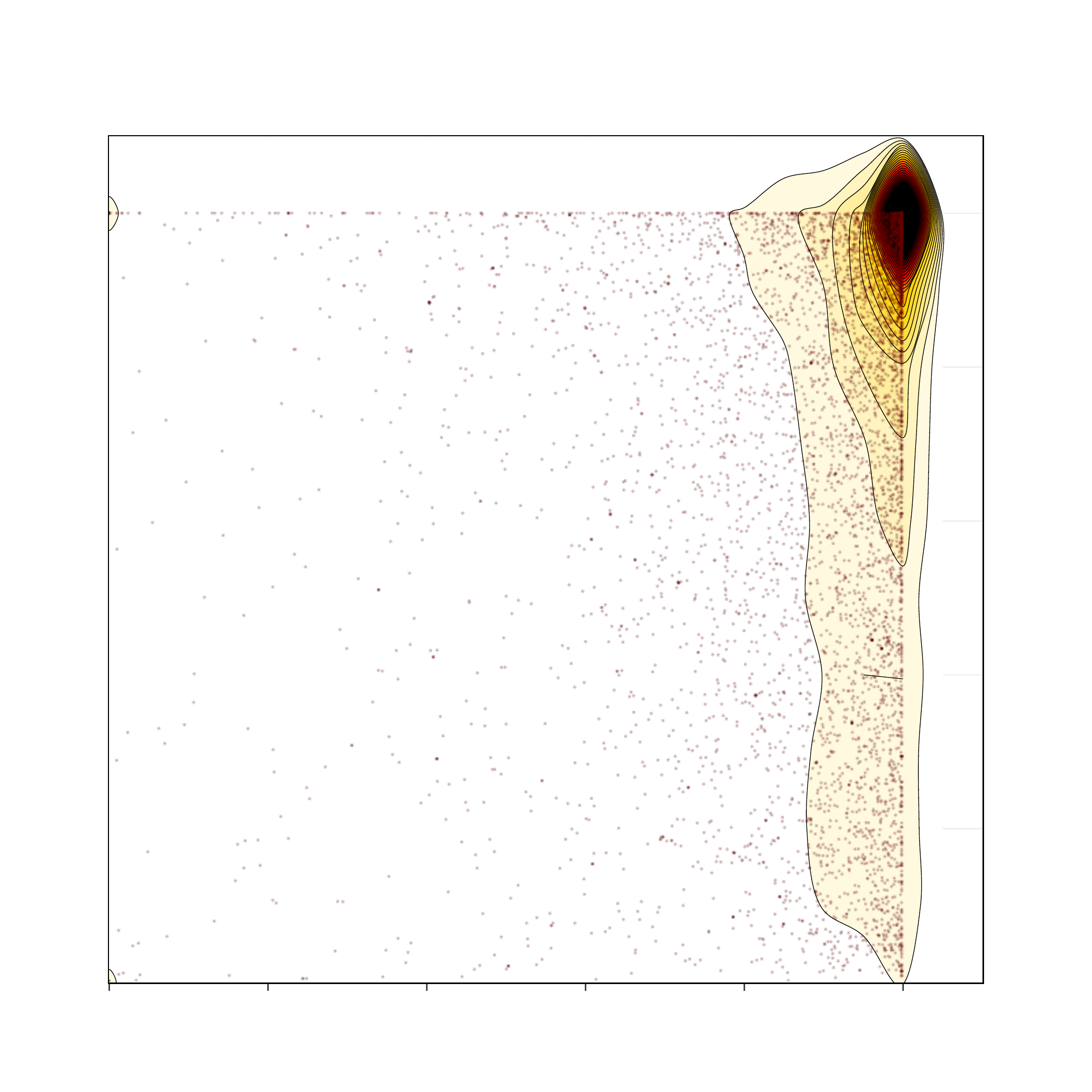
    \caption{{
    Distribution of the Precision \lo{(abscissa)}/Recall \lo{(ordinate)} scores \lo{displayed} as a scatter plot (each point cloud is a sample) and its associated density function for the \dtsABC dataset, for the different number of input scales in \pced (from left to right : $4, 8, 16, 32, 64$ and $128$ input scales)}
    { (except when explicitly stated, we use 16 scales for all our experiments).}
    }
    \label{fig:precRecallDifferentScales}
\end{figure*}

We now discuss the performance\nico{s} in training \nico{and} processing times\nico{,} and edge detection capabilities of our \pced for $4$, $8$, $16$, $32$, $64$ and $128$ input scales. 
We modify our architecture by \nico{changing}
the number of pairwise concatenation/processing layers (respectively $1, 2, 3, 4, 5, 6$) in its first stage (Figure~\ref{fig:arch8}-left, green dense layers). 
For all configurations, the training curves of our network on \dtsDFLT and \dtsABC are similar to those presented in Figure~\ref{fig:lc}.
We report the timings obtained with the different number of input scales in Table~\ref{tab:timeDiffScales} and we illustrate the results in Figure~\ref{fig:differentScales} on the \dataPisa and on a model of \lo{the} \dtsABC dataset.
In Figure~\ref{fig:precRecallDifferentScales} we also show the distribution of precision/recall scores for each number of input scales. 
Not surprisingly, computation times increase with the number of input scales. 
Depending on the number of different scales representing important features in a model, $8$ to $32$ scales may be used. 
For instance, as illustrated in Figures~\ref{fig:differentScales}-bottom and~\ref{fig:precRecallDifferentScales}, on the \dtsABC dataset where edges are of similar scale and sharpness, $8$ scales may be enough while Figure~\ref{fig:differentScales}-top shows on the \dataPisa model that $32$ scales would improve the results on large models with several scales of interest. 
As a default architecture, we present \pced with $16$ scales as a good compromise between speed and detection rate. 
It provides both accurate detection on all the models we tested while enabling interactive training sessions (as presented in Section~\ref{subsec:interactive}), however, 
if less emphasis is put on computation times, the input configuration with $32$ scales may be considered. 
In our experiments, $64$ and $128$ input scales did not improve over $32$ scales while being significantly slower. 
}

\begin{table*}[t]
\centering
\footnotesize
\setlength\tabcolsep{1pt} 
{
\begin{tabular}{|l|c|c|c|c|c|c|}
\cline{2-7}
 \multicolumn{1}{c|}{}   & 4      &  8   &   16    & 32     & 64     & 128     \\ \hline
Training on \dtsDFLT     & \twodurations{4.32}{42.85}  & \twodurations{6.45}{84.23} & \twodurations{19.32}{172.4} & \twodurations{239.34}{345.32} & \twodurations{250.27}{687.05} &  \twodurations{378.56}{1385.32}     \\ \hline
\hline
\dtsABC                  &\twodurations{410.4}{2268.53} &  \twodurations{680.38}{4732.25} &  \twodurations{1530}{9300} & \twodurations{3062}{18600}  &  \twodurations{6000}{37200}   &  
\color{colorHours}{3:20:00 (20:40:00) h} \\ \hline
\dataPisa                & \twodurations{1.34}{12.464} & \twodurations{2.23}{25.42} & \twodurations{4.12}{54.49} & \twodurations{7.32}{108.08} & \twodurations{15.1}{209.66} & \twodurations{31.64}{407.42}     \\ \hline
\end{tabular}
}
\caption{{Timings for  training (first row) and processing (second and third) point clouds using \pced for the different number of input scales (from $4$ to $128$). We report the network training/classification time followed in brackets by the \texttt{total\_time} with \texttt{total\_time = GLS\_precomputation+processing\_time}. For the \dtsABC dataset (second row) we provide the times for processing the $7167$ models.}
\label{tab:timeDiffScales}
}
\end{table*}

\subsection{Training and classification times}
\label{subsec:timings}

In this section we compare the learning and evaluation speed of our specialized approach to more versatile networks based on PointNet~\cite{qi2016pointnet}, the state of the art solution for point-based learning and semantic shape analysis.
Among the different possibilities, we selected \ecnet~\cite{ecnet2018} 
\nico{, } 
\pcp~\cite{GuerreroEtAl:PCPNet:EG:2018} as they can be easily adapted to our more geometric and focused problem.
\lo{We also include the evaluation of the edge classification presented on \pienet on the \dtsABC dataset by~\citet{PieNet2020}. }

\paragraph{Experimental setup for compared approaches}
For \pcp, we \lo{ran} a specialization training in order to adapt its original output from point normals estimation to point classification. 
We removed the output normalization from their architecture, and trained the last layers of the network (using the default hyperparameters provided by the authors) using our data (400 epochs on \dtsDFLT, and 50 epochs on \dtsABC).
This allows us to take advantage of \pcp original training for the first layers.
%
\lo{For \ecnet, we use the pre-trained version provided by the authors (denoted \ecnet(\dtsEC)), and we report the evaluation provided in~\citet{PieNet2020} for \ecnet(\dtsABC) and \pienet(\dtsABC), respectively the \ecnet network and \pienet trained on \dtsABC, when applied on the \dtsABC dataset.} 
\nico{Note that \ecnet}
oversamples the input point cloud, and label each generated point as edge or non-edge.
We retrieve the classification of the input point cloud by assigning to each input point the label of the closest output sample.
\nico{\pcp, \ecnet and \pienet approaches}
require more computation power and memory than provided by our 10-cores Intel Xeon-E5 to handle large data sets as \dtsABC.
For training \lo{\pcp, and evaluating both \pcp and \ecnet}, we thus used {an} NVIDIA Quadro RTX 6000 with 24GB of G-RAM 
\lo{ and \pienet 
was trained on an NVIDIA TITAN X GPU in~\cite{PieNet2020}}.
This is reminded by a $^*$ in the text \lo{and in tables} when we report timings.

\begin{figure}[b]
    \centering
    \footnotesize
    \def\svgwidth{0.25\linewidth} 
    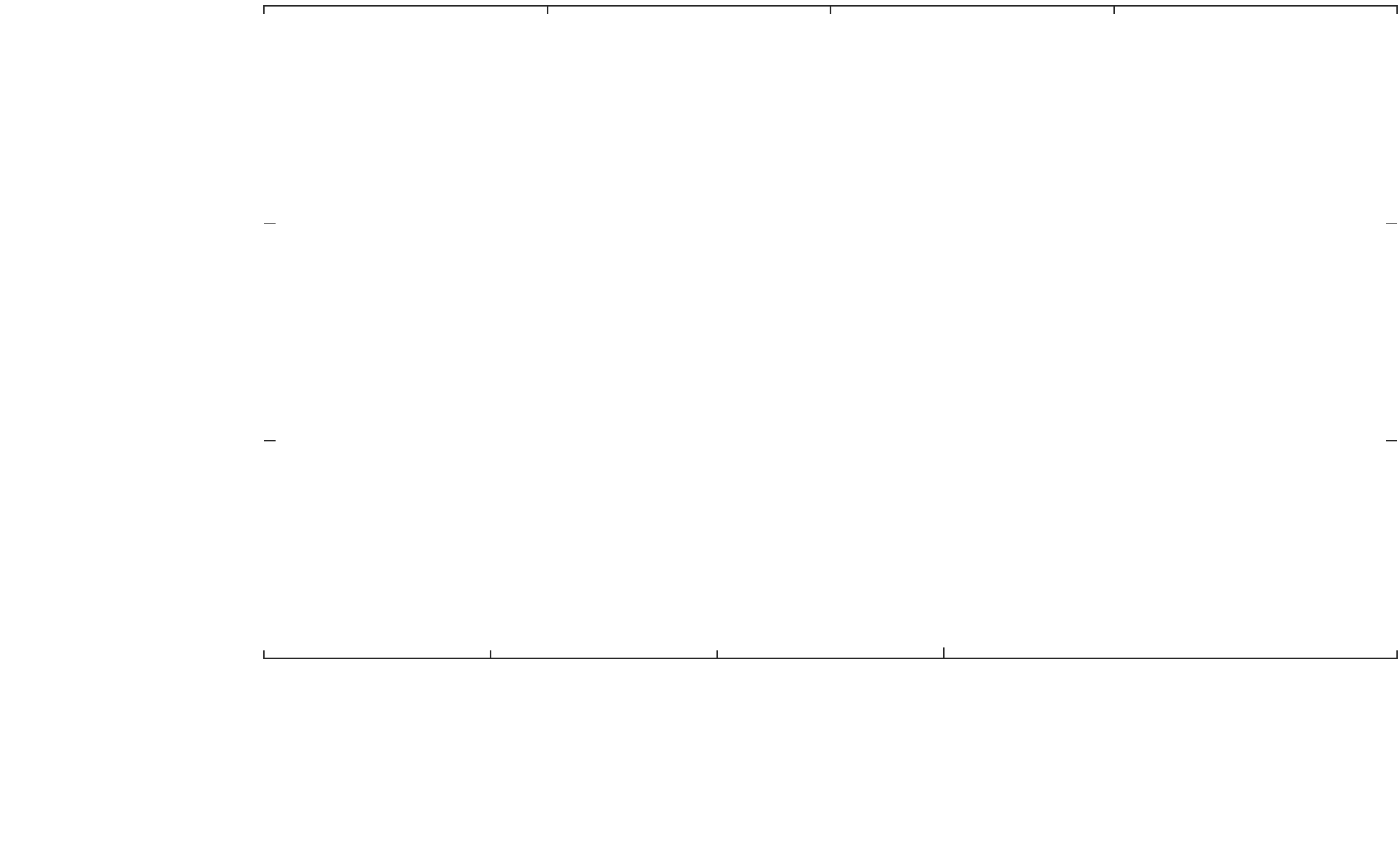 
    \hspace{5mm}
    \def\svgwidth{0.25\linewidth} 
    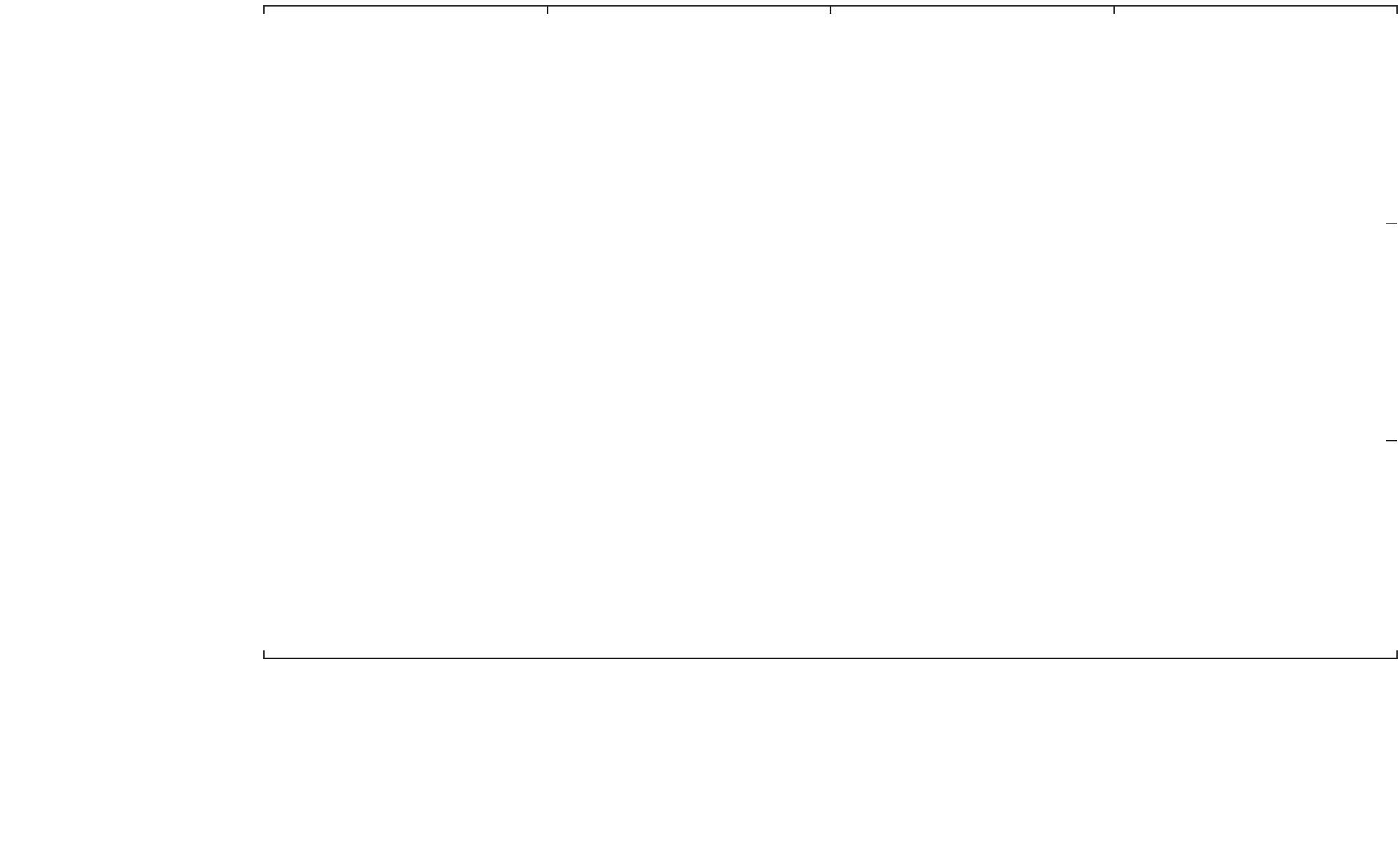 
    \caption{\pcp: learning curves measured on Training (T) and Validation (V) sets on the (left) \dtsDFLT and (right) \dtsABC datasets.
    \label{fig:pcptraining}
    }
\end{figure}

\paragraph{Training}
We trained \pcp with three classes on our \dtsDFLT dataset (denoted \pcp(\dtsDFLT)) and with two classes on \dtsABC (denoted \pcptc(\dtsABC)). 
As shown by the loss plot in Figure~\ref{fig:pcptraining}-left, training of the deep architecture of \pcp converge to relatively high loss on our \dtsDFLT training set. 
Also, despite several attempts, \pcp couldn't stabilise on \dtsSHREC due to the high variability of the annotations.
Training on a larger dataset such as \dtsABC (see the loss plot in Figure~\ref{fig:pcptraining}-right) is better adapted and enables the convergence of \pcp training with low loss.
%
Regarding the timings, \pcp  ~\lo{and \pienet} require significantly longer training than our approach (more than $500$ times slower on \dtsABC), as reported in Table~\ref{table:trainingtime}.

\begin{table}[t]
\centering
\footnotesize
\begin{tabular}{|l|c|c|c|}
\cline{2-4}
\multicolumn{1}{c|}{} & \pcp                    & \nico{\pienet(\dtsABC)}  & \pced         \\ 
\hline
\dtsDFLT      & \duration{2400}*                       & -                 & \twodurations{19.32}{172.4}  \\ \hline
\dtsABC       & \duration{68400}*                     & \duration{82800}* & \twodurations{131}{1200}     \\ \hline
\end{tabular}
\caption{
{Timings for training \pcp on the \dtsDFLT dataset with three classes and on the \dtsABC dataset with two classes. For \pced, we report first the \nico{training} time, followed in brackets by the \texttt{total\_time = GLS+\nico{training}\_time}. We also remind the \pced training times with the same number of classes for comparisons.}\label{table:trainingtime}
\nico{The training time of \pienet is reported from~\citet{PieNet2020}.
Timings with a $^*$ have been performed on recent high performance GPUs (e.g., NVIDIA TITAN X).}
}
\end{table}

\begin{figure}[htb]
    \centering
{
    \footnotesize
    \includegraphics[width = 0.5\linewidth]{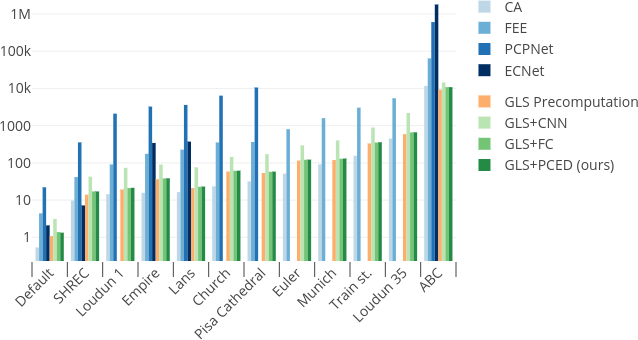}
}
    \caption{Time (in seconds) required by the approaches presented in Table~\ref{tab:visualTime} to classify different models.
    \label{fig:PTPC}
    }
\end{figure}

\paragraph{Classification}
In addition to 
\nico{\pcp, \ecnet and \pienet,}
we compare our approach with \ca~\cite{Bazazian15} (for all our experiments, we used 10 neighbors and threshold=0.65, except for \dataMunich for which we used 20 neighbors to handle sparse sampling), 
\fee~\cite{Merigot2011} (we specifically adjusted parameters for each dataset),
our baselines \cnn and \fc.
We report in Figure~\ref{fig:PTPC} the evolution of the classification timings (in logarithmic scale) when increasing the point cloud size.
We observe for \pced, \fc and \cnn that scale-space calculation (which can be done in pre-process) requires more time than the classification itself. 
It also illustrates 
how fast our compact network architecture is, compared to more complex and computationally intensive architectures as \pcp, \ecnet, \lo{and \pienet}.
Finally, \pcp and \ecnet require more than $24$ GB G-RAM to handle large point clouds and we thus could not provide the timings for the classification of our larger models by these networks.

\begin{figure*}
    \centering
    \footnotesize
    \def\svgwidth{0.18\linewidth} 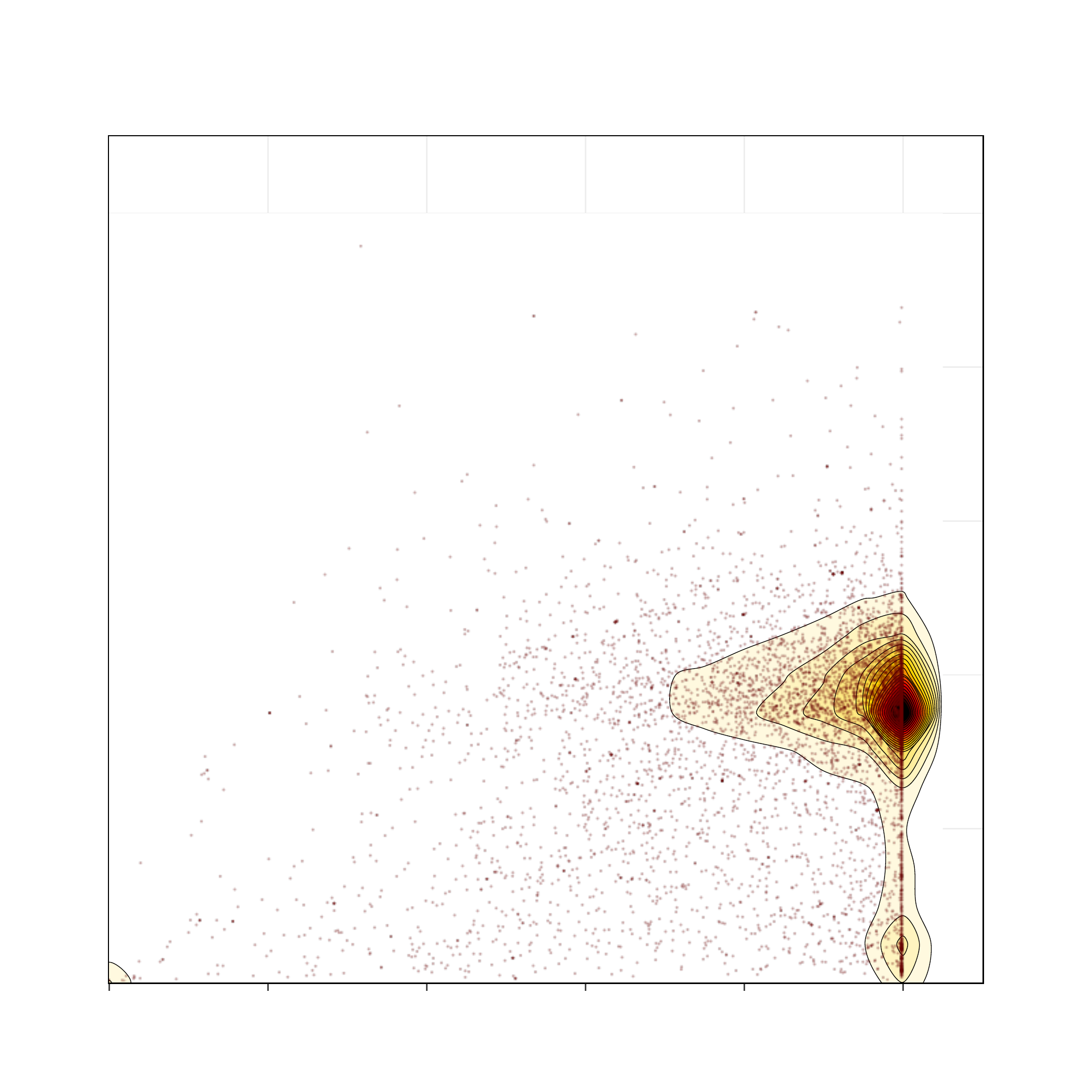
    \def\svgwidth{0.18\linewidth} 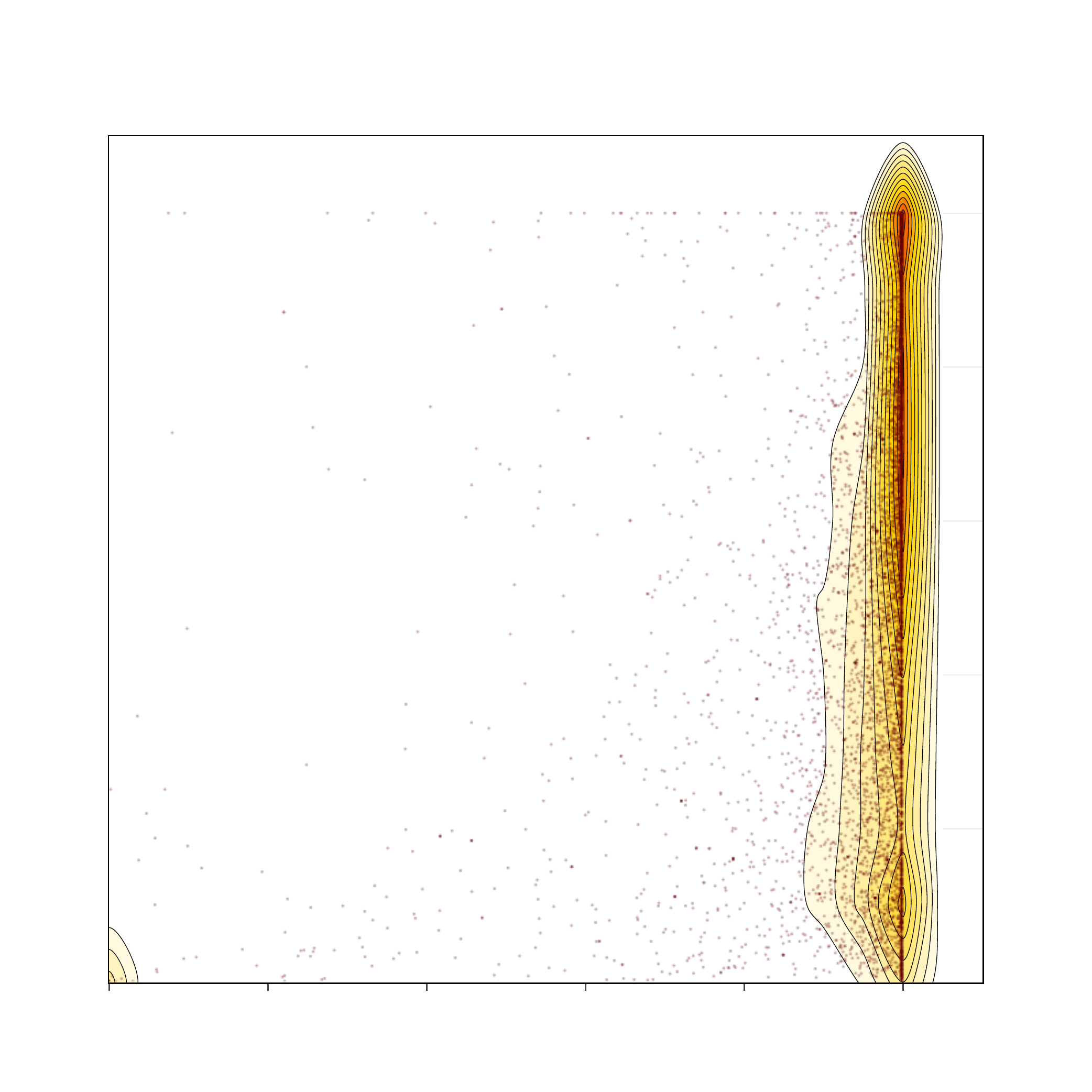
    \def\svgwidth{0.18\linewidth} 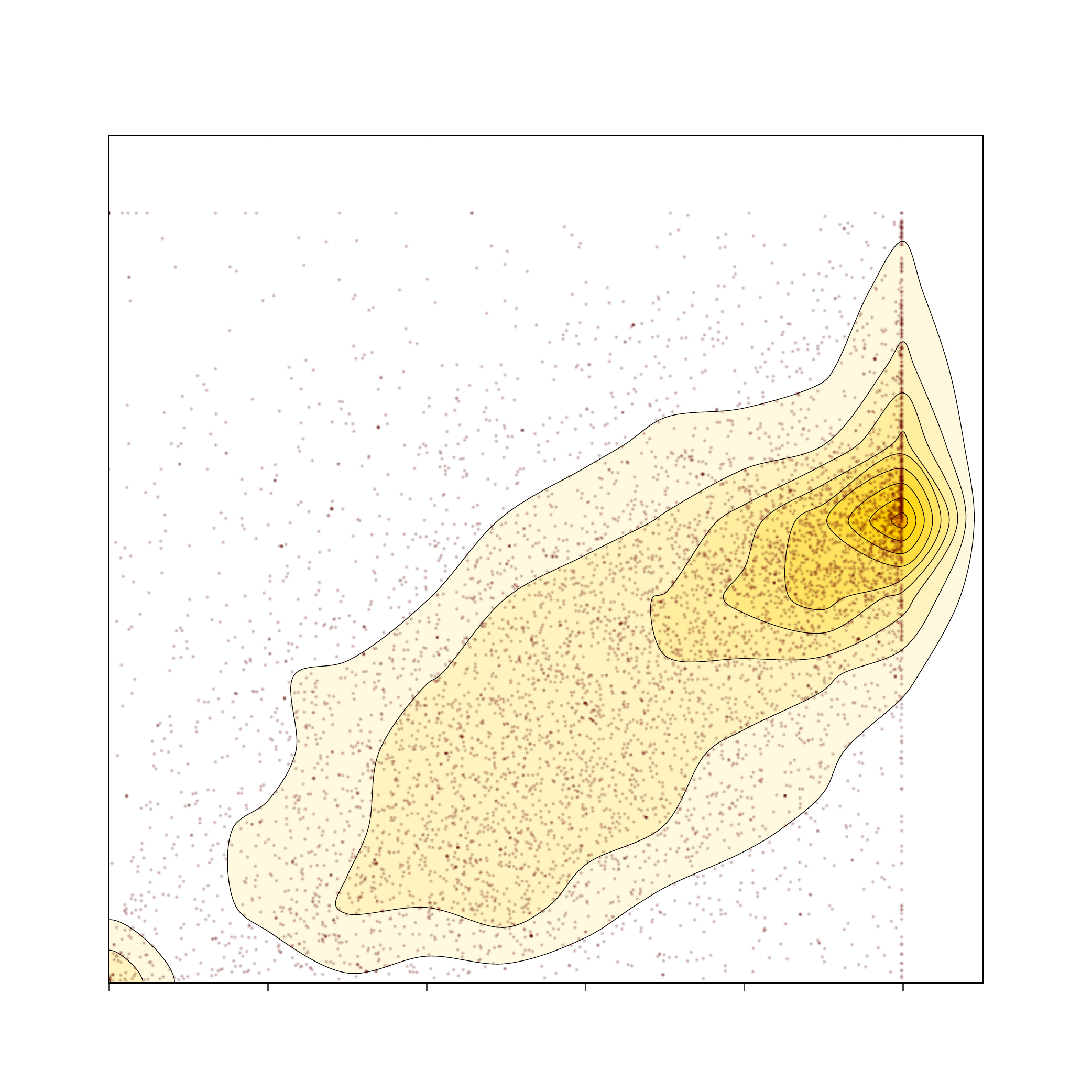
    \def\svgwidth{0.18\linewidth} 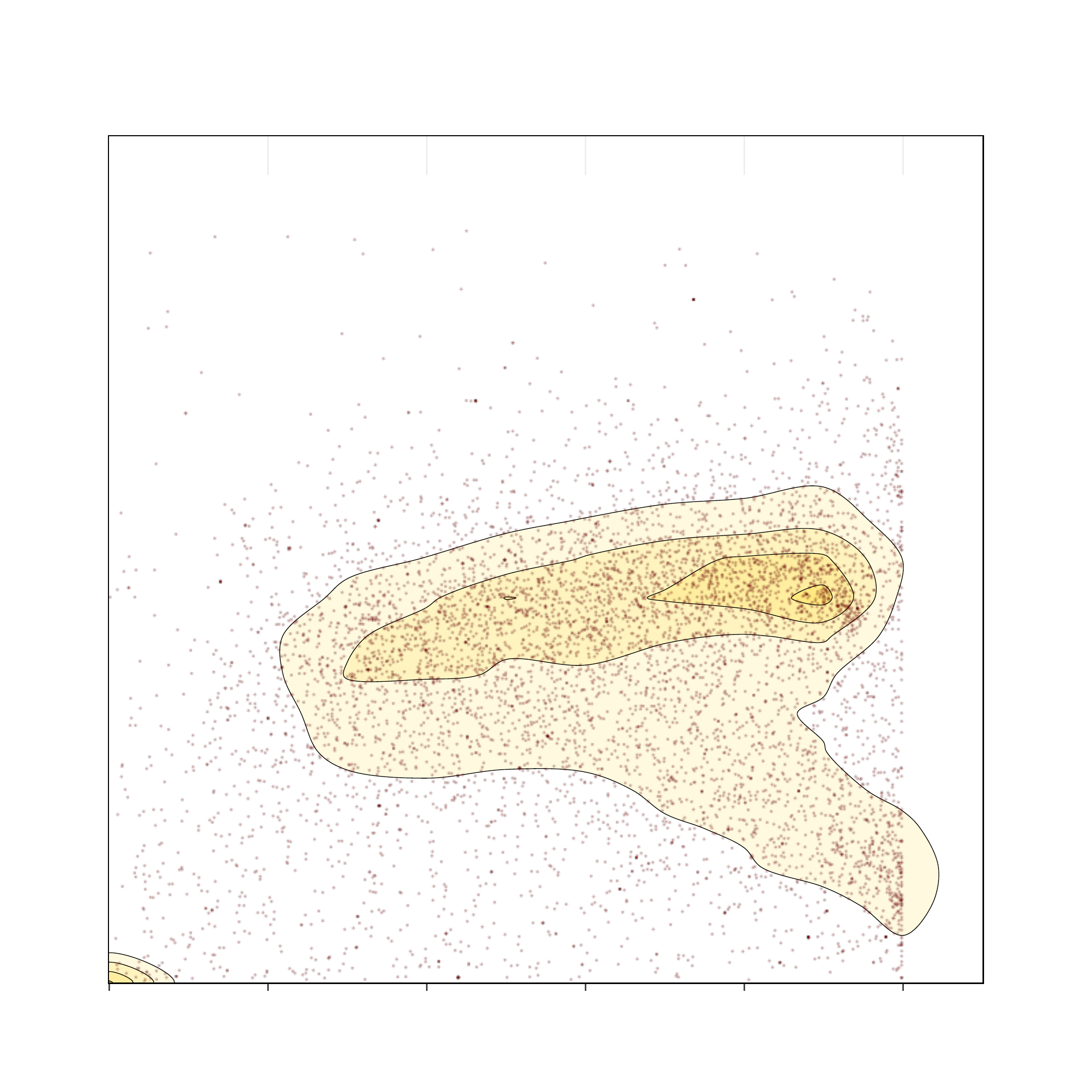
    \fbox[trl]{
    \def\svgwidth{0.18\linewidth} \import{./images/heatmaps/}{pced-default.pdf_tex}
    }
    \\
    \def\svgwidth{0.18\linewidth} 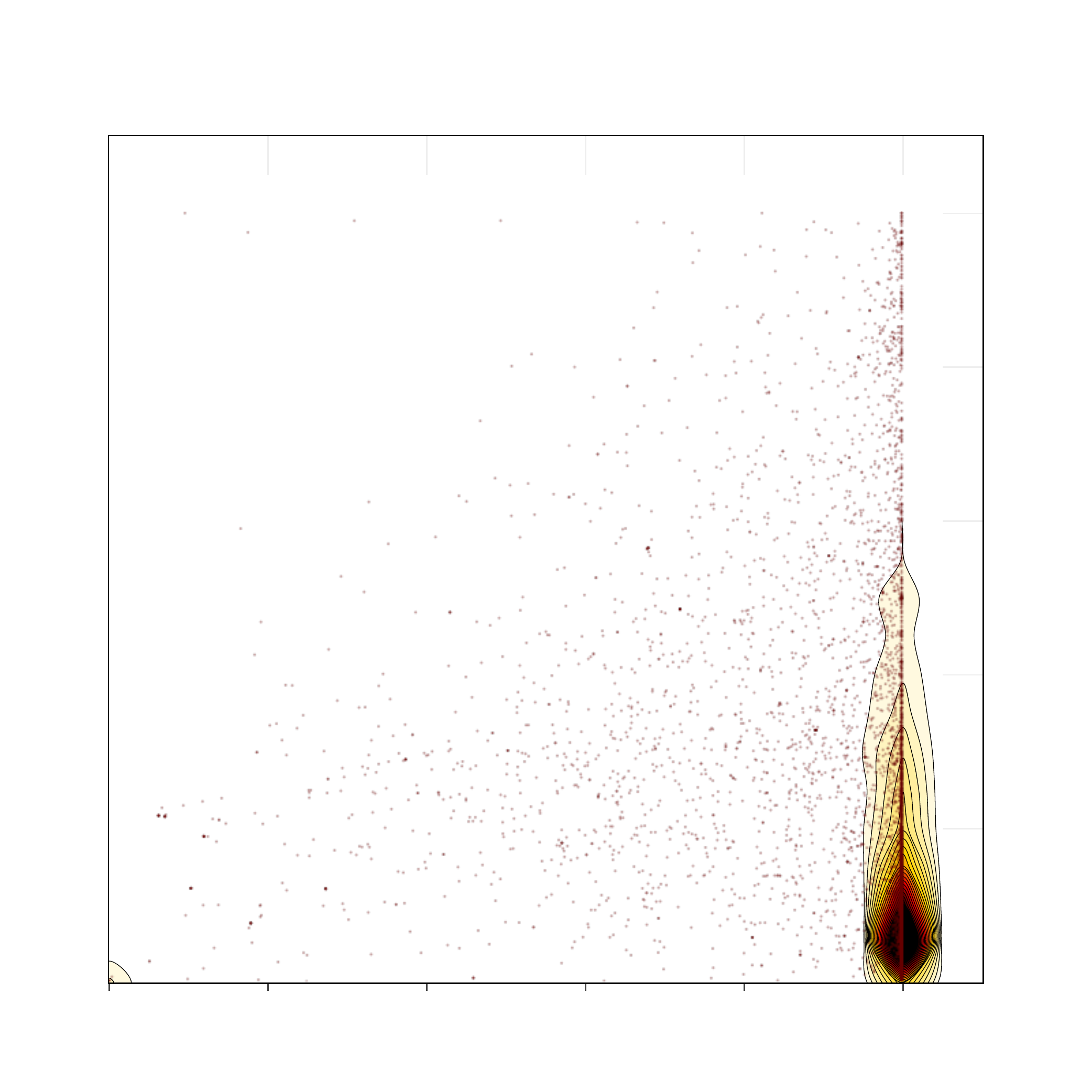
    \def\svgwidth{0.18\linewidth} 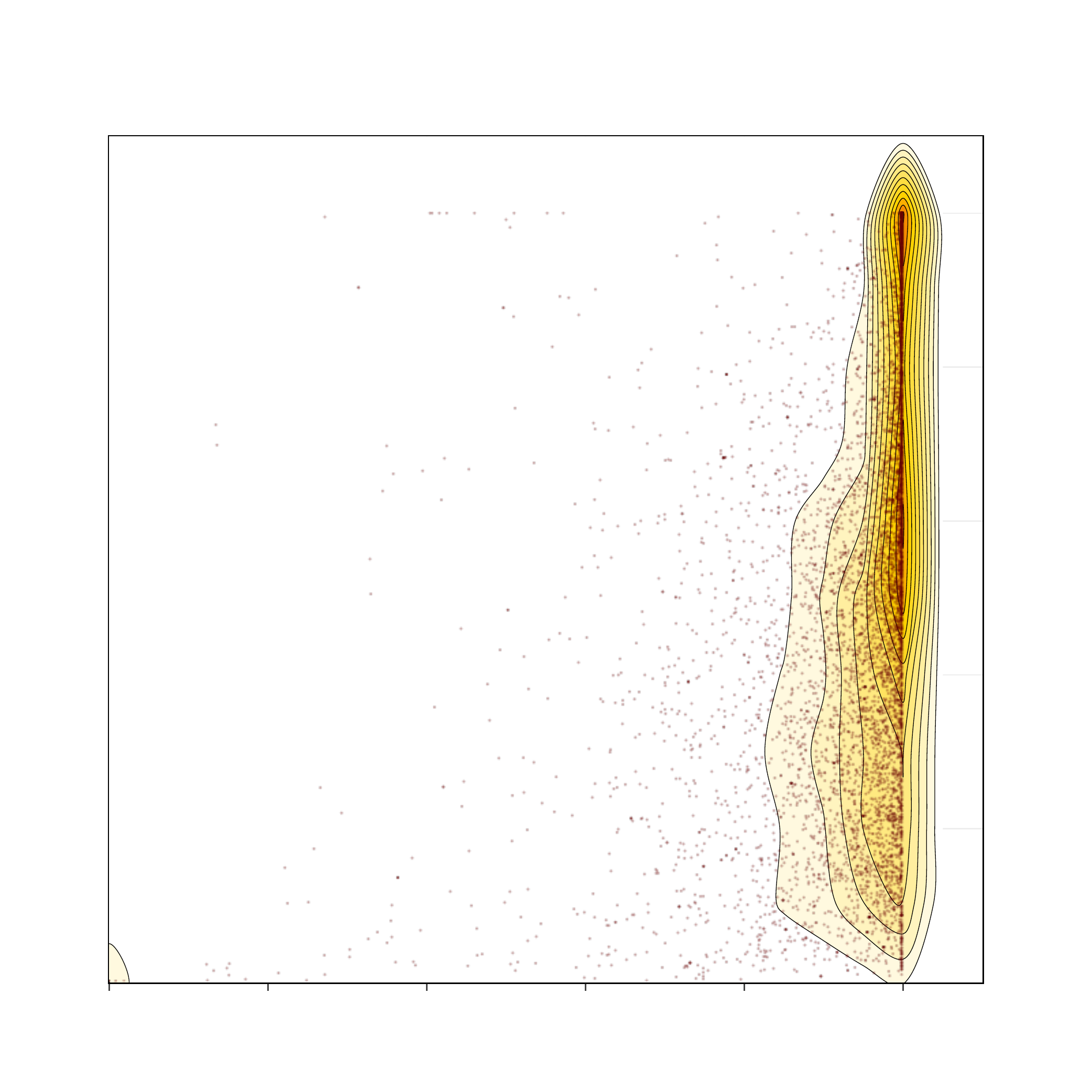
    \def\svgwidth{0.18\linewidth} 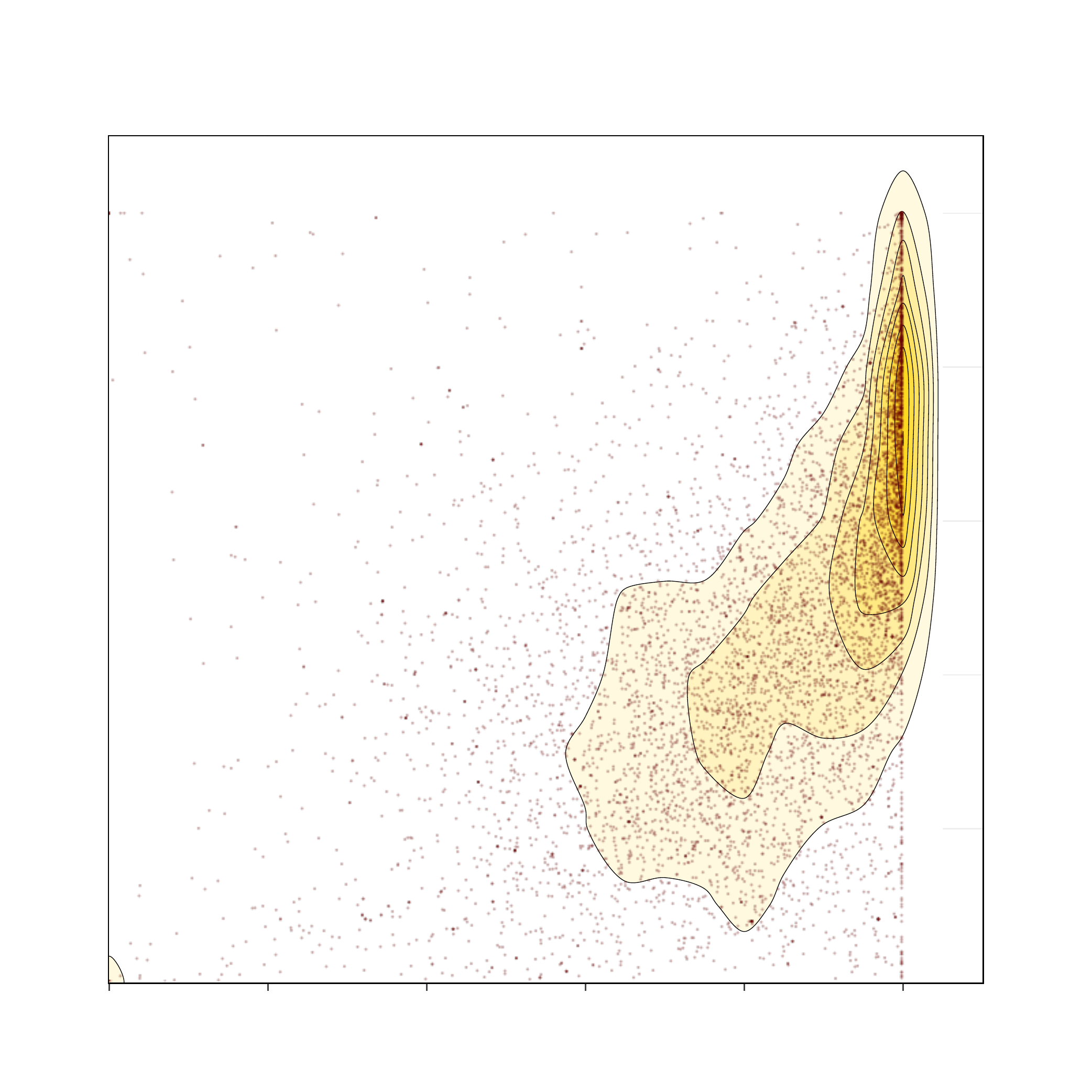
    \def\svgwidth{0.18\linewidth} 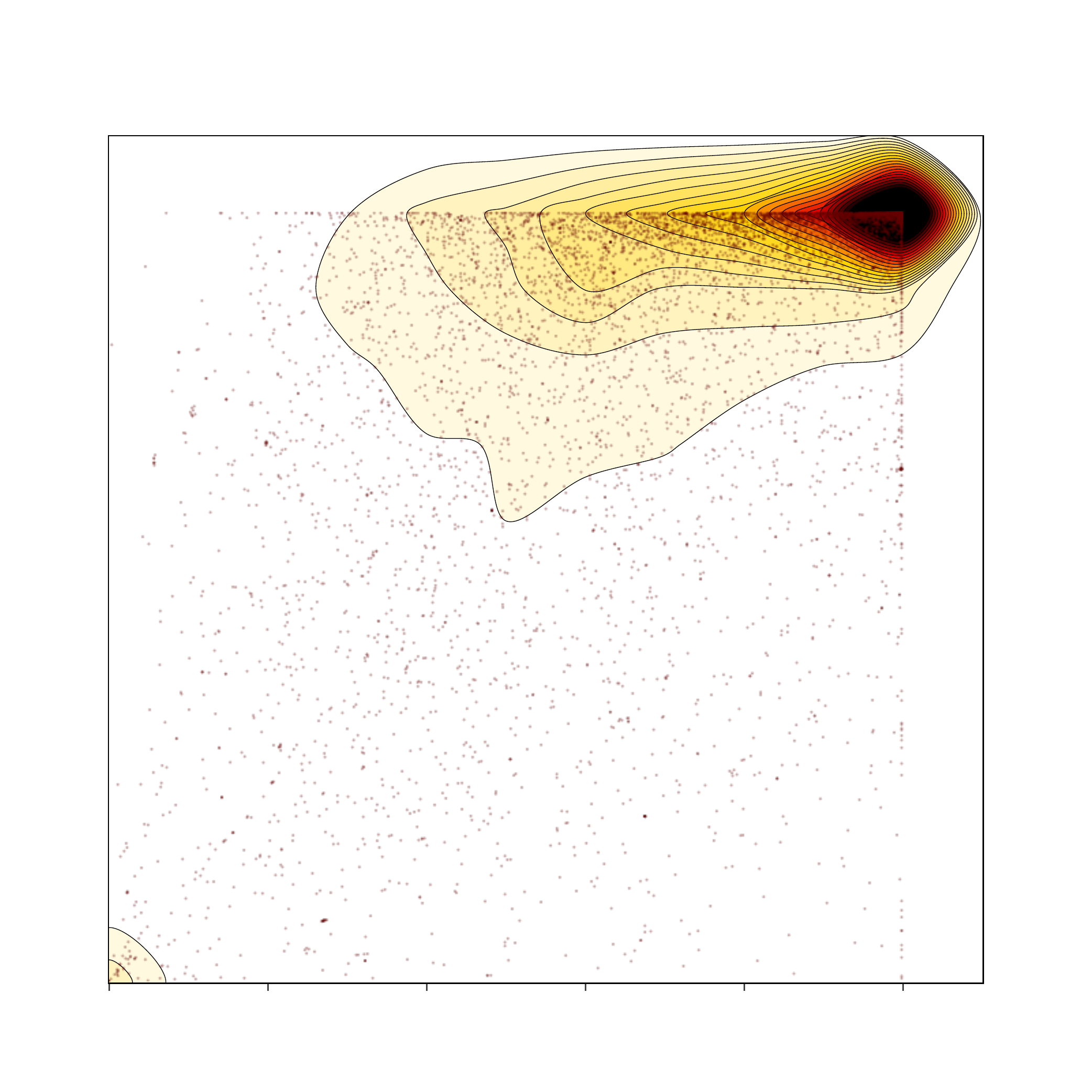
    \fbox[brl]{
    \def\svgwidth{0.18\linewidth} 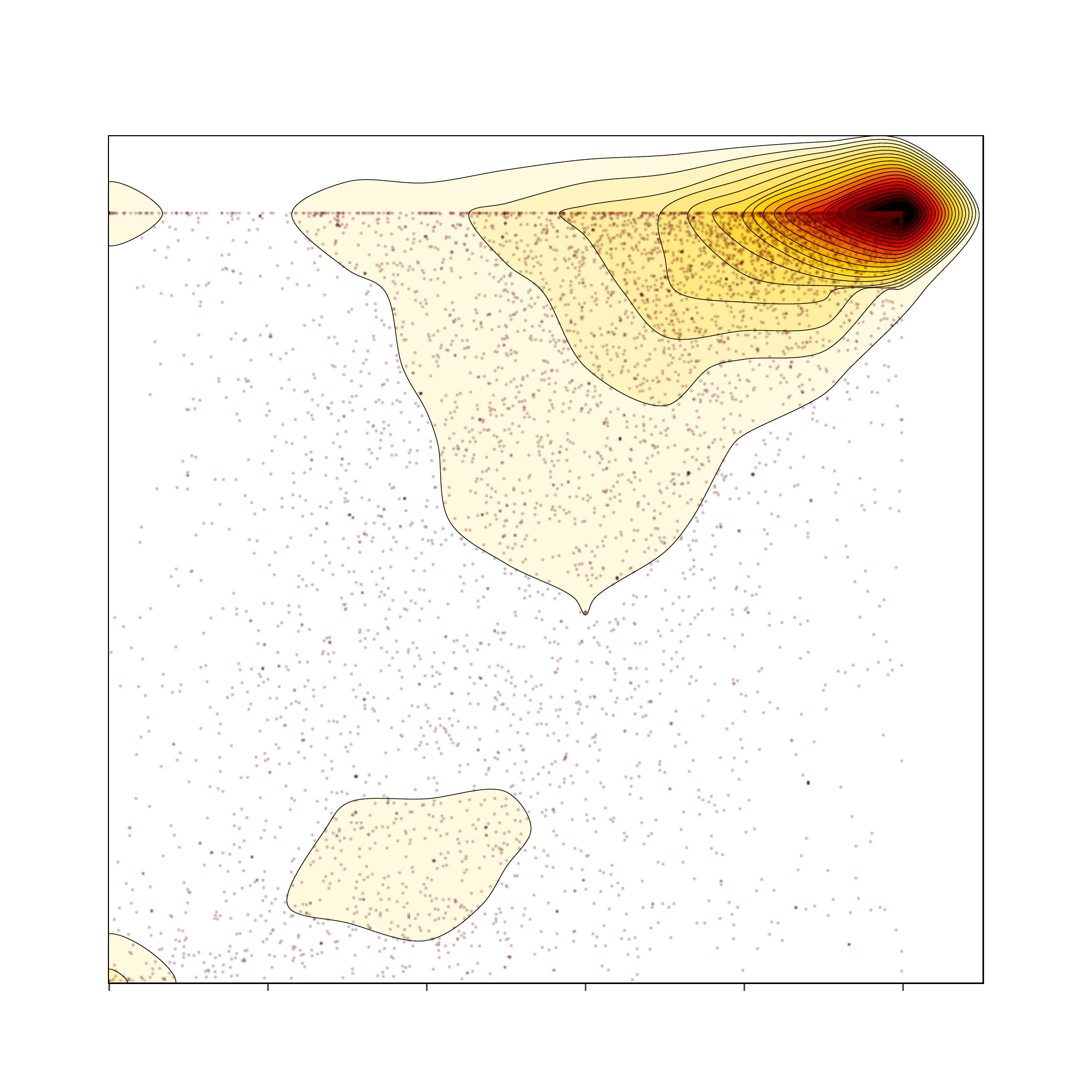
    }
    \caption{
    Distribution of the Precision \lo{(abscissa)}/Recall \lo{(ordinate)} scores \lo{displayed} as a scatter plot (each point cloud is a sample) and its associated density function for the \dtsABC dataset.
    Interactive plots are provided in the joined website.
    \nico{All maps share the same color scale.}
    \label{fig:scp_comp}
    }
\end{figure*}

\subsection{Quantitative evaluation}
\label{subsec:comparisons}
We now compare the classification produced by the aforementioned approaches with ground truth classification on datasets \dtsDFLT, \dtsABC and \dtsSHREC.

\begin{figure}[htb]
\footnotesize
\def\svgwidth{\linewidth}
\begingroup%
  \makeatletter%
  \providecommand\color[2][]{%
    \errmessage{(Inkscape) Color is used for the text in Inkscape, but the package 'color.sty' is not loaded}%
    \renewcommand\color[2][]{}%
  }%
  \providecommand\transparent[1]{%
    \errmessage{(Inkscape) Transparency is used (non-zero) for the text in Inkscape, but the package 'transparent.sty' is not loaded}%
    \renewcommand\transparent[1]{}%
  }%
  \providecommand\rotatebox[2]{#2}%
  \newcommand*\fsize{\dimexpr\f@size pt\relax}%
  \newcommand*\lineheight[1]{\fontsize{\fsize}{#1\fsize}\selectfont}%
  \ifx\svgwidth\undefined%
    \setlength{\unitlength}{7688.25bp}%
    \ifx\svgscale\undefined%
      \relax%
    \else%
      \setlength{\unitlength}{\unitlength * \real{\svgscale}}%
    \fi%
  \else%
    \setlength{\unitlength}{\svgwidth}%
  \fi%
  \global\let\svgwidth\undefined%
  \global\let\svgscale\undefined%
  \makeatother%
  \begin{picture}(1,0.36989317)%
    \lineheight{1}%
    \setlength\tabcolsep{0pt}%
    \put(0,0){\includegraphics[width=\unitlength,page=1]{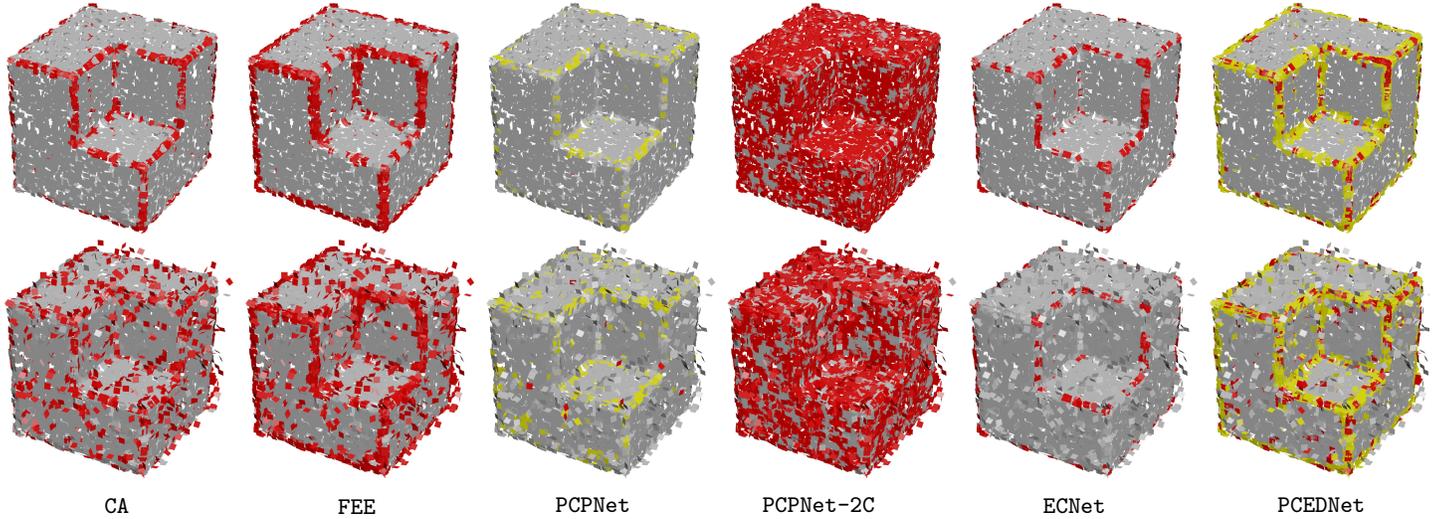}}%
    \put(0.07599936,0.00860773){\makebox(0,0)[lt]{\lineheight{1.25}\smash{\begin{tabular}[t]{l}\ca\end{tabular}}}}%
    \put(0.23908457,0.00831964){\makebox(0,0)[lt]{\lineheight{1.25}\smash{\begin{tabular}[t]{l}\fee\end{tabular}}}}%
    \put(0.39082527,0.00968689){\makebox(0,0)[lt]{\lineheight{1.25}\smash{\begin{tabular}[t]{l}\pcp\end{tabular}}}}%
    \put(0.53523384,0.00968689){\makebox(0,0)[lt]{\lineheight{1.25}\smash{\begin{tabular}[t]{l}\pcptc\end{tabular}}}}%
    \put(0.73123016,0.00860773){\makebox(0,0)[lt]{\lineheight{1.25}\smash{\begin{tabular}[t]{l}\ecnet\end{tabular}}}}%
    \put(0.89485353,0.00939885){\makebox(0,0)[lt]{\lineheight{1.25}\smash{\begin{tabular}[t]{l}\pced\end{tabular}}}}%
  \end{picture}%
\endgroup%

\caption{Classification on \dtsDFLT, model \dataTwoCubes: $\sigma=0$ (top) and $0.14$ (bottom). See results of the other approaches in the accompanying website.}
\label{fig:cubes}
\end{figure}

\begin{table}[b]
\centering
\footnotesize
\setlength\tabcolsep{4pt} 
\begin{tabular}{|l|c|c|c|c|c|c|}
\hline
Method           & Precision      & Recall         & \textbf{MCC}   & F1             & Accuracy      & IoU \\ \hline
\ca              & 0.490          & 0.880          & 0.506          & 0.628          & 0.752         & \nico{0.457} \\ \hline
\fee             & 0.341          & 0.814          & 0.471          & 0.480          & 0.879         & \nico{0.316} \\ \hline
\hline
\fc (\dtsDFLT)   & 0.618          &  0.958.        & 0.682          & 0.753          & 0.849         & \nico{0.604} \\ \hline
\cnn (\dtsDFLT)  & 0.546          & 0.955          & 0.623          & 0.694          & 0.807         & \nico{0.532} \\ \hline
\pcp (\dtsDFLT)  & 0.722          & 0.198          & 0.301          & 0.310          & 0.789         & \nico{0.183} \\ \hline
\textbf{\pced} (\dtsDFLT) & 0.826          & 0.952          & \textbf{0.857} & \textbf{0.890} & 0.946         & \nico{\textbf{0.802}} \\ \hline
\nico{\textbf{\pcedtc} (\dtsDFLT)} & \nico{0.364}  & \nico{0.611}  & \nico{0.402}    & \nico{0.430}   & \nico{0.908} & \nico{0.274} \\ \hline
\hline
\ecnet (\dtsEC)      & \textbf{1.000} & 0.457          & 0.656          & 0.620          & \textbf{0.960} & \nico{0.450}\\ \hline
\lo{\textbf{\pcedtc} (\dtsEC)}  & \lo{0.201}      & \lo{\textbf{0.991}} & \lo{0.383} & \lo{0.334}    & \lo{0.759} & \nico{0.200}      \\ \hline
\end{tabular}
\caption{Quantitative evaluation on \dtsDFLT: median scores (see scores of the other approaches in the accompanying website). 
\label{tab:quant_default}
}
\end{table}

\begin{table}[b]
\centering
\footnotesize
\setlength\tabcolsep{4pt} 
\begin{tabular}{|l|c|c|c|c|c|c|}
\hline
Method             & Precision      & Recall         & \textbf{MCC}   & F1             & Accuracy    & IoU   \\ \hline
\ca                & 0.348          & 0.944          & 0.520          & 0.504          & 0.881       & \nico{0.338}      \\ \hline
\fee               & 0.135          & \textbf{1.000} & 0.062          & 0.235          & 0.278       & \nico{0.132}      \\ \hline
\hline
\fc (\dtsDFLT)     & 0.452          & 0.679          & 0.469          & 0.513          & 0.921       & \nico{0.327}      \\ \hline
\cnn (\dtsDFLT)    & 0.530          & 0.995          & 0.648          & 0.689          & 0.922       & \nico{0.519}      \\ \hline
\textbf{\pced} (\dtsDFLT)   & 0.746          & 0.745         & 0.688          & 0.713          & 0.966        & \nico{0.548}      \\ \hline
\textbf{\pcedtc} (\dtsDFLT) & 0.662          & 0.936         & 0.708          & 0.730          & 0.958        & \nico{0.574}     \\ \hline
\hline
\ecnet (\dtsEC)        & 0.425          & 0.648          & 0.423          & 0.460          & 0.910       & \nico{0.294}      \\ \hline
\nico{\textbf{\pcedtc} (\dtsEC)} & \lo{0.378}          & \lo{0.849}         & \lo{0.480}          & \lo{0.503}          & \lo{0.888}        & \nico{0.334}      \\ \hline
\hline
\nico{\ecnet (\dtsABC)}    & \lo{0.487}          & \lo{0.573}          & -              & \lo{0.526}          &  -  & \nico{0.356}          \\ \hline
\nico{\pienet (\dtsABC)}   & \lo{0.692}          & \lo{0.858}          & -              & \lo{0.766}          &  -  & \nico{0.622}          \\ \hline
\fctc (\dtsABC)    & 0.470          & 0.871          & 0.555          & 0.581          & 0.920         & \nico{0.408}    \\ \hline
\cnntc (\dtsABC)   & 0.507          & 0.983          & 0.646          & 0.662          & 0.928           & \nico{0.491}  \\ \hline
\pcptc (\dtsABC)   & \textbf{0.954} & 0.756          & 0.797          & 0.807          & \textbf{0.979} & \nico{\textbf{0.668}}  \\ \hline
\textbf{\pcedtc} (\dtsABC)  & 0.735          & 0.984         & \textbf{0.808} & \textbf{0.822} & 0.970          & \nico{0.597}   \\ \hline
\end{tabular}
\caption{Quantitative evaluation on \dtsABC: median scores \nico{(see scores of the other approaches in the accompanying website)}. 
\nico{\ecnet (\dtsABC) and \pienet (\dtsABC) are reported from \citet{PieNet2020}}
\label{tab:quant_abc}
}
\end{table}

\paragraph{\dtsDFLT dataset}
Results are reported in Table~\ref{tab:quant_default} and illustrated in both Figure~\ref{fig:cubes} and the accompanying website.
Both \cnn and \fc are strong baselines getting significantly higher MCC scores than most methods from previous work.
\lo{\pcp(\dtsDFLT) shows a very limited detection of sharp edges.}
\ecnet(\dtsEC) produces very precise classifications, 
but overall misses to classify 50\% of the points labeled as edges in ground truth (high precision and low recall).
{This classification by \ecnet(\dtsEC) of a small subset of the edge points is emphasized by the low F1 and MCC scores that aggregate precision and recall.}
Regarding single-scale fitting-based approaches (i.e. \ca and \fee), both approaches fail at handling outliers (for \fee, we used the following parameters: r$=0.025$, R$=0.05$ and th$=0.16$).
\pced(\dtsDFLT) produces the best results on this dataset, getting significantly higher indicators than other approaches.
Overall, \pced trained on the \dtsDFLT dataset provides the best balance between precision and recall with the highest F1 and, more importantly, MCC scores.

\lo{Finally, we observe that \pcedtc(\dtsEC) provides a better recall and a lower precision than \ecnet(\dtsEC), 
with a lower MCC on \dtsDFLT. 
The same behavior is observed on the \dtsSHREC (Table~\ref{tab:quant_shrec}) and \dtsABC (Table~\ref{tab:quant_abc}) datasets, with a higher MCC for \pcedtc on the \dtsABC dataset.
It illustrates the more conservative behavior of \pcedtc(\dtsEC) while \ecnet(\dtsEC) tends to detect subsets of correct edges.}

\begin{figure*}[t]
\footnotesize
\def\svgwidth{\linewidth}
\import{./images/figures/}{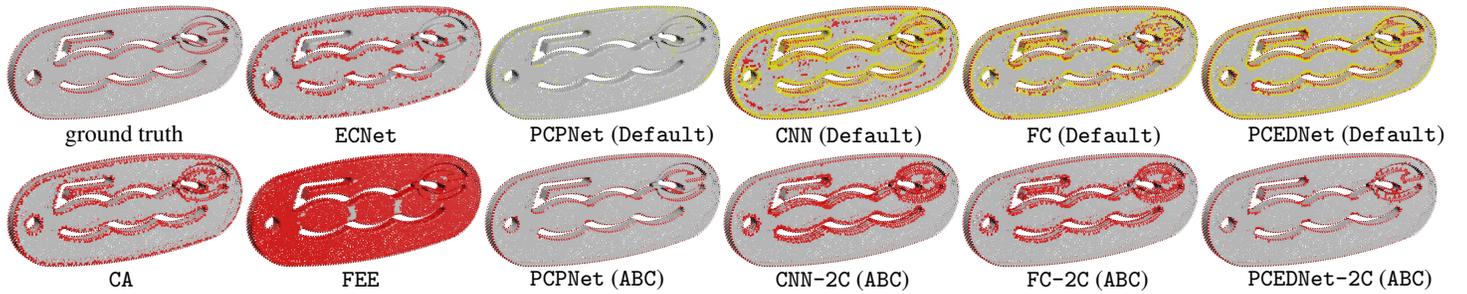}
\caption{Model \texttt{7029} of the \dtsABC dataset.}
\label{fig:abc}
\end{figure*}

\paragraph{\dtsABC dataset}
Due to the large size of \dtsABC (7167 models) we present in Figure~\ref{fig:scp_comp} (and in the accompanying website) the results as a precision/recall scatter plots, which provide a more readable overall view of the performance of the different approaches. 
For each approach, we plot each 3D model as a point sample, and display the resulting density maps as 2D-contour maps. 
We use the \texttt{Hot} colormap, with density values ranging from 50 (light yellow) to 1000 (dark red). 
A perfect classification result would lead to a Dirac distribution centered at location (1,1) and with magnitude 7167.
%
Most of the analysed approaches are 
producing classifications with high recall, but differ by their capacity to reach high precision.
Table~\ref{tab:quant_abc} shows the median statistical scores obtained by each method on the whole dataset.
Single-scale fitting-based approaches, e.g., \ca and \fee (we use r$=0.4$, R$=0.8$ and th$=0.16$ for this dataset) exhibit very high recall but low precision (lower than $0.4$) with the classification of a lot of FP.
As shown in supplementary materials, these approaches produce good results for very clean and simple geometries, but fail at analyzing thin objects (i.e. neighborhood might include two pieces of opposite surfaces, see \fee in Figure~\ref{fig:abc}) and other intricate shapes.
\cnn follows a similar pattern, but is able to produce better classifications with high precision (MCC $\approx0.65$).
\fc produces results with lower quality (MCC $<0.55$) and higher dependency to the training set. 
Both \fc(\dtsDFLT) and \fctc(\dtsABC) are however competitive compared to state of the art methods.

Overall, 
\lo{the evaluation of \ecnet(\dtsEC) (in terms of MCC, F1 and Accuracy) is better than fitting-based approaches. 
\ecnet(\dtsABC) exhibits a higher precision and a lower recall than \ecnet(\dtsEC), resulting in better general behavior.
It remains however less efficient than our baselines trained on \dtsABC, which, in terms of F1 score, are also less effective than \pienet(\dtsABC).}
\lo{All the variants of our \pced produce varying precision and recall but similar F1 scores above $0.7$, and \pcedtc(\dtsABC) shows the best results with an MCC and F1 scores above $0.8$.}
Despite the large difference between the training sets of \dtsDFLT and \dtsABC, our approach \pced(\dtsDFLT) produces classifications with better MCC than competitors (MCC $=0.688$), but at the cost of a recall loss.
As shown in the joined website, the classifications produced by both \pced(\dtsDFLT) and \pcedtc(\dtsABC) remain visually convincing despite quantitative differences against the ground truth.
A notable exception to the global trend high-recall/variable-precision is \pcptc(\dtsABC), which exhibits higher precision/lower recall than any other approach.
It also reaches comparable scores as our approach (e.g. MCC, F1, Accuracy), but visual inspection reveals that \pcptc(\dtsABC) tends to miss some edges entirely (lower recall), while our approach finds relatively thicker edges (lower precision).
Also, \pcptc(\dtsABC) requires 3 days to process the entire dataset (instead of 3 hours for \pcedtc(\dtsABC)).


\begin{table}[b]
\centering
\footnotesize
\setlength\tabcolsep{4.7pt} 
\begin{tabular}{|l|c|c|c|c|c|c|}
\hline
Method                           & Prec.          & Recall         & \textbf{MCC}   & F1             & Accuracy & IoU      \\ \hline
\ca                              & 0.434          & 0.449          & 0.390          & 0.442          & 0.876    & \nico{0.284}        \\ \hline
\fee                             & 0.191          & 0.527          & 0.151          & 0.278          & 0.727    & \nico{0.160}        \\ \hline
\hline
\pcp (\dtsDFLT)                  & 0.000          & 0.000          & 0.000          & 0.009          & 0.893    & \nico{0.000}        \\ \hline
\fc (\dtsDFLT)                   & 0.392          & 0.538          & 0.391          & 0.455          & 0.880    & \nico{0.294}        \\ \hline
\fc (\dtsDFLT)$^+$               & 0.313          & 0.909          & 0.428          & 0.461          & 0.793    & \nico{0.300}        \\ \hline
\cnn (\dtsDFLT)                  & 0.468          & 0.928          & \textbf{0.570} & \textbf{0.611} & 0.911    & \nico{\textbf{0.440}} \\ \hline
\cnn (\dtsDFLT)$^+$              & 0.299          & \textbf{0.945} & 0.429          & 0.446          & 0.802    & \nico{0.287}        \\ \hline
\textbf{\pcedtc (\dtsDFLT)}      & 0.462          & 0.623          & 0.455          & 0.510          & 0.868    & \nico{0.342}        \\ \hline
\textbf{\pced (\dtsDFLT)}        & \textbf{0.495} & 0.377          & 0.344          & 0.437          & 0.916    & \nico{0.280}        \\ \hline
\textbf{\pced (\dtsDFLT)$^+$}    & 0.349          & 0.898          & 0.444          & 0.489          & 0.814    & \nico{0.323}        \\ \hline
\hline
\ecnet (\dtsEC)                  & 0.365          & 0.503          & 0.341          & 0.397          & 0.845    & \nico{0.248}        \\ \hline
\lo{\textbf{\pcedtc} (\dtsEC)}  & \lo{0.254}      & \lo{0.747}     & \lo{0.285}     & \lo{0.365}     & \lo{0.676} & \nico{0.223}      \\ \hline
\hline
\fctc (\dtsSHREC)                & 0.406          & 0.715          & 0.392          & 0.480          & 0.849    & \nico{0.316}        \\ \hline
\cnntc (\dtsSHREC)               & 0.207          & 0.881          & 0.202          & 0.248          & 0.622    & \nico{0.142}        \\ \hline
\textbf{\pcedtc (\dtsSHREC)}     & 0.441          & 0.872          & 0.426          & 0.522          & 0.874    & \nico{0.353}        \\ \hline
\end{tabular}
\caption{Quantitative evaluation on \dtsSHREC: median scores. For 3-classes methods marked with $^+$, we compare both the smooth and sharp edges classes with the ground truth, i.e. Positives=Sharp+Smooth.
\label{tab:quant_shrec}
}
\end{table}

\begin{table*}[t]
\begin{center}
\small
\setlength\tabcolsep{1pt} 
\begin{tabular}{|c||c|c||c|c|c|c||c||c|c|c|} 
\hline 
Dataset         & \#obj & \#vert. & \ca               & \fee              & \pcp               &  \ecnet                & \textit{GLS}     & \cnn (total)                 & \fc (total)                & \pced (total)  \\ 
\hline 
\dataLoudunSmall        & 1     & 1M      & \duration{14.3}   & \duration{90.88}  & \duration{2100}*   & -                  & \duration{19.3}  & \twodurations{54.39}{19.3}   & \twodurations{1.8}{19.3}   & \twodurations{2.1}{19.3}   \\
\hline 
\dataEmpire          & 1     & 1.2M    & \duration{15.7}   & \duration{175.16} & \duration{3262.67}*& \duration{343.19}* & \duration{36.4}  & \twodurations{53.38}{36.4}   & \twodurations{1.8}{36.4}   & \twodurations{2.1}{36.4}   \\
\hline 
\dataLans            & 1     & 1.23M   & \duration{16.4}   & \duration{226.62} & \duration{3608.33}*& \duration{373.33}* & \duration{21.0}  & \twodurations{55.06}{21.0}   & \twodurations{1.8}{21.0}   & \twodurations{2.1}{21.0}   \\
\hline 
\dataChurch          & 1     & 1.9M    & \duration{23.4}   & \duration{353.97} & \duration{6400}*   & -                  & \duration{58.4}  & \twodurations{86.19}{58.4}   &  \twodurations{2.9}{58.4}   &   \twodurations{3.4}{58.4}   \\
\hline 
\dataPisa  & 1     & 2.5M    & \duration{31.8}   &  \duration{363.247} & \duration{10600}*  & -                  & \duration{53.6}  & \twodurations{117.26}{53.6}  &  \twodurations{3.9}{53.6}   &  \twodurations{4.7}{53.6}   \\
\hline 
\dataEuler           & 1     & 3.9M    & \duration{51.2}   & \duration{802.647 } &  -                 & -                  & \duration{115.2} & \twodurations{180.32}{115.2} &  \twodurations{6}{115.2}    &  \twodurations{7.1}{115.2}   \\
\hline 
\dataMunich          & 1     & 6M      & \duration{91}     & \duration{1593.529} &  -                 & -                  & \duration{119.2} & \twodurations{281.84}{119.2} &  \twodurations{9.9}{119.2} &  \twodurations{11.6}{119.2}    \\
\hline
\dataTrainSt       & 1     & 12.45M  & \duration{155}    & \duration{3044.943 } & -                  & -                  & \duration{331.9} & \twodurations{557.92}{331.9} &  \twodurations{19}{331.9}  &  \twodurations{25.7}{331.9}    \\
\hline 
\dataLoudunBig       & 1     & 35M     & \duration{449}    &  \duration{5425.904}&  -                 & -                  & \duration{592.6} & \twodurations{1598.33}{592.6}&  \twodurations{66.5}{592.6}  &   \twodurations{69.8}{592.6}   \\
\hline 
{\dataRueMadame}       & {1}     & 20M     & -    &  - &  -                 & -                  & \duration{683.518} & - & -  &   \twodurations{34.21}{717.728}   \\
\hline 
\hline 
\dtsDFLT        & 8     &  55k    &  \duration{0.533} & \duration{4.41}   & \duration{22.11}*  & \duration{2.09}*   & \duration{1.08}  & \twodurations{2.03}{1.08} & \twodurations{0.302}{1.08}  & \twodurations{0.254}{1.08}  \\
\hline 
\dtsSHREC       & 15    & 654k    &  \duration{9.7}   & \duration{41.47}  & \duration{354.97}* & \chems{\duration{168.92}*}   & \duration{14}    & \twodurations{28.58}{14}  & \twodurations{3.1}{14}  & \twodurations{3.1}{14}    \\
\hline 
\dtsABC         & 7167  & 312.3M  &  \duration{11700} & \duration{64200} & \duration{604800}*  & \duration{1800000}*& \duration{9300}  & \twodurations{5100}{9300}    & \twodurations{1540}{9300} & \twodurations{1530}{9300}  \\
\hline
\end{tabular} 
\caption{
Timing comparison for classification, where (s) stands for seconds, {\color{colorMinutes}(m)} for minutes, {\color{colorHours}(h)} for hours, and {\color{colorDays}(d)} for days. 
For datasets (e.g., \dtsDFLT, \dtsSHREC and \dtsABC) we report the time needed to process all the models. 
The column \textit{GLS} corresponds to the precomputation of the GLS descriptors.
For columns CNN, \fc and PCED, we report first the classification time, followed in brackets by the \texttt{total\_time = GLS+classification\_time}.
Timings marked with * have been obtained using dedicated hardware ({NVIDIA} RTX 6000).
}
\label{tab:visualTime}
\end{center}
\end{table*}

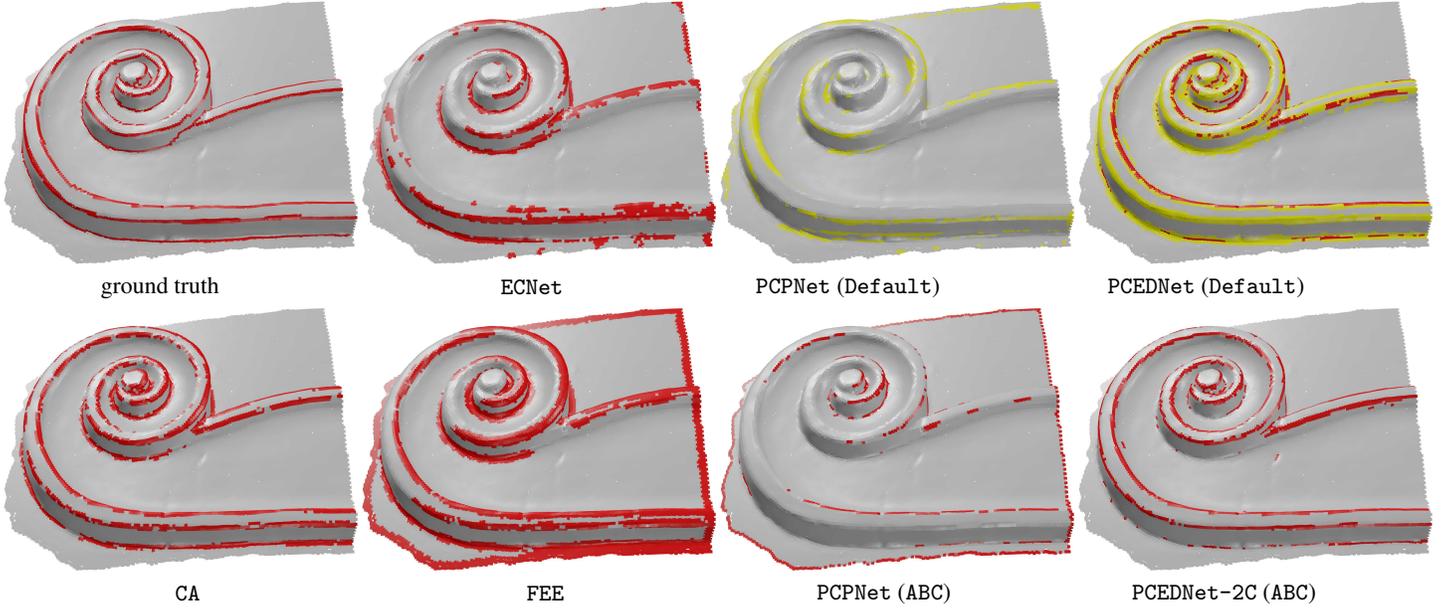
\begin{figure}[t]
\footnotesize
\def\svgwidth{\linewidth}
\begingroup%
  \makeatletter%
  \providecommand\color[2][]{%
    \errmessage{(Inkscape) Color is used for the text in Inkscape, but the package 'color.sty' is not loaded}%
    \renewcommand\color[2][]{}%
  }%
  \providecommand\transparent[1]{%
    \errmessage{(Inkscape) Transparency is used (non-zero) for the text in Inkscape, but the package 'transparent.sty' is not loaded}%
    \renewcommand\transparent[1]{}%
  }%
  \providecommand\rotatebox[2]{#2}%
  \newcommand*\fsize{\dimexpr\f@size pt\relax}%
  \newcommand*\lineheight[1]{\fontsize{\fsize}{#1\fsize}\selectfont}%
  \ifx\svgwidth\undefined%
    \setlength{\unitlength}{5679.75bp}%
    \ifx\svgscale\undefined%
      \relax%
    \else%
      \setlength{\unitlength}{\unitlength * \real{\svgscale}}%
    \fi%
  \else%
    \setlength{\unitlength}{\svgwidth}%
  \fi%
  \global\let\svgwidth\undefined%
  \global\let\svgscale\undefined%
  \makeatother%
  \begin{picture}(1,0.42836392)%
    \lineheight{1}%
    \setlength\tabcolsep{0pt}%
    \put(0,0){\includegraphics[width=\unitlength,page=1]{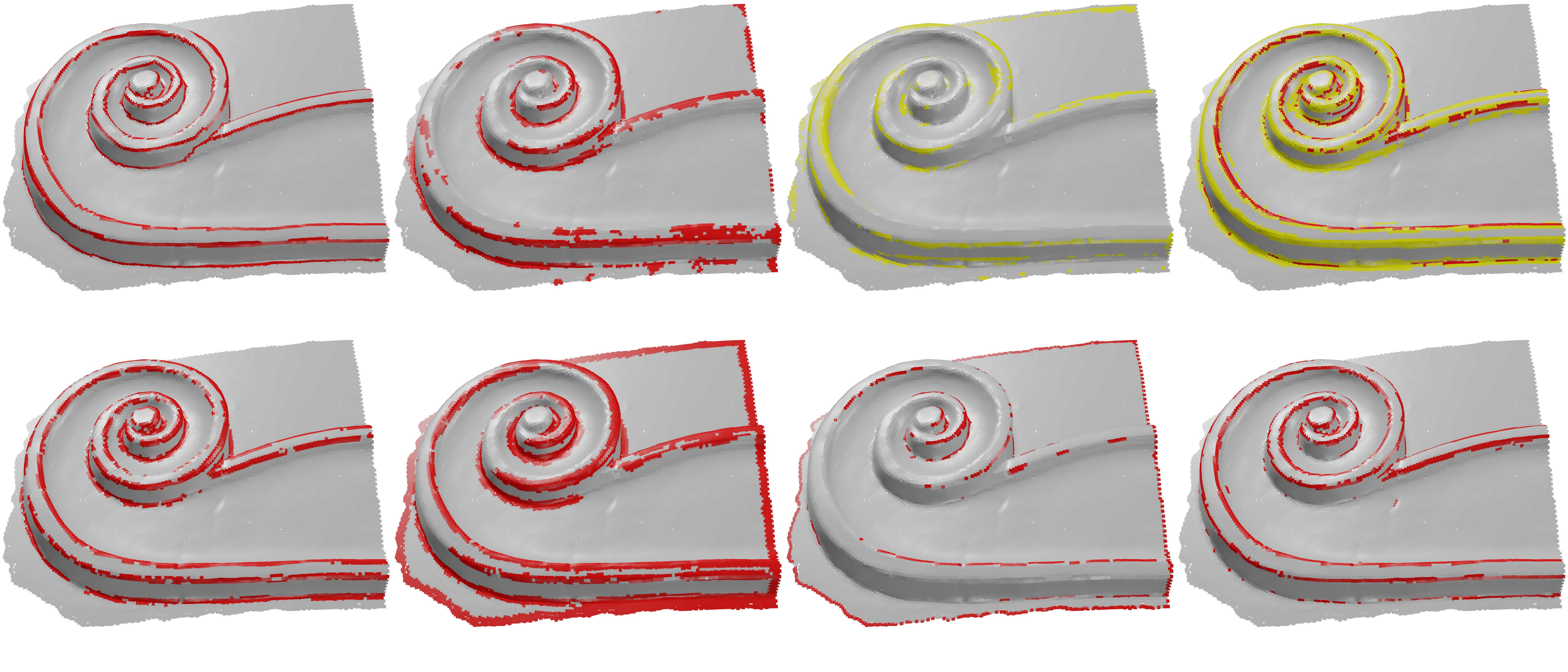}}%
    \put(0.06939941,0.22212287){\makebox(0,0)[lt]{\lineheight{1.25}\smash{\begin{tabular}[t]{l}ground truth\end{tabular}}}}%
    \put(0.34841482,0.22137924){\makebox(0,0)[lt]{\lineheight{1.25}\smash{\begin{tabular}[t]{l}\ecnet\end{tabular}}}}%
    \put(0.52647835,0.22212287){\makebox(0,0)[lt]{\lineheight{1.25}\smash{\begin{tabular}[t]{l}\pcp (\dtsDFLT)\end{tabular}}}}%
    \put(0.77203185,0.22212287){\makebox(0,0)[lt]{\lineheight{1.25}\smash{\begin{tabular}[t]{l}\pced (\dtsDFLT)\end{tabular}}}}%
    \put(0.7889242,0.00835739){\makebox(0,0)[lt]{\lineheight{1.25}\smash{\begin{tabular}[t]{l}\pcedtc (\dtsABC)\end{tabular}}}}%
    \put(0.56899673,0.00835739){\makebox(0,0)[lt]{\lineheight{1.25}\smash{\begin{tabular}[t]{l}\pcp (\dtsABC)\end{tabular}}}}%
    \put(0.36647065,0.00734295){\makebox(0,0)[lt]{\lineheight{1.25}\smash{\begin{tabular}[t]{l}\fee\end{tabular}}}}%
    \put(0.12114819,0.00761375){\makebox(0,0)[lt]{\lineheight{1.25}\smash{\begin{tabular}[t]{l}\ca\end{tabular}}}}%
  \end{picture}%
\endgroup%

\caption{Model \texttt{7} of the \dtsSHREC dataset. Results for \cnn and \fc can be found in the joined website.}
\label{fig:shrec-comp}
\end{figure}

\paragraph{\dtsSHREC dataset}
This dataset has been originally designed to evaluate curve detection algorithms on organic and relatively smooth objects with a comparison to human labelling.
In addition, some objects have strong semantics (e.g. human face), and we observe that 
the ground-truth classification 
seems to take into account this semantic rather than strictly respecting  geometric features.
This makes this \dtsSHREC dataset very challenging to 
\lo{classify},
as illustrated in Table~\ref{tab:quant_shrec} by the lower scores obtain by all approaches, in comparison to the other datasets.
%
In particular, \pcp(\dtsDFLT) fails at classifying edge points (MCC $=0$).
On this dataset, \ecnet tends to produce noisy and thick edges, getting lower score than \ca.
\ca and \fee (we use r$=2$, R$=4$ and th$=0.16$ for this dataset) produce more convincing results, with \ca reaching MCC$=0.39$.
\cnn(\dtsDFLT) get the best scores on this dataset (both MCC and F1), but the same architecture trained on \dtsSHREC gets lower detection rate.
Both \fc and \pced show better stability w.r.t. the training set (MCC $\approx0.4$ for all variants), but lower score than \cnn.
Regardless to the quantitative analysis, the results produced by \pced (\dtsDFLT) are very interesting on this dataset.
By enforcing the detection of both smooth and sharp edges, it better adapts to the smooth nature of the analysed objects and it provides very pertinent results as illustrated in Figure~\ref{fig:shrec-comp}.
\begin{figure*}
\footnotesize
\def\svgwidth{\linewidth}
\begingroup%
  \makeatletter%
  \providecommand\color[2][]{%
    \errmessage{(Inkscape) Color is used for the text in Inkscape, but the package 'color.sty' is not loaded}%
    \renewcommand\color[2][]{}%
  }%
  \providecommand\transparent[1]{%
    \errmessage{(Inkscape) Transparency is used (non-zero) for the text in Inkscape, but the package 'transparent.sty' is not loaded}%
    \renewcommand\transparent[1]{}%
  }%
  \providecommand\rotatebox[2]{#2}%
  \newcommand*\fsize{\dimexpr\f@size pt\relax}%
  \newcommand*\lineheight[1]{\fontsize{\fsize}{#1\fsize}\selectfont}%
  \ifx\svgwidth\undefined%
    \setlength{\unitlength}{4234.5bp}%
    \ifx\svgscale\undefined%
      \relax%
    \else%
      \setlength{\unitlength}{\unitlength * \real{\svgscale}}%
    \fi%
  \else%
    \setlength{\unitlength}{\svgwidth}%
  \fi%
  \global\let\svgwidth\undefined%
  \global\let\svgscale\undefined%
  \makeatother%
  \begin{picture}(1,0.20750632)%
    \lineheight{1}%
    \setlength\tabcolsep{0pt}%
    \put(0,0){\includegraphics[width=\unitlength,page=1]{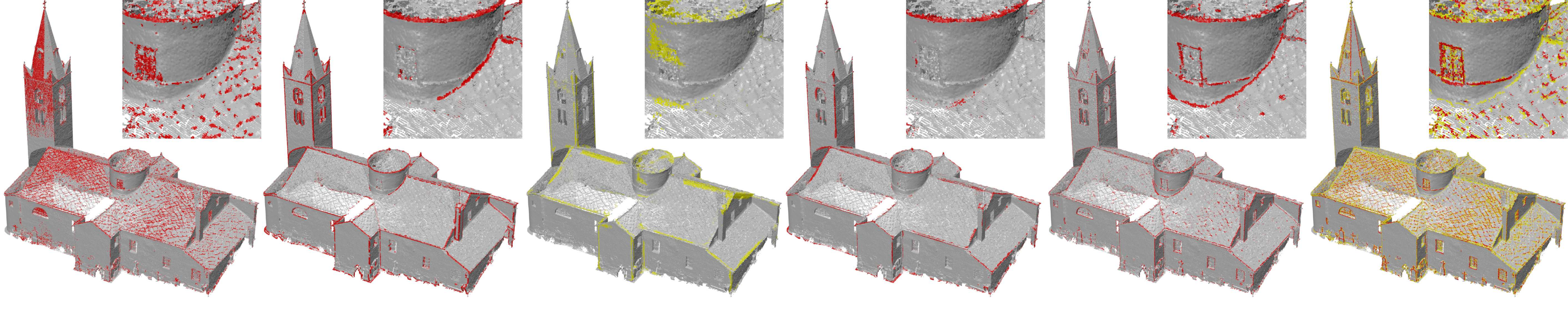}}%
    \put(0.07259618,0.00283866){\makebox(0,0)[lt]{\lineheight{1.25}\smash{\begin{tabular}[t]{l}\ca\end{tabular}}}}%
    \put(0.2375938,0.00310019){\makebox(0,0)[lt]{\lineheight{1.25}\smash{\begin{tabular}[t]{l}\fee\end{tabular}}}}%
    \put(0.36136581,0.00381834){\makebox(0,0)[lt]{\lineheight{1.25}\smash{\begin{tabular}[t]{l}\pcp (\dtsDFLT)\end{tabular}}}}%
    \put(0.5356035,0.00381834){\makebox(0,0)[lt]{\lineheight{1.25}\smash{\begin{tabular}[t]{l}\pcp (\dtsABC)\end{tabular}}}}%
    \put(0.72574187,0.00310019){\makebox(0,0)[lt]{\lineheight{1.25}\smash{\begin{tabular}[t]{l}\ecnet\end{tabular}}}}%
    \put(0.85495245,0.00381834){\makebox(0,0)[lt]{\lineheight{1.25}\smash{\begin{tabular}[t]{l}\pced (\dtsDFLT)\end{tabular}}}}%
  \end{picture}%
\endgroup%

\caption{\dataLans model. Results for \cnn and \fc can be found in the joined website.}
\label{fig:lans}
\end{figure*}

\subsection{Visual evaluation}
\label{subsec:visualEvaluation}
In this section we present visual results on acquired datasets without ground truth labelling.
All these scenes are presented in more details in the accompanying video
and in the joined website.
For all the results presented in this section, we compare \pcp(\dtsDFLT), \pcp(\dtsABC) and \lo{\ecnet(\dtsEC)} (on smallest scenes only), \fc(\dtsDFLT), \cnn(\dtsDFLT), \pced(\dtsDFLT), \ca and \fee.
Table~\ref{tab:visualTime} shows precomputation (GLS) and classification times for all approaches.
We see that processing time for \pced (including precomputation) remains of the same order of magnitude than less robust approaches as \ca{} and \fee, and outperforms \pcp and \ecnet (up to two orders of magnitude). 
Once trained, our approach also avoids the tedious parameter tuning of geometric methods.

Figure~\ref{fig:lans} represents a part of \dataLans without outliers and with very few noise. 
In this example,
\fee detects larger scale edges cleanly, yet it misses fine details.
\ca detects edges, but it still struggles with surface noise and irregularities. 
\pcp has an improved behavior over the noisy model and is able to detect smooth-edges. 
However, it still fails at classifying sharp-edges. 
Similar to \fee and \pcp, \fc extracts large scale edges leaving out finer details, which highlights the importance of scale separation layers in \pced network architecture.
Both \cnn and \pced produce visually convincing results, but \cnn does not detect some fine details that are adequately detected by \pced.
\ecnet also produces good results on this example, detecting most of the edges and some of the thin details.

\begin{figure}[htb]
\footnotesize
\def\svgwidth{\linewidth}
\begingroup%
  \makeatletter%
  \providecommand\color[2][]{%
    \errmessage{(Inkscape) Color is used for the text in Inkscape, but the package 'color.sty' is not loaded}%
    \renewcommand\color[2][]{}%
  }%
  \providecommand\transparent[1]{%
    \errmessage{(Inkscape) Transparency is used (non-zero) for the text in Inkscape, but the package 'transparent.sty' is not loaded}%
    \renewcommand\transparent[1]{}%
  }%
  \providecommand\rotatebox[2]{#2}%
  \newcommand*\fsize{\dimexpr\f@size pt\relax}%
  \newcommand*\lineheight[1]{\fontsize{\fsize}{#1\fsize}\selectfont}%
  \ifx\svgwidth\undefined%
    \setlength{\unitlength}{2916bp}%
    \ifx\svgscale\undefined%
      \relax%
    \else%
      \setlength{\unitlength}{\unitlength * \real{\svgscale}}%
    \fi%
  \else%
    \setlength{\unitlength}{\svgwidth}%
  \fi%
  \global\let\svgwidth\undefined%
  \global\let\svgscale\undefined%
  \makeatother%
  \begin{picture}(1,0.70427789)%
    \lineheight{1}%
    \setlength\tabcolsep{0pt}%
    \put(0,0){\includegraphics[width=\unitlength,page=1]{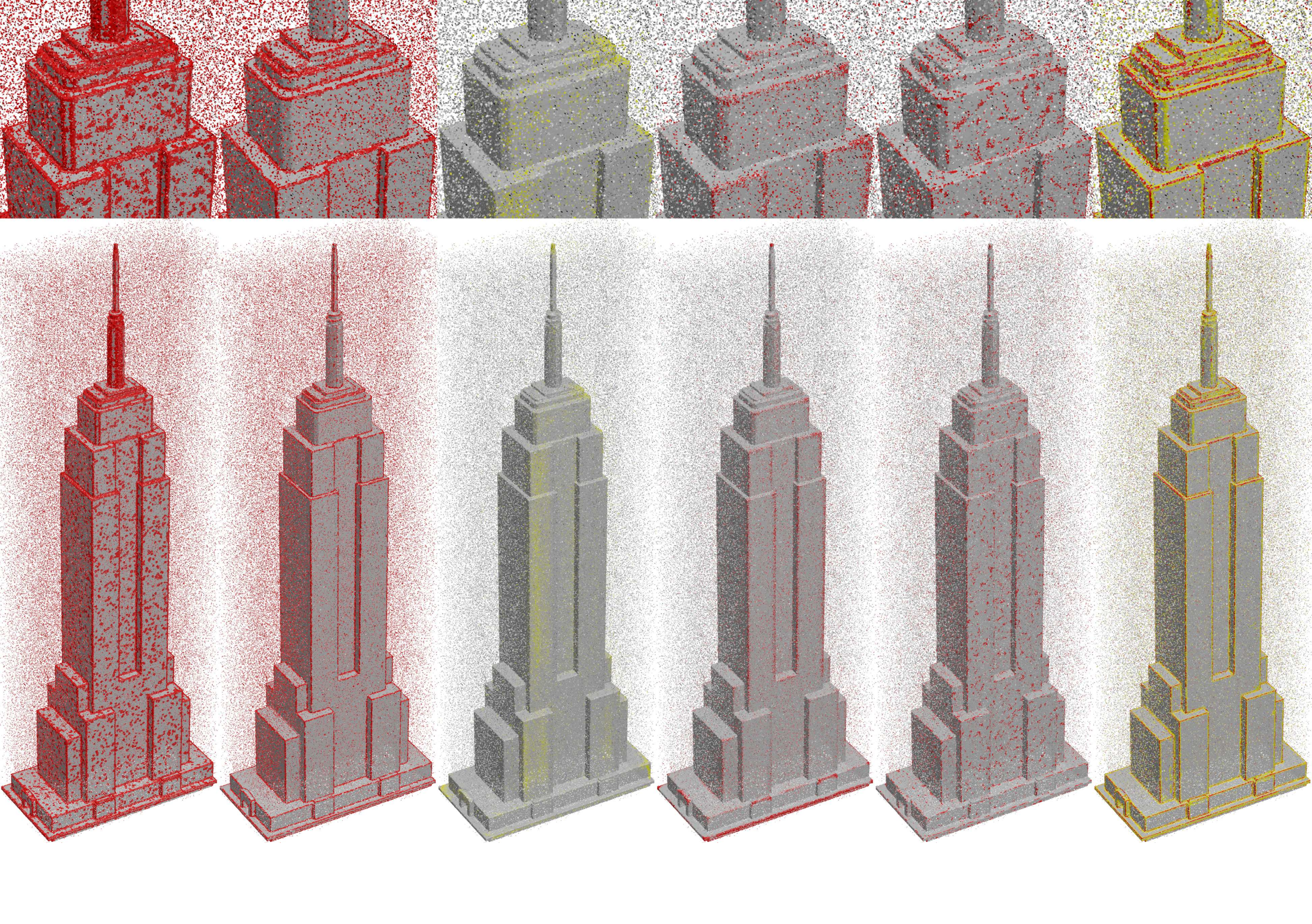}}%
    \put(0.05376718,0.03497459){\makebox(0,0)[lt]{\lineheight{1.25}\smash{\begin{tabular}[t]{l}\ca\end{tabular}}}}%
    \put(0.22914811,0.03535436){\makebox(0,0)[lt]{\lineheight{1.25}\smash{\begin{tabular}[t]{l}\fee\end{tabular}}}}%
    \put(0.3778739,0.03639724){\makebox(0,0)[lt]{\lineheight{1.25}\smash{\begin{tabular}[t]{l}\pcp\end{tabular}}}}%
    \put(0.55193684,0.03639724){\makebox(0,0)[lt]{\lineheight{1.25}\smash{\begin{tabular}[t]{l}\pcp \end{tabular}}}}%
    \put(0.71379761,0.03535436){\makebox(0,0)[lt]{\lineheight{1.25}\smash{\begin{tabular}[t]{l}\ecnet\end{tabular}}}}%
    \put(0.87346232,0.03639724){\makebox(0,0)[lt]{\lineheight{1.25}\smash{\begin{tabular}[t]{l}\pced \end{tabular}}}}%
    \put(0.36178711,0.00671795){\makebox(0,0)[lt]{\lineheight{1.25}\smash{\begin{tabular}[t]{l}(\dtsDFLT)\end{tabular}}}}%
    \put(0.55915895,0.00671795){\makebox(0,0)[lt]{\lineheight{1.25}\smash{\begin{tabular}[t]{l}(\dtsABC)\end{tabular}}}}%
    \put(0.86429536,0.00671795){\makebox(0,0)[lt]{\lineheight{1.25}\smash{\begin{tabular}[t]{l}(\dtsDFLT)\end{tabular}}}}%
  \end{picture}%
\endgroup%

\caption{\dataEmpire model. This point cloud contains a clean structure with a severe amount of outliers. Results for \cnn and \fc can be found in the joined website.}
\label{fig:empire}
\end{figure}

Figure~\ref{fig:empire} presents the behavior of each method when processing models with important noise and 
\nico{a lot}
of outliers. 
As observed on synthetic datasets, \ecnet is strongly penalized by the outliers, and most of the flat areas are misclassified as edges.
\ca detects all the edge points on the synthetic model\nico{s} but 
also considers the outliers as edges. 
\pcp classification is weakened by the outliers and does not provide any positive results.
\fc shows results similar to \ca, the network is capable of extracting different edges, yet it is still very sensitive to outliers.
\cnn and \pced both obtain a high recall on edges, with \pced being the least sensitive to noise.

\begin{figure}[htb]
\footnotesize
\def\svgwidth{\linewidth}
\begingroup%
  \makeatletter%
  \providecommand\color[2][]{%
    \errmessage{(Inkscape) Color is used for the text in Inkscape, but the package 'color.sty' is not loaded}%
    \renewcommand\color[2][]{}%
  }%
  \providecommand\transparent[1]{%
    \errmessage{(Inkscape) Transparency is used (non-zero) for the text in Inkscape, but the package 'transparent.sty' is not loaded}%
    \renewcommand\transparent[1]{}%
  }%
  \providecommand\rotatebox[2]{#2}%
  \newcommand*\fsize{\dimexpr\f@size pt\relax}%
  \newcommand*\lineheight[1]{\fontsize{\fsize}{#1\fsize}\selectfont}%
  \ifx\svgwidth\undefined%
    \setlength{\unitlength}{4314bp}%
    \ifx\svgscale\undefined%
      \relax%
    \else%
      \setlength{\unitlength}{\unitlength * \real{\svgscale}}%
    \fi%
  \else%
    \setlength{\unitlength}{\svgwidth}%
  \fi%
  \global\let\svgwidth\undefined%
  \global\let\svgscale\undefined%
  \makeatother%
  \begin{picture}(1,0.40547144)%
    \lineheight{1}%
    \setlength\tabcolsep{0pt}%
    \put(0,0){\includegraphics[width=\unitlength,page=1]{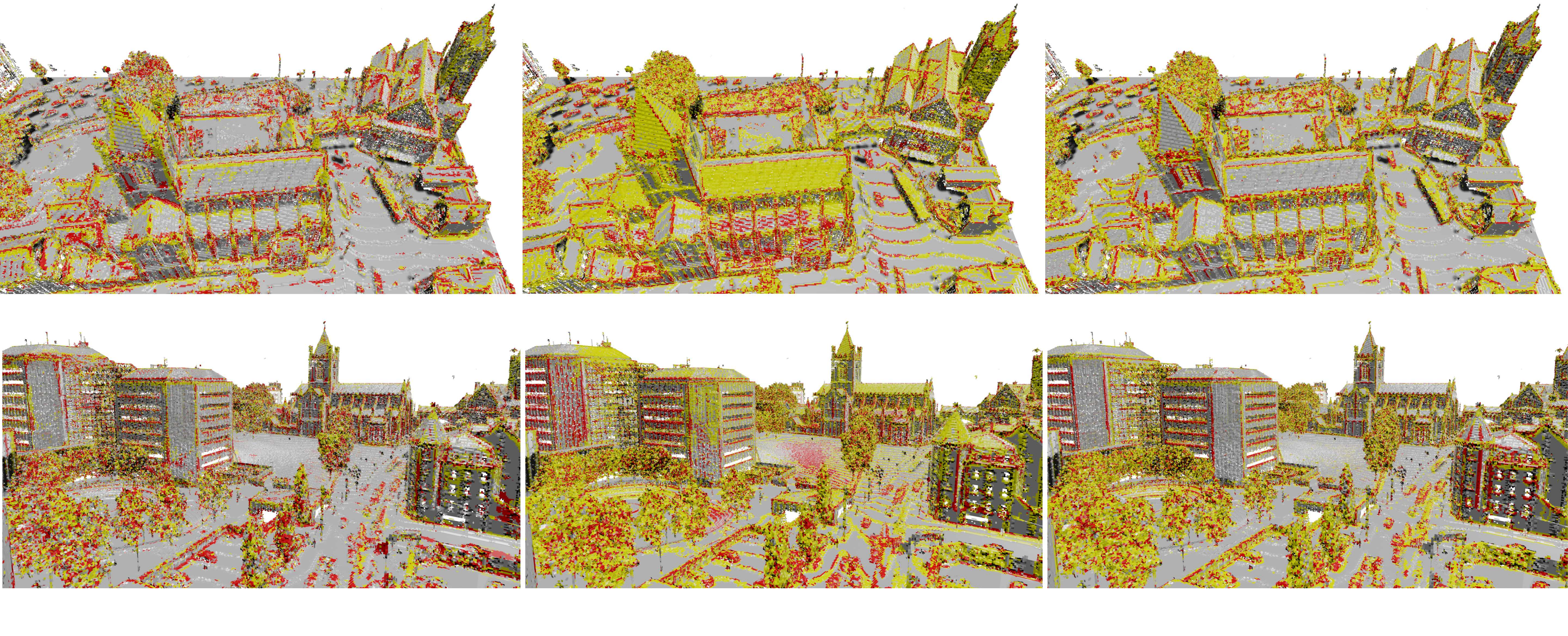}}%
    \put(0.07633939,0.00695898){\makebox(0,0)[lt]{\lineheight{1.25}\smash{\begin{tabular}[t]{l}\cnn (\dtsDFLT)\end{tabular}}}}%
    \put(0.41287227,0.00731552){\makebox(0,0)[lt]{\lineheight{1.25}\smash{\begin{tabular}[t]{l}\fc (\dtsDFLT)\end{tabular}}}}%
    \put(0.7293438,0.00829458){\makebox(0,0)[lt]{\lineheight{1.25}\smash{\begin{tabular}[t]{l}\pced (\dtsDFLT)\end{tabular}}}}%
  \end{picture}%
\endgroup%

\caption{\dataChurch. Results for \fee, \fc and \pcp can be found in the joined website.}
\label{fig:CC}
\end{figure}

\cnn, \fc and \pced rely on the same parameterization. 
We thus provide in Figures~\ref{fig:CC} a closer comparison of the results produced by these architectures. 
In Figures~\ref{fig:CC}, even though all networks generate similar results, the \cnn and \fc become less accurate when processing irregular surfaces as trees, cars, light poles, etc.

\begin{figure}[htb]
\footnotesize
\def\svgwidth{\linewidth}
\begingroup%
  \makeatletter%
  \providecommand\color[2][]{%
    \errmessage{(Inkscape) Color is used for the text in Inkscape, but the package 'color.sty' is not loaded}%
    \renewcommand\color[2][]{}%
  }%
  \providecommand\transparent[1]{%
    \errmessage{(Inkscape) Transparency is used (non-zero) for the text in Inkscape, but the package 'transparent.sty' is not loaded}%
    \renewcommand\transparent[1]{}%
  }%
  \providecommand\rotatebox[2]{#2}%
  \newcommand*\fsize{\dimexpr\f@size pt\relax}%
  \newcommand*\lineheight[1]{\fontsize{\fsize}{#1\fsize}\selectfont}%
  \ifx\svgwidth\undefined%
    \setlength{\unitlength}{4357.5bp}%
    \ifx\svgscale\undefined%
      \relax%
    \else%
      \setlength{\unitlength}{\unitlength * \real{\svgscale}}%
    \fi%
  \else%
    \setlength{\unitlength}{\svgwidth}%
  \fi%
  \global\let\svgwidth\undefined%
  \global\let\svgscale\undefined%
  \makeatother%
  \begin{picture}(1,0.57699048)%
    \lineheight{1}%
    \setlength\tabcolsep{0pt}%
    \put(0,0){\includegraphics[width=\unitlength,page=1]{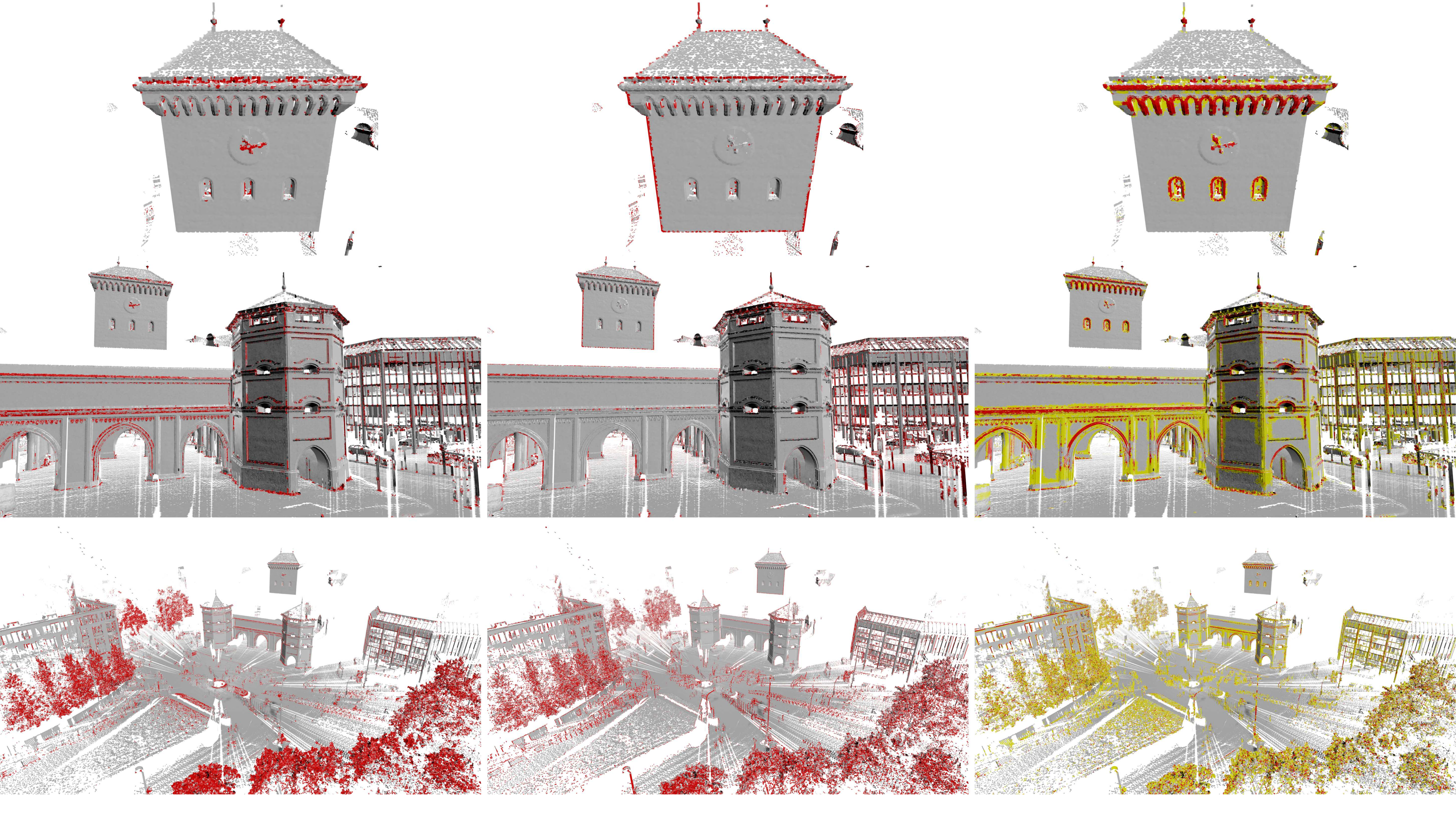}}%
    \put(0.14120906,0.006695){\makebox(0,0)[lt]{\lineheight{1.25}\smash{\begin{tabular}[t]{l}\ca\end{tabular}}}}%
    \put(0.48356001,0.00704797){\makebox(0,0)[lt]{\lineheight{1.25}\smash{\begin{tabular}[t]{l}\fee\end{tabular}}}}%
    \put(0.73568823,0.00801727){\makebox(0,0)[lt]{\lineheight{1.25}\smash{\begin{tabular}[t]{l}\pced (\dtsDFLT)\end{tabular}}}}%
  \end{picture}%
\endgroup%

\caption{\dataMunich model (see Results for \cnn and \fc in the joined website).}
\label{fig:MS}
\end{figure}
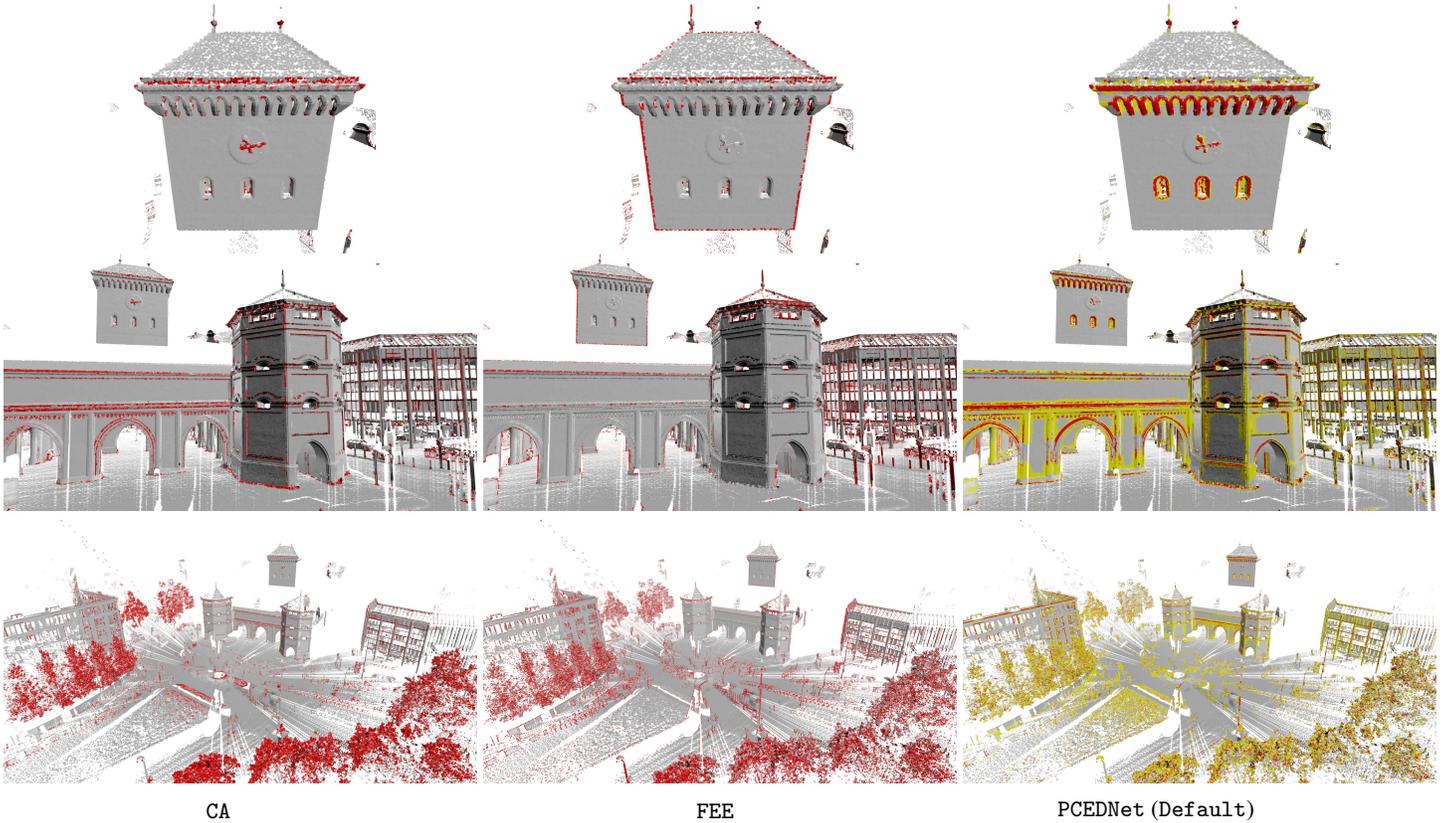

Figure~\ref{fig:MS} shows how our approach performs on a large point cloud (6M points) in comparison with \ca and \fee (other approaches require too much memory/time).
\dataMunich exhibits common irregularities of acquired data, i.e. variable point densities, large gaps, scan noise, and volumetric objects (e.g., trees).
\pced(\dtsDFLT) produces cleaner detection and is less affected by acquisition artefacts than the other approaches.

\begin{figure*}[t]
\centering
\footnotesize
\begin{tabular}{ccccc}
\setlength\tabcolsep{0pt} 
    \includegraphics[width = 0.175\linewidth]{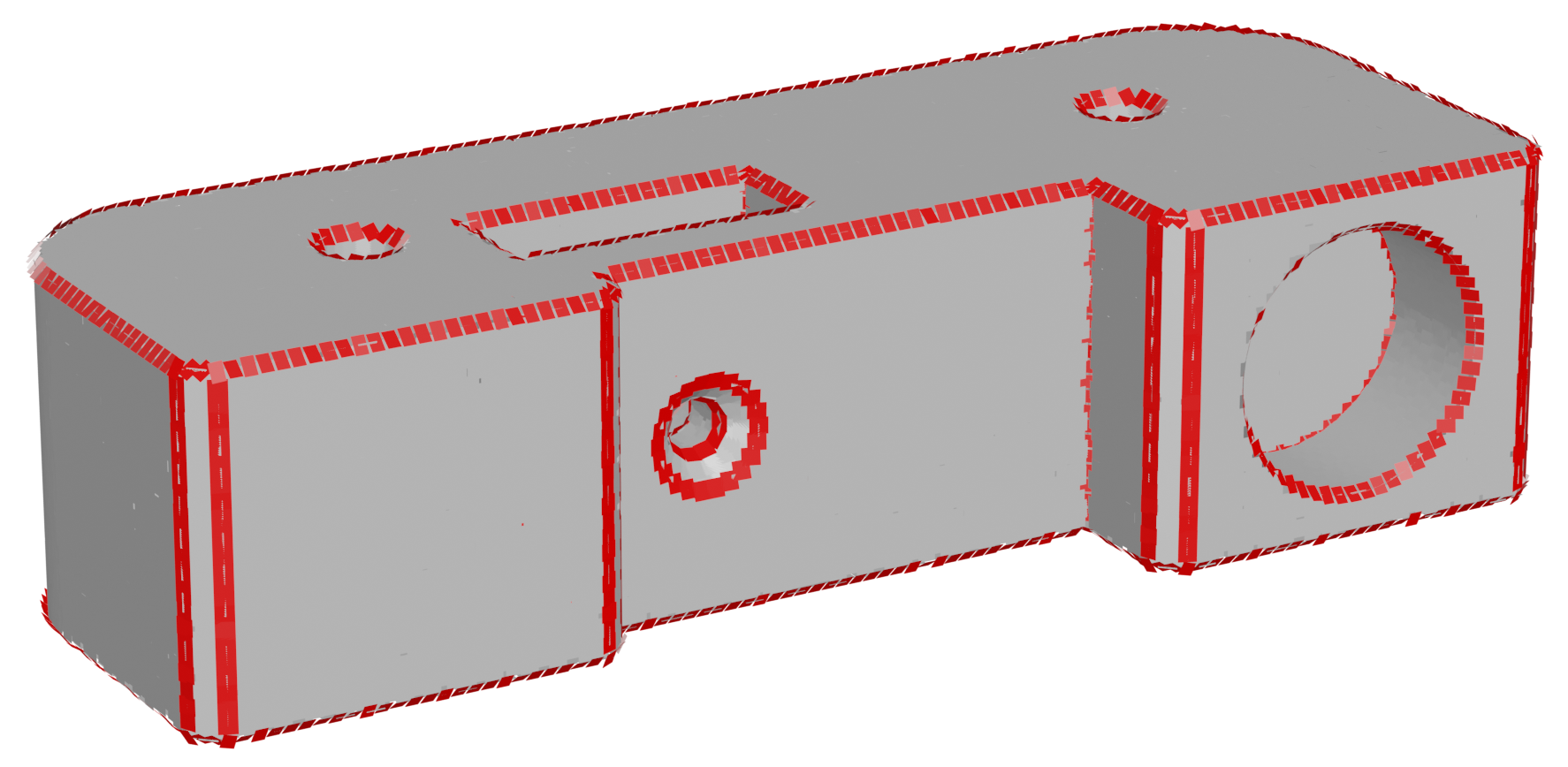}&
    \includegraphics[width = 0.175\linewidth]{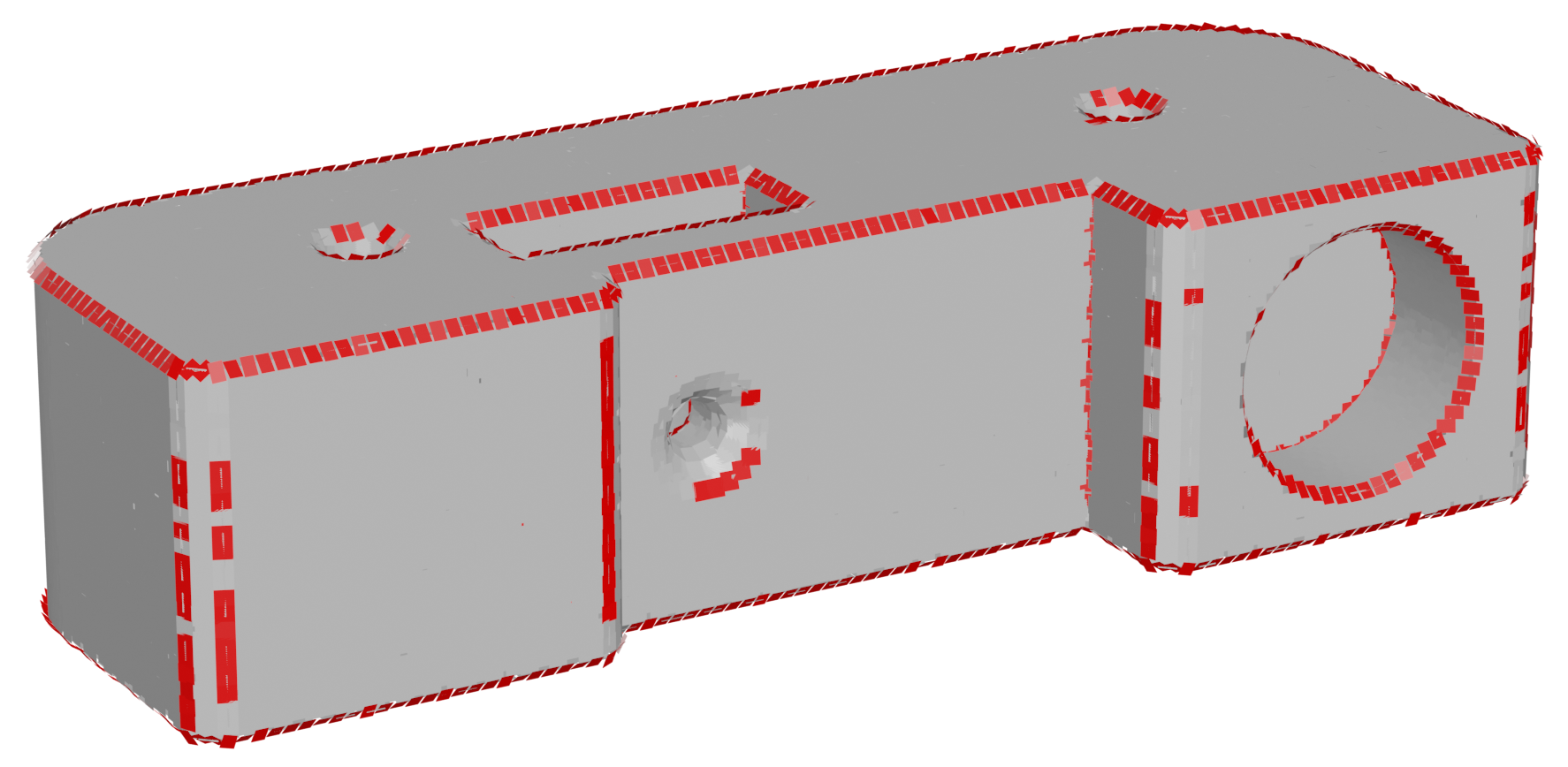}&
    \includegraphics[width = 0.175\linewidth]{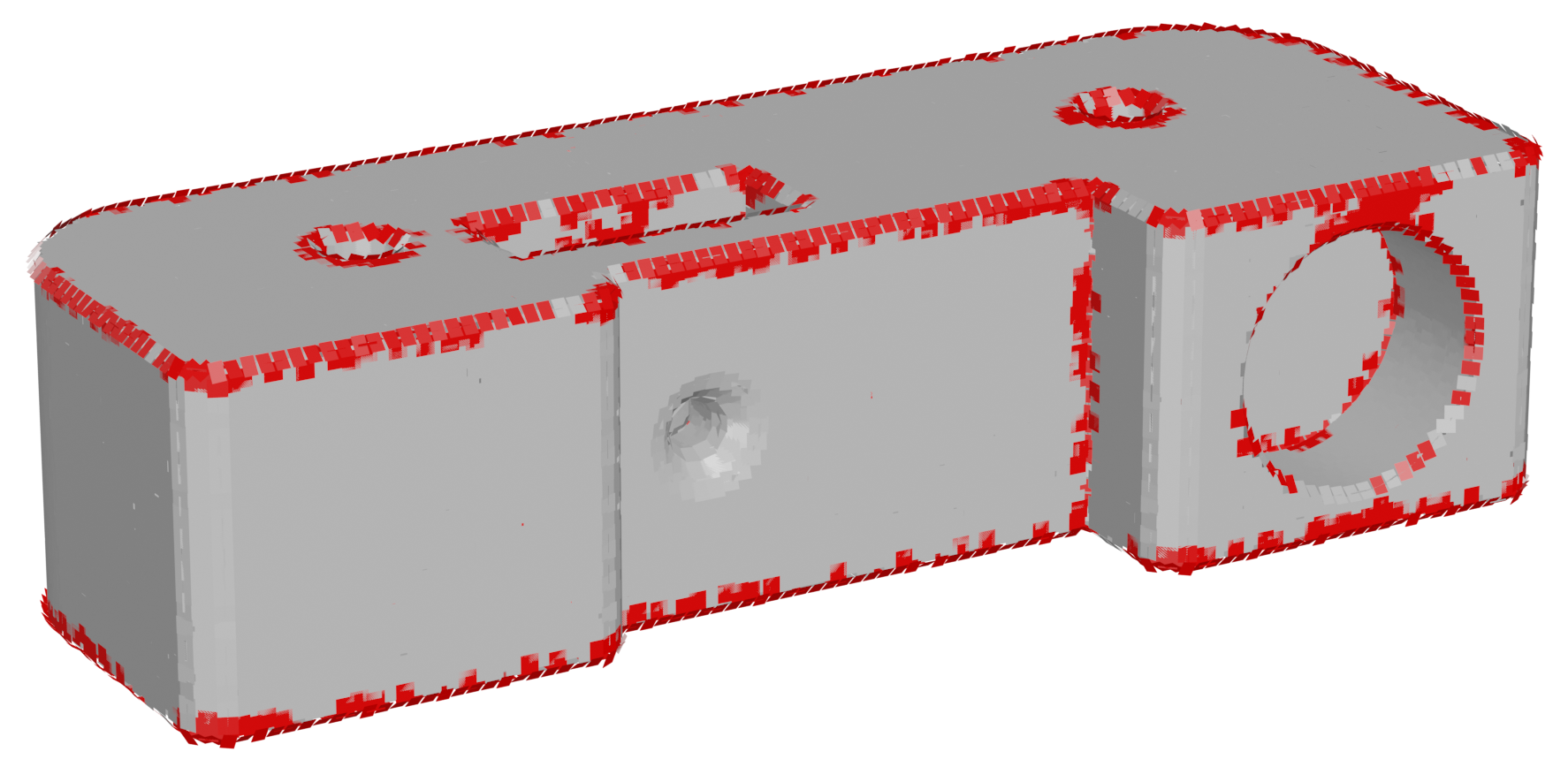}&
    \fbox[trl]{\includegraphics[width = 0.19\linewidth]{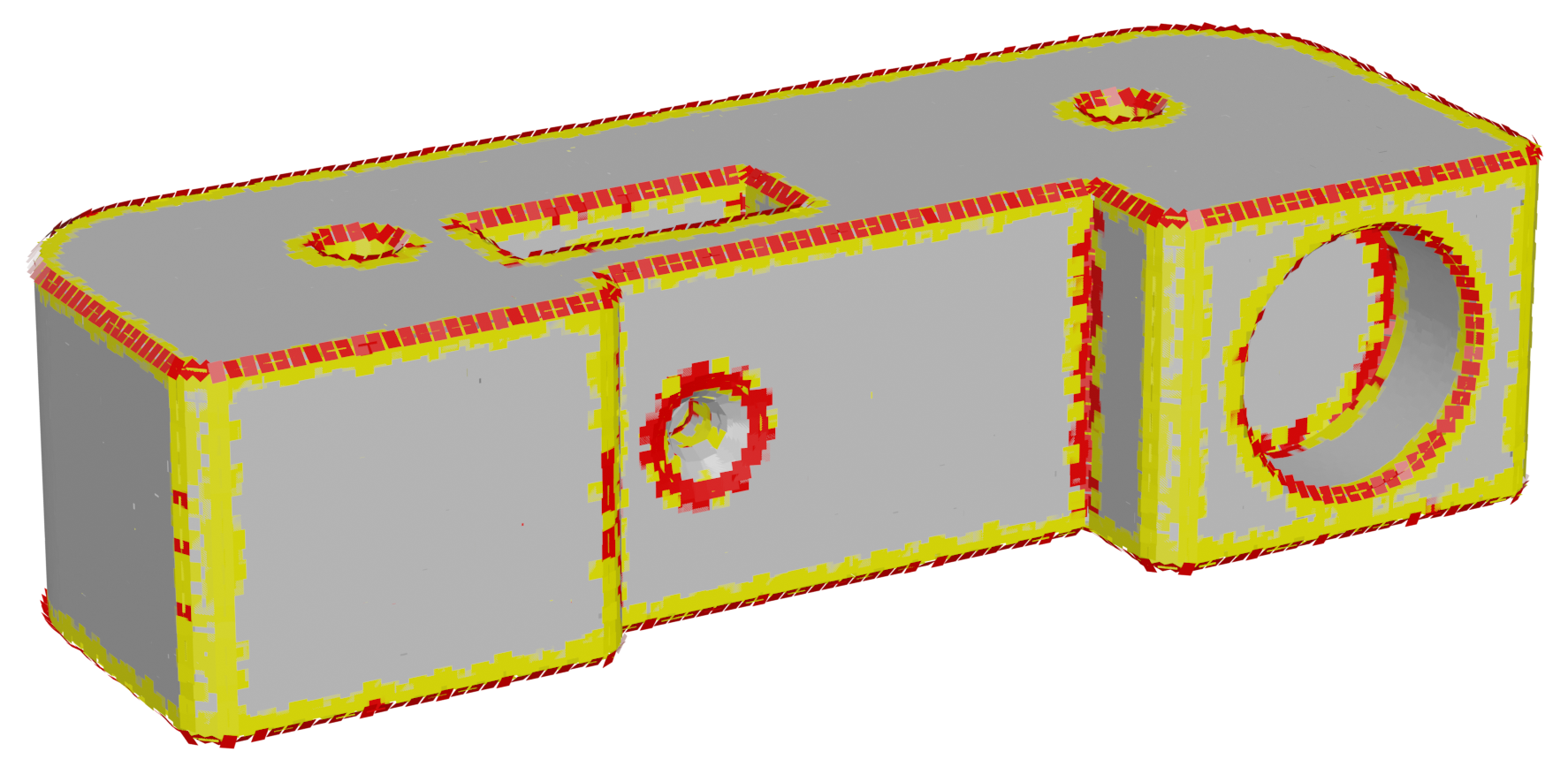}}&
    \includegraphics[width = 0.175\linewidth]{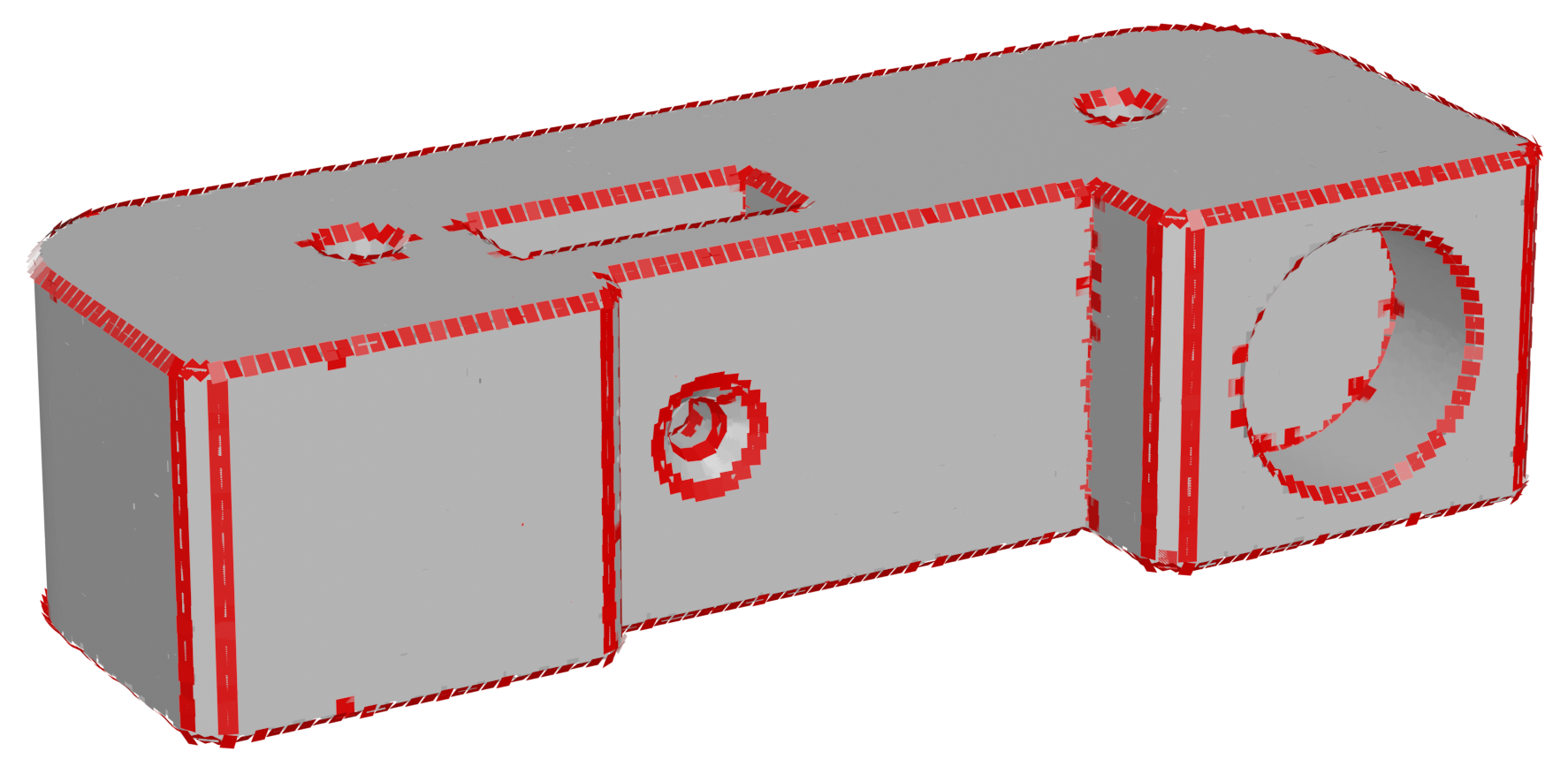}\\
    \includegraphics[width = 0.175\linewidth]{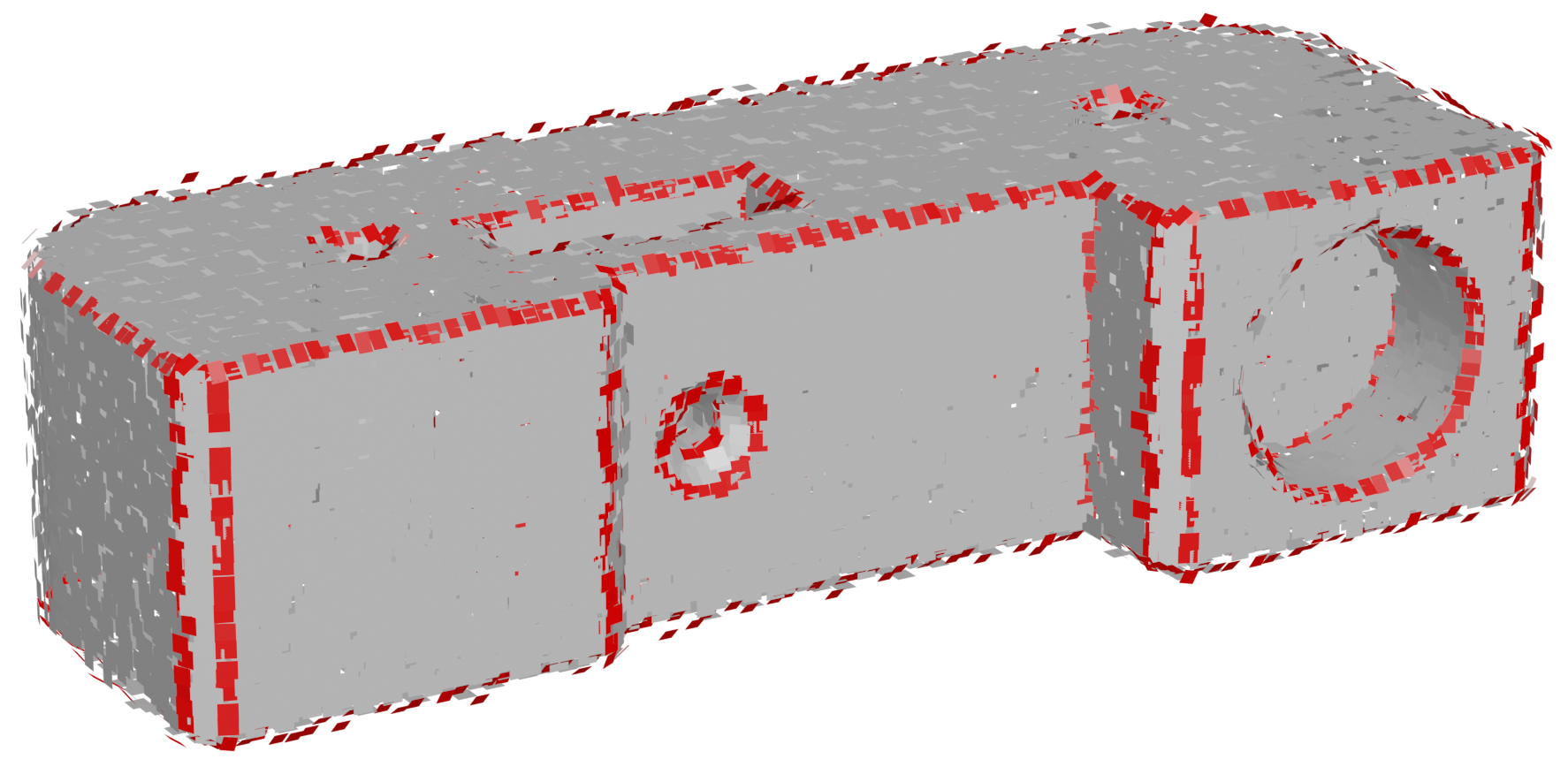}&
    \includegraphics[width = 0.175\linewidth]{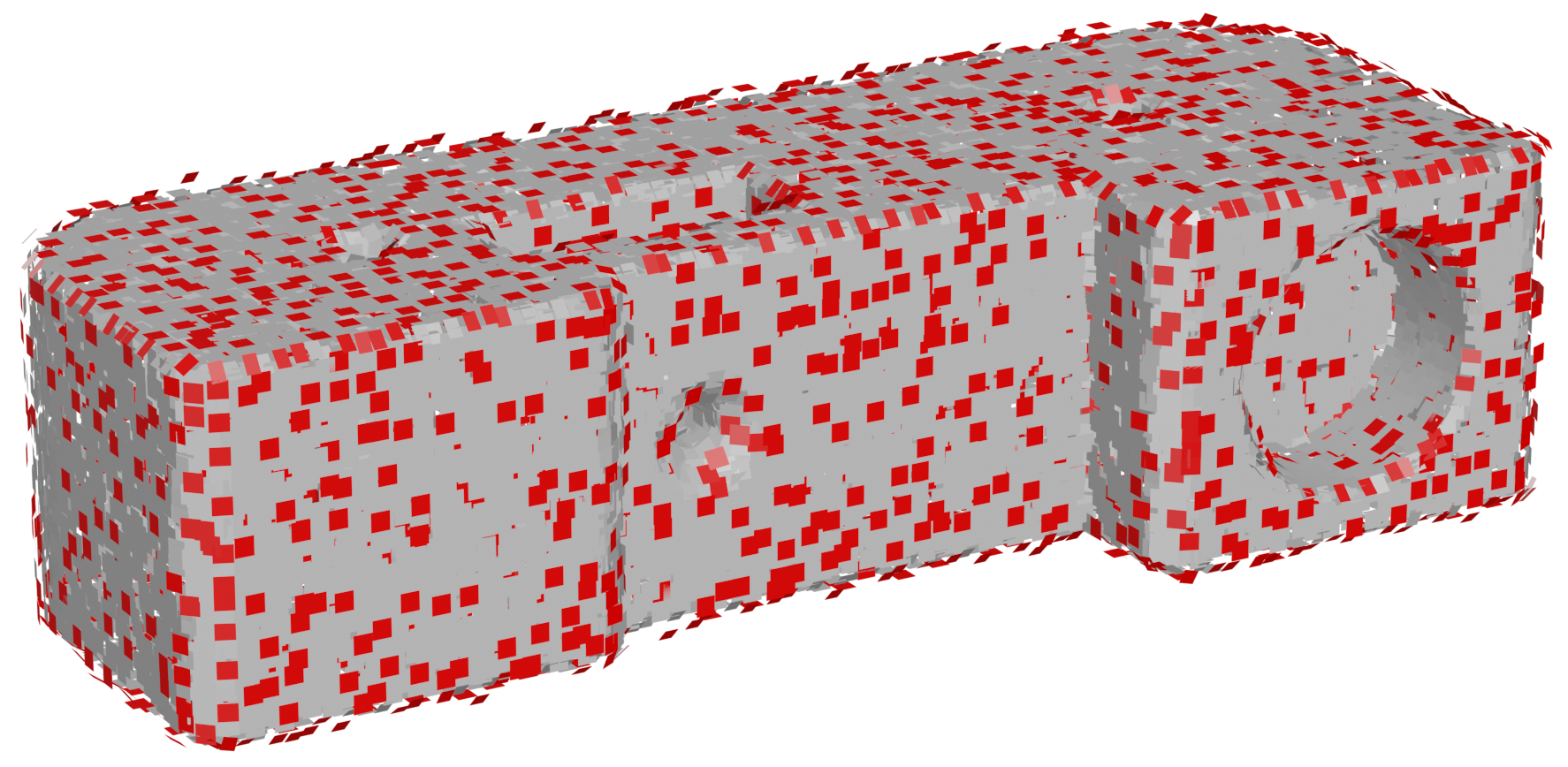}&
    \includegraphics[width = 0.175\linewidth]{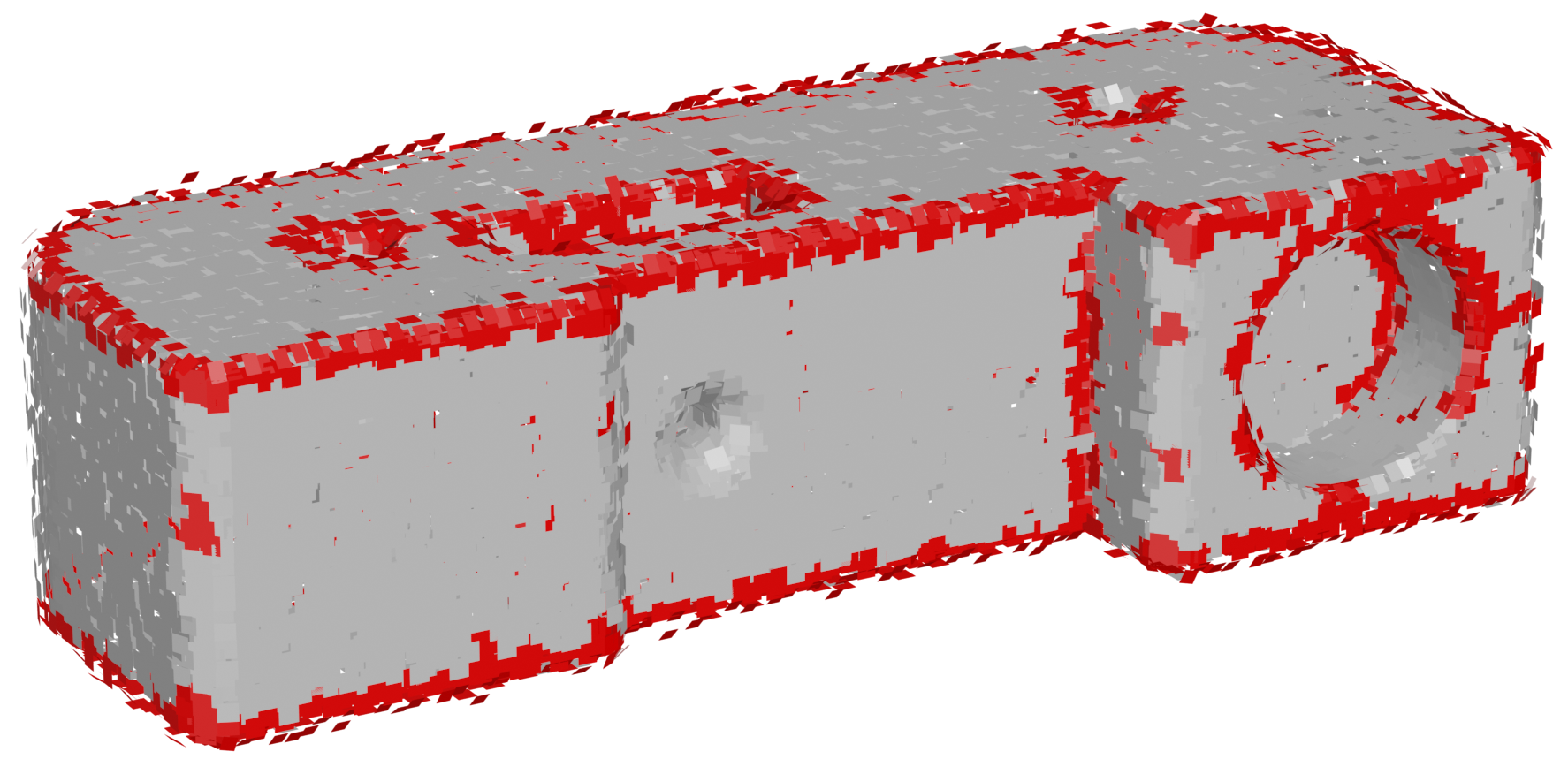}&
    \fbox[brl]{\includegraphics[width = 0.19\linewidth]{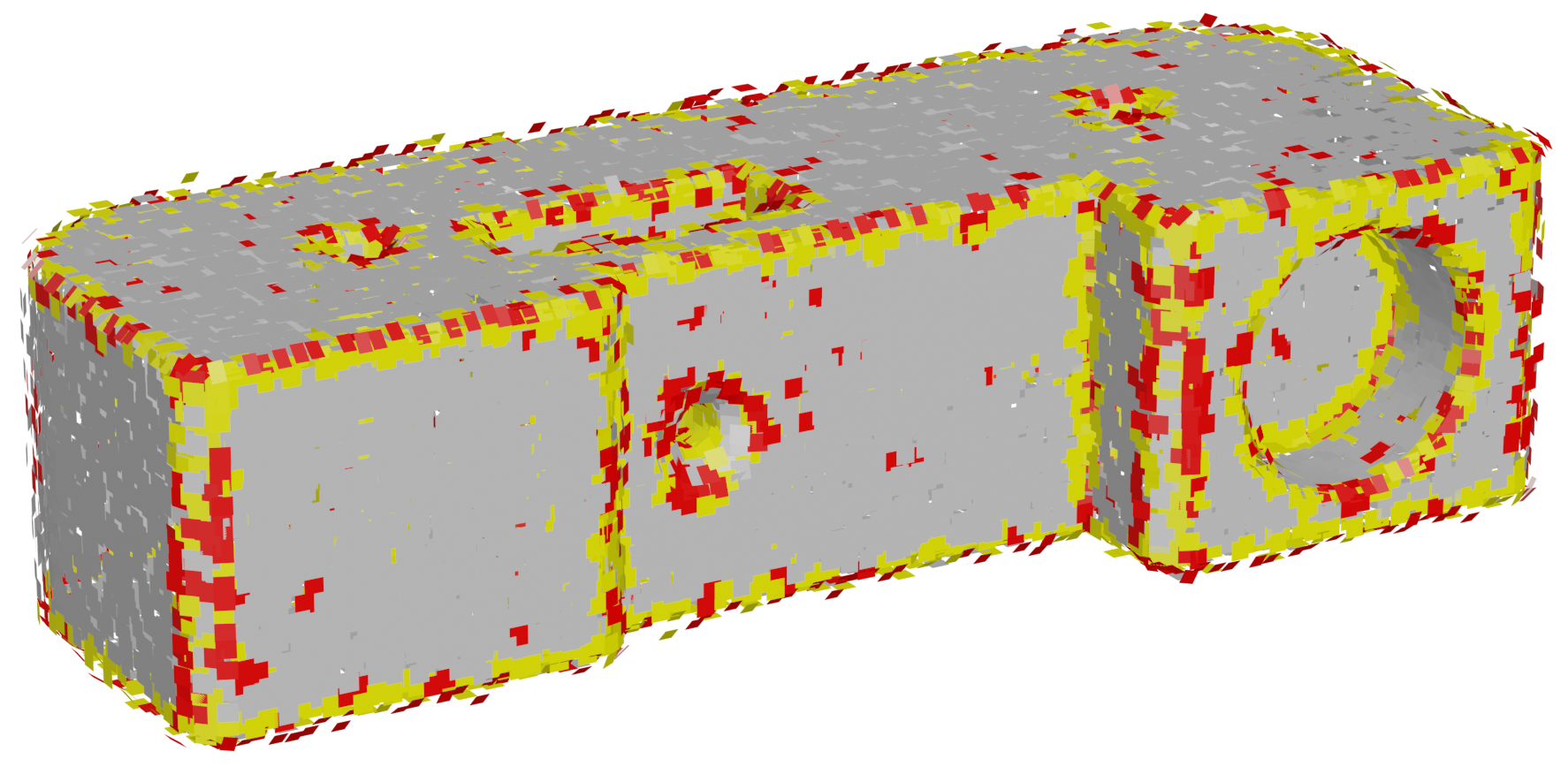}}&
    \includegraphics[width = 0.175\linewidth]{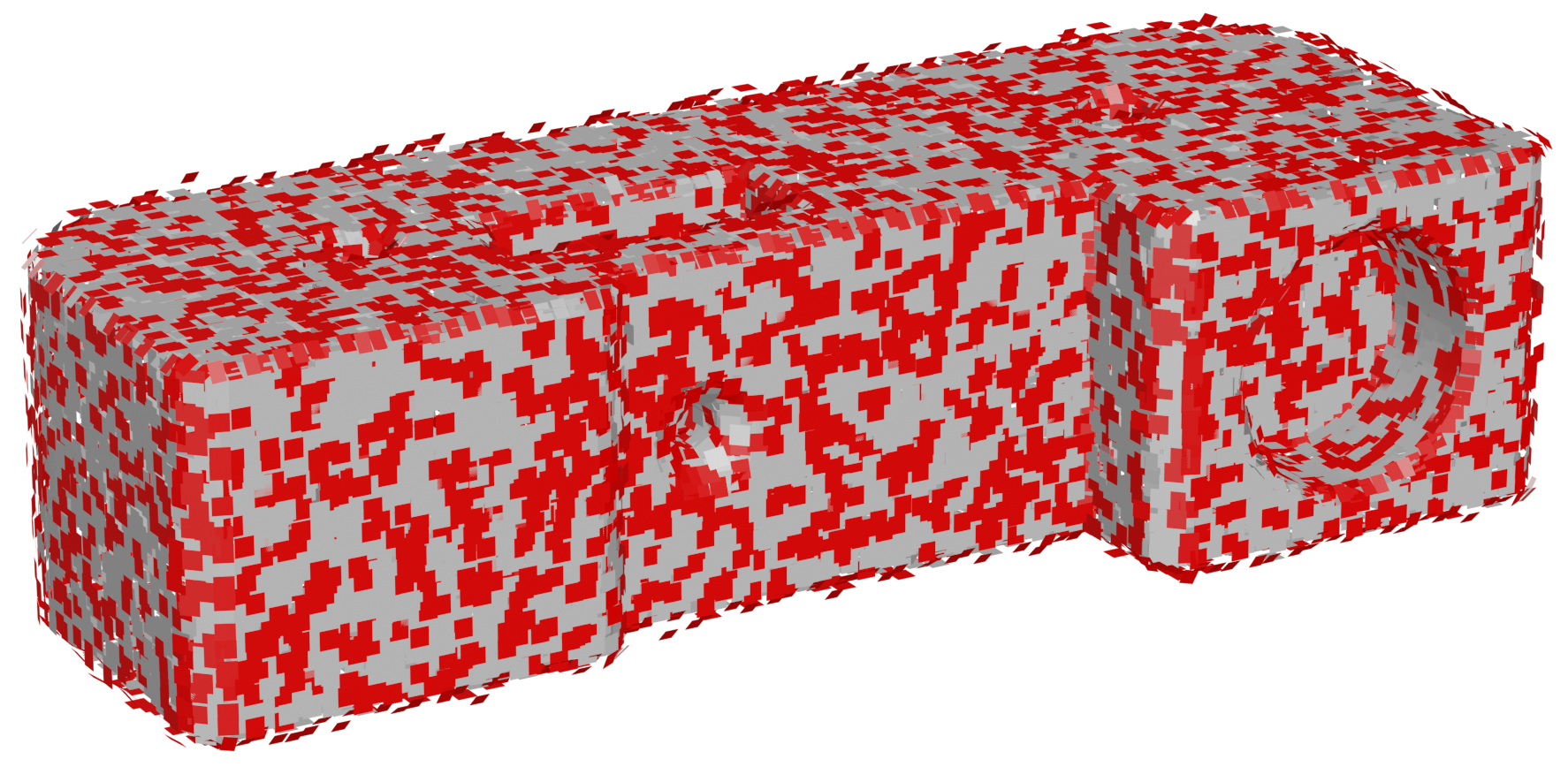}\\
    Ground Truth & \pcptc & \ecnet & \pced (\dtsDFLT) & \pcedtc (\dtsABC)  
    \end{tabular}
\caption{{Top: Classification produced by the different networks on model $0133$ of the \dtsABC dataset. Bottom: Same model altered with a random Gaussian noise of deviation $\sigma=0.04$. }}
\label{fig:NoiseABC}
\end{figure*}

\subsection{Behavior on noisy data}
\label{subsec:noise}

\begin{figure*}[htb]
    \centering
    \footnotesize
    \centering
    \footnotesize
    \def\svgwidth{0.196\linewidth}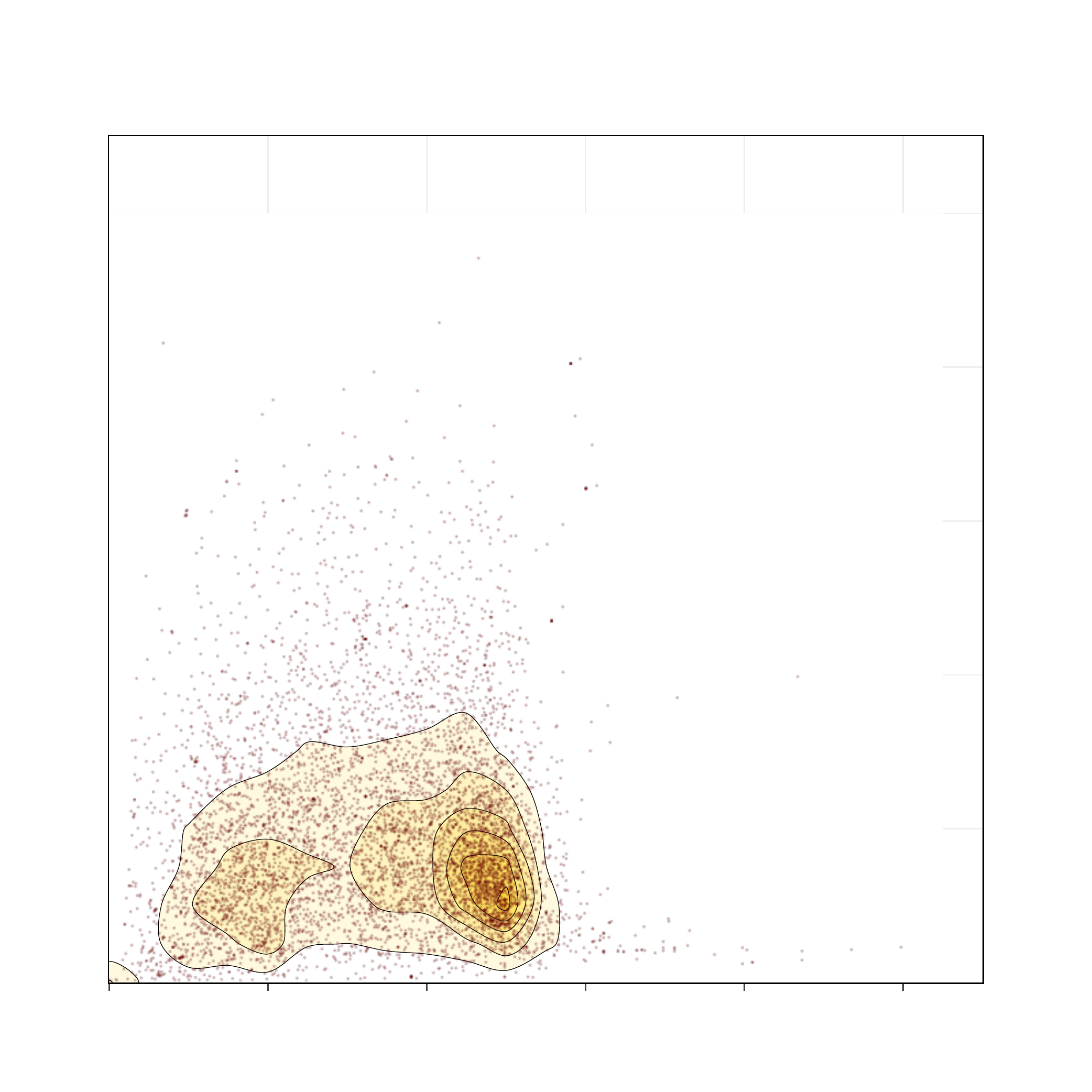
    \def\svgwidth{0.196\linewidth}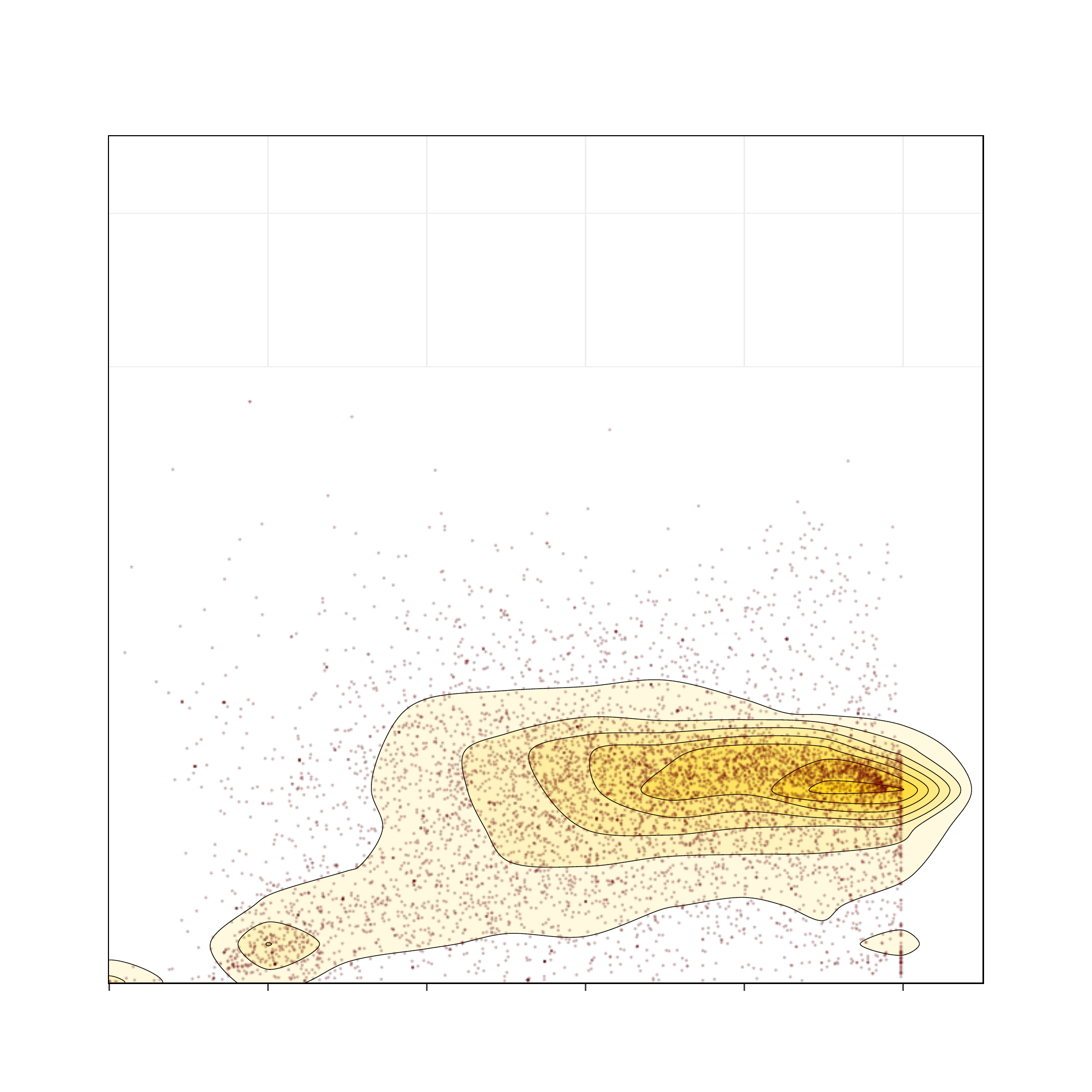
    \def\svgwidth{0.196\linewidth}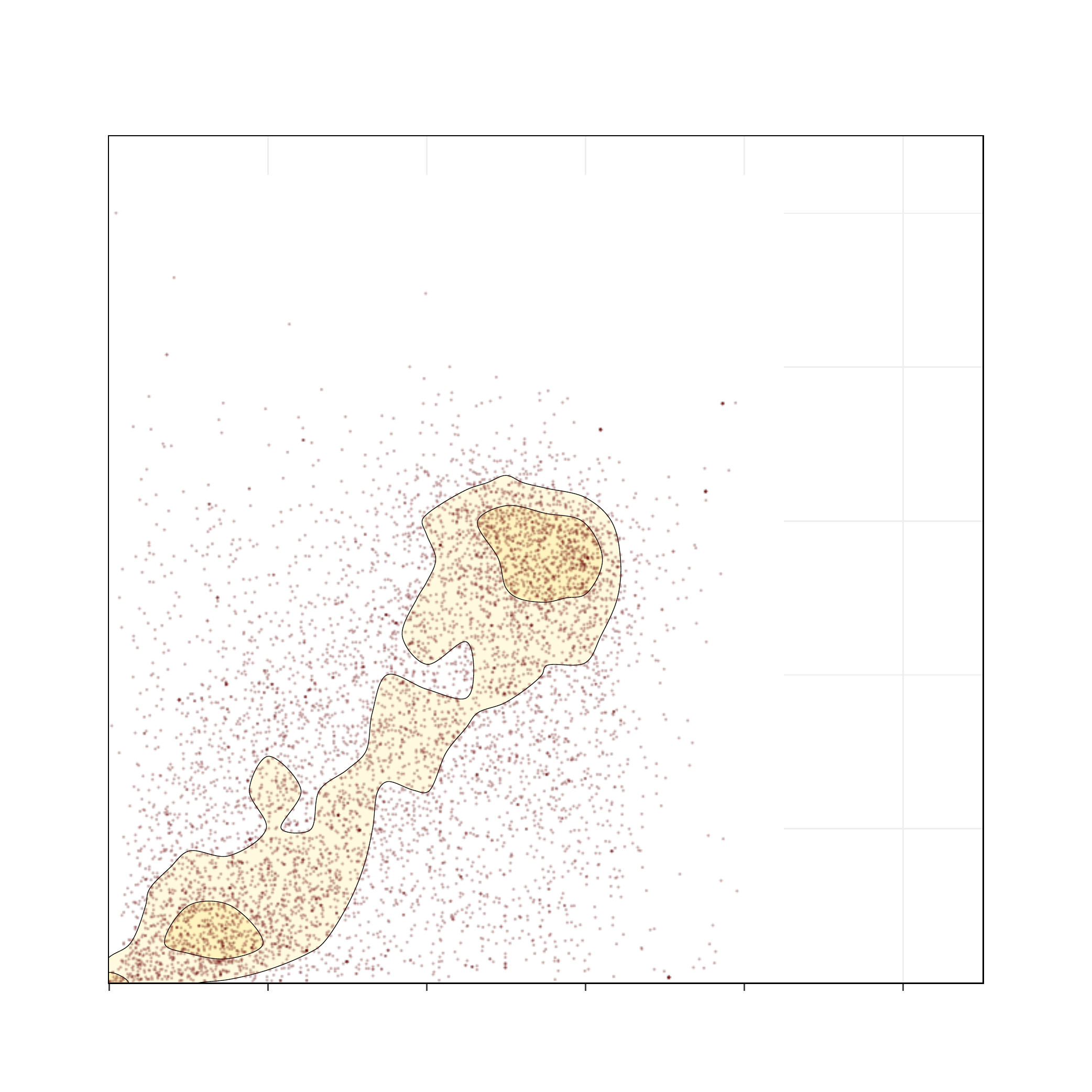
    \def\svgwidth{0.196\linewidth}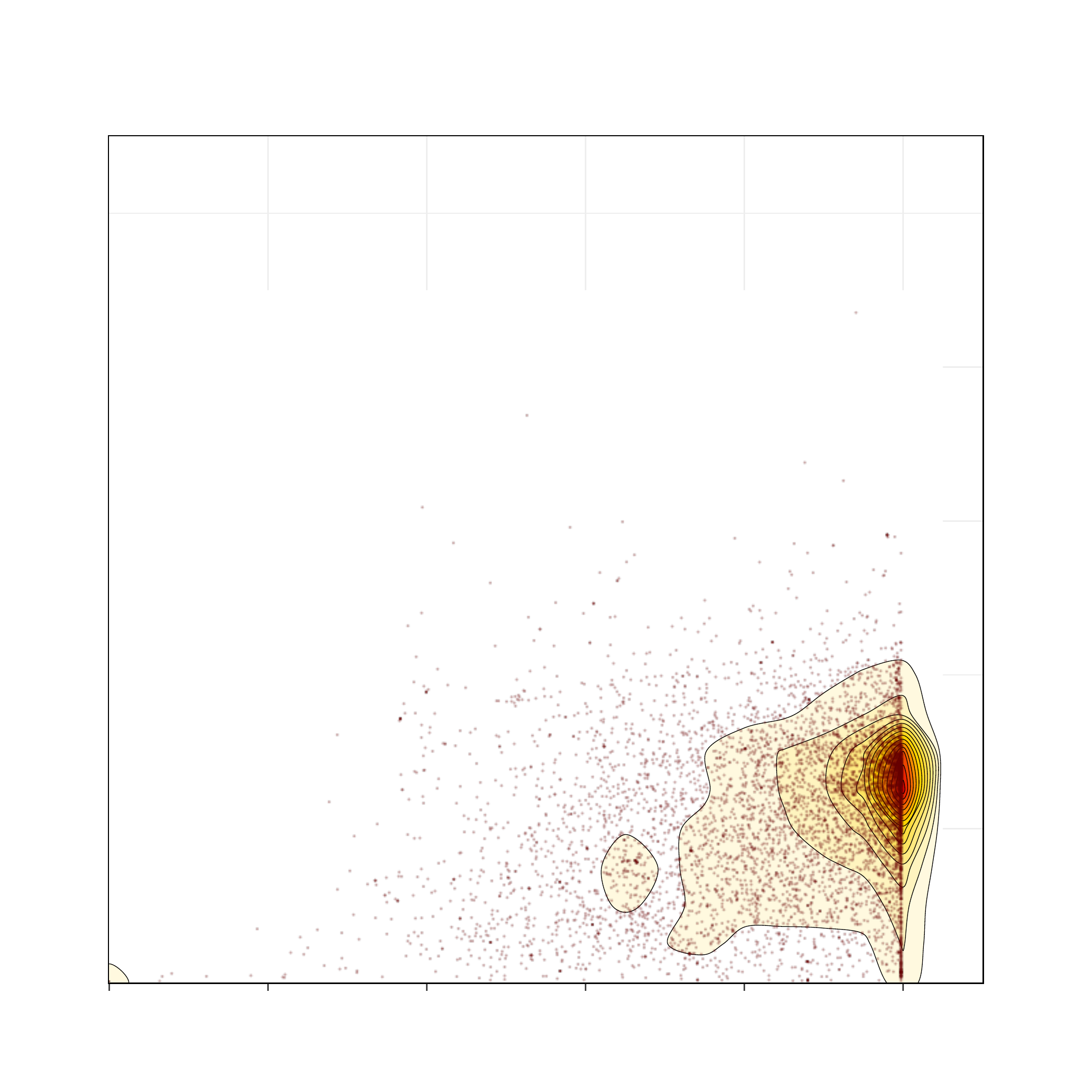
    \def\svgwidth{0.196\linewidth}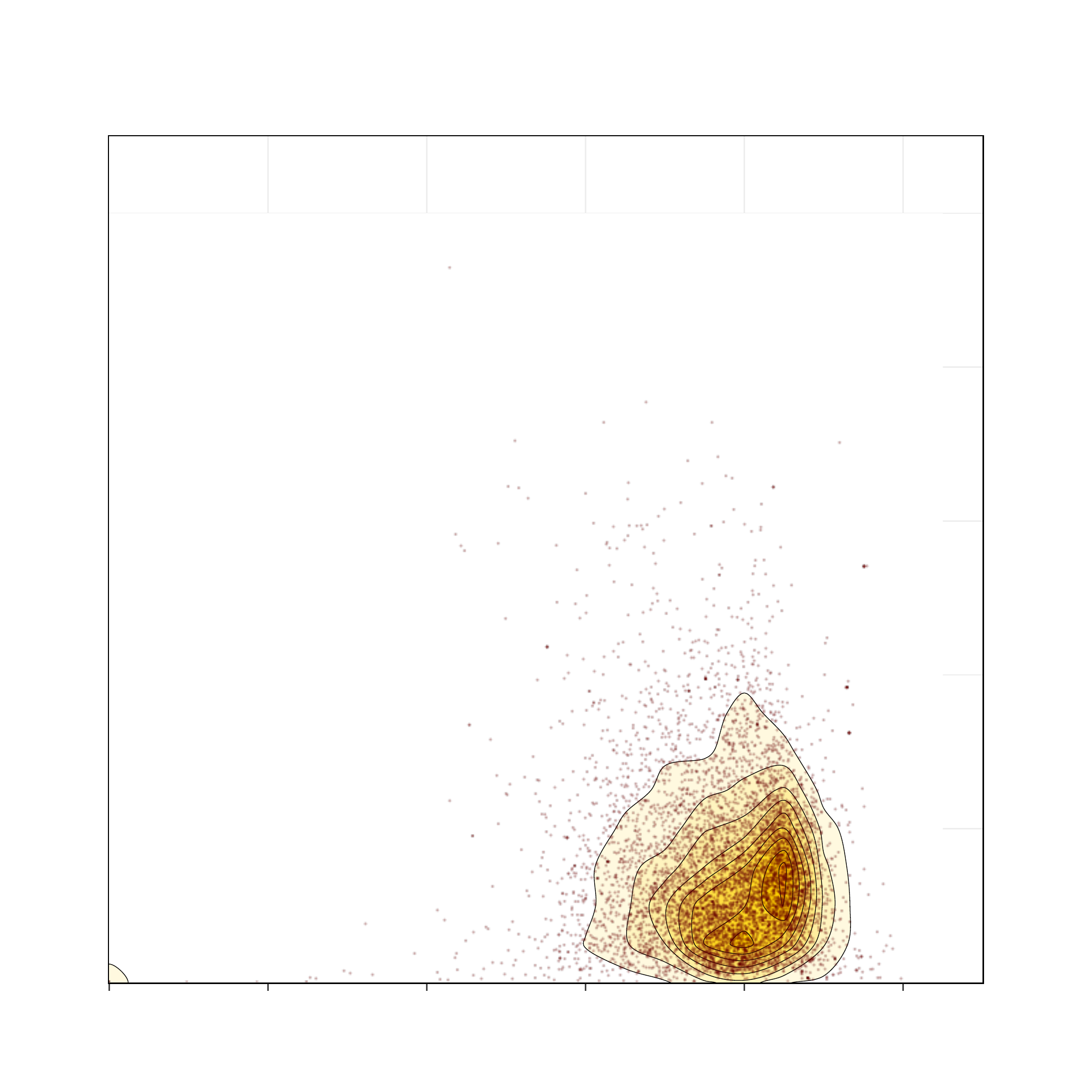
    \caption{{
    Distribution of the Precision \lo{(abscissa)}/Recall \lo{(ordinate)} scores \lo{displayed} as a scatter plot and its associated density function for the \dtsABC dataset altered with a Gaussian noise of deviation $\sigma=0.04$. From left to right, the detection produced by \pcptc, \ecnet, \pced, \pced (\dtsDFLT)$^+$ (see Table~\ref{tab:quant_default}) and \pcedtc trained on \dtsABC.}}
    \label{fig:NoiseABCprecRecall}
\end{figure*}

\begin{figure*}[htb]
\centering
\begin{tabular}{cccc}
\includegraphics[height=0.17\textheight]{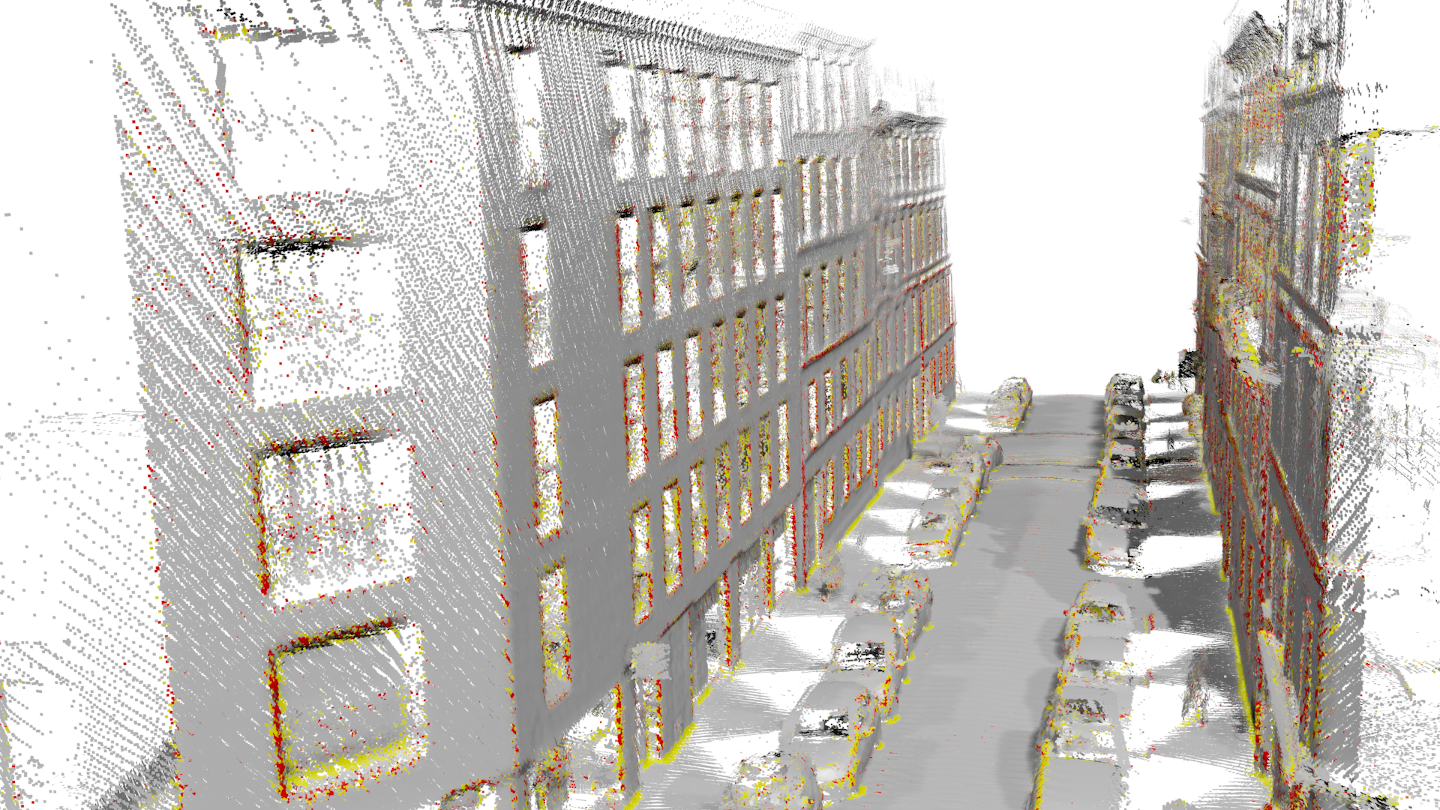}&  
\includegraphics[height=0.17\textheight]{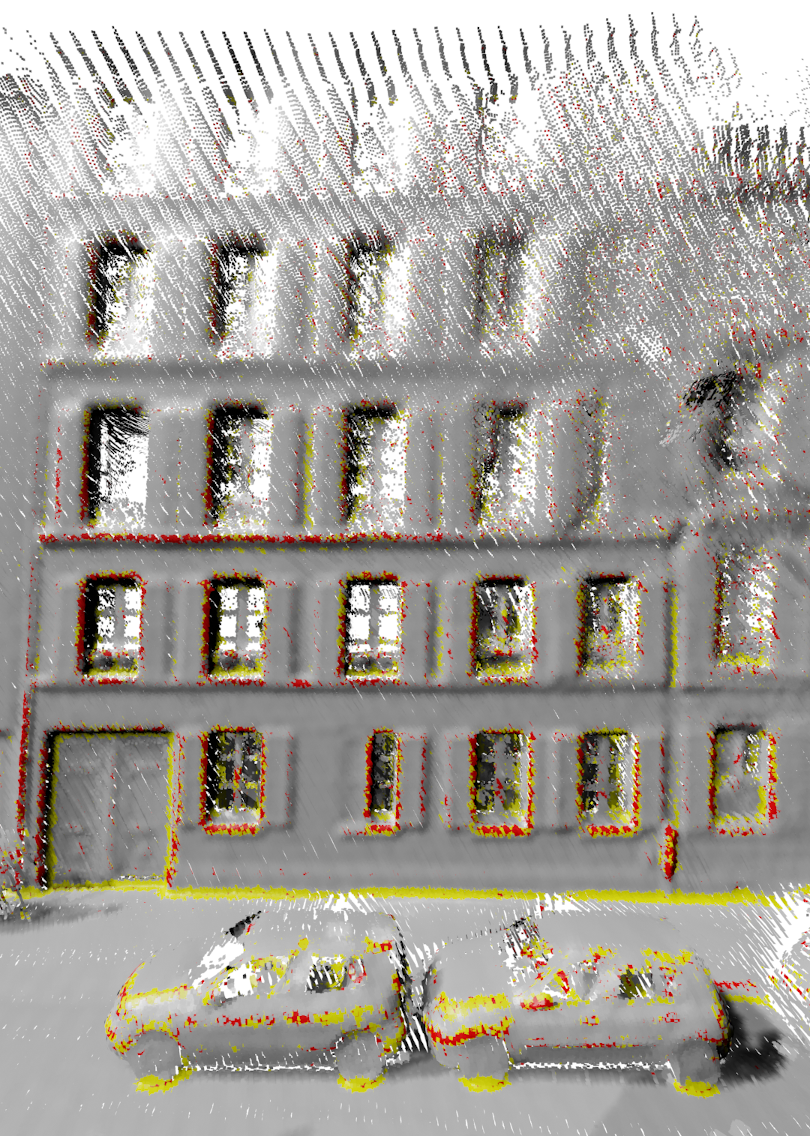}&
\includegraphics[height=0.17\textheight]{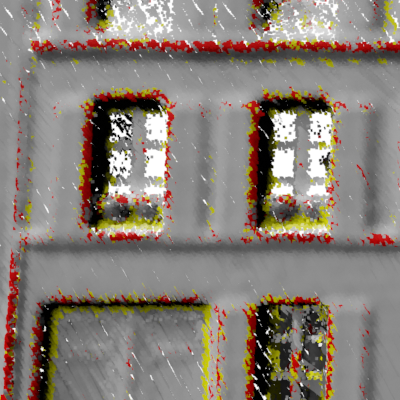}&
\includegraphics[height=0.17\textheight]{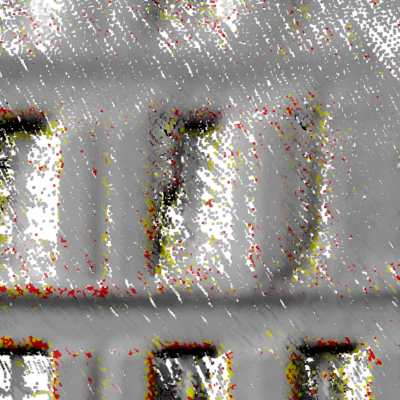}
\end{tabular}
\caption{{\lo{Illustration of the classification produced by \pced on the noisy acquired} model \dataRueMadame.}}
\label{fig:NoisePRM}
\end{figure*}

\begin{table}[h]
{
\setlength\tabcolsep{2.7pt} 
\centering
\small
\begin{tabular}{|l|c|c|c|c|c|c|}
\hline
Method           & Precision      & Recall         & \textbf{MCC}   & F1             & Accuracy  & IoU     \\ \hline
\ecnet (\dtsEC)  &  0.247    &      0.687   & 0.317   &     0.358   &   0.824  & \nico{0.216}\\ \hline
\lo{\textbf{\pcedtc} (\dtsEC)} & \lo{0.163} & \lo{0.950} & \lo{0.236}  & \lo{0.275}  & \lo{0.633} & \nico{0.158}   \\ \hline
\hline
\textbf{\pced} (\dtsDFLT) &   \textbf{0.323} & 0.375      & 0.267  & 0.319  &    \textbf{0.903}   & \nico{0.185}  \\ \hline
\textbf{\pced} (\dtsDFLT)$^+$ &   0.231 & \textbf{0.920}      &  \textbf{0.358}  & \textbf{0.366}  &    0.768  & \nico{\textbf{0.222}}   \\ \hline
\hline
\pcptc (\dtsABC) &  0.159        &  0.363     &    0.155       &     0.204     &   0.833    & \nico{0.113}   \\ \hline
\textbf{\pcedtc} (\dtsABC) &  0.121   &  0.793        &    0.181    &  0.213       & 0.626       & \nico{0.117} \\ \hline
\end{tabular}
\caption{{Quantitative evaluation on \dtsABC dataset altered with a Gaussian noise of deviation $\sigma=0.04$ (median scores). For \pced trained on the \dtsDFLT dataset (with three classes), we show the results when considering the points of both the sharp and smooth edge classes as edges ($4^{th}$ row, denoted with $^+$), and when considering only the sharp-edge class as edges \lo{($3^{rd}$ row)}.} 
\label{tab:quant_abc_noise}
}
}
\end{table}

{We compare the classification produced by \pcptc, \ecnet, \pced and \pcedtc on the \dtsABC dataset altered in the normal direction by a Gaussian noise of deviation $\sigma = 0.04$. See Figure~\ref{fig:NoiseABC}-bottom for classification results on model $0133$ (other models are included in joined website), and Figure~\ref{fig:NoiseABCprecRecall} for scatter plots illustrating the distribution of the (Precision, Recall) scores on the entire dataset.
As illustrated on Figure~\ref{fig:NoiseABC}-bottom, the two approaches trained on \dtsABC (\pcptc and \pcedtc) cannot disambiguate between noise and edges. 
This is due to the lack of noise in the \dtsABC training set.
Trained on more versatile data, both \ecnet and \lo{\pced~(\dtsEC)} are more robust to noise and still exhibit high recall, despite a loss in precision.
In presence of noise, \pced tends to classify inaccurate edge points as smooth edges. 
We thus propose to also evaluate the classification results considering the edge and smooth edge classes as positive matches (denoted with $^+$ in Table~\ref{tab:quant_abc_noise} and Figure~\ref{fig:NoiseABCprecRecall}).
With this setting, \pced~(\dtsDFLT)$^+$ produces the best quantitative scores in Recall, MCC and F1, slightly overpassing \ecnet, while being trained on smaller dataset and requiring 3 hours of computation instead of 20 days.
\pced~(\dtsDFLT)$^+$, however, detects smoother edges, as on chamfers, that are avoided by \ecnet, as illustrated in Figure~\ref{fig:NoiseABC} $1^{st}$, $3^{rd}$ and $4^{th}$ columns.

Figure~\ref{fig:NoisePRM} shows the capability of \pced to classify edges on an acquired noisy model \dataRueMadame (we could not process this 12M-points dataset with \pcptc and \ecnet because of their hardware requirements). 
Figure~\ref{fig:cubes}-bottom provides a qualitative evaluation of the different classification methods on the model \dataTwoCubes of the \dtsDFLT dataset with a severe Gaussian noise (deviation $\sigma = 0.14$). 
In this case also, \ecnet and \pced exhibit a more convincing behavior. 
}

\subsection{Interactive learning}
\label{subsec:interactive}
As shown in Table~\ref{table:trainingtime}, \pced requires very low training time, and is stable for very small training sets as our \dtsDFLT.
We illustrate how such a flexible network can be trained interactively to better adapt to user wishes and data specificity.

Our interactive training system performs as follows:
the user loads a point cloud on which GLS descriptors are precomputed. 
This is done in less than 20 seconds for 1M points.
Then, the user manually label{s} some points of the two classes (sharp edge and non-edge) or the three classes (sharp edge, smooth edge and non-edge).
We observed that the non-edge class should contain more points than the two others.
This training sets are provided to the \pced network initialized with random values, which learns for 5k epochs in around 10 seconds for approximately 10k input points. 
As for any other dataset, the number of points per class is automatically balanced during training.
Once trained, the network classifies the whole point cloud in around 2 seconds per 1M points.
Upon classification, the user can refine the training set according to the network output.
If he does so, the network is trained again {from a random initialization} with the updated learning sets (as training is fast enough and modifying the previous training is less efficient).

\begin{figure}[t]
    \centering
{
    \footnotesize
    \def\svgwidth{\linewidth}
    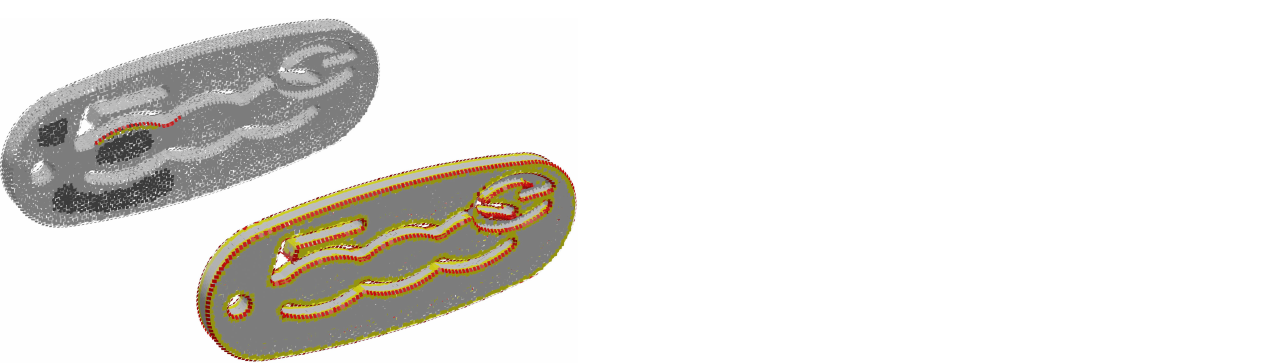
}
    \caption{
    Interactive training: user inputs and classification results.
    }
    \label{fig:interactive}
\end{figure}

{We illustrate in Figures~\ref{fig:teaser}-c and~\ref{fig:interactive}, and in the accompanying video how a user can generate a high-quality classification corresponding to a specific edge definition defined by the annotations he provides during \lo{an} interactive training session. For instance, in Figure~\ref{fig:teaser}-c-top, edges are defined 
sharp, while in Figure~\ref{fig:teaser}-c-bottom, edges are defined as a large scale feature including chamfers. }

\begin{figure}[b]
    \centering
    \includegraphics[width=0.5\linewidth]{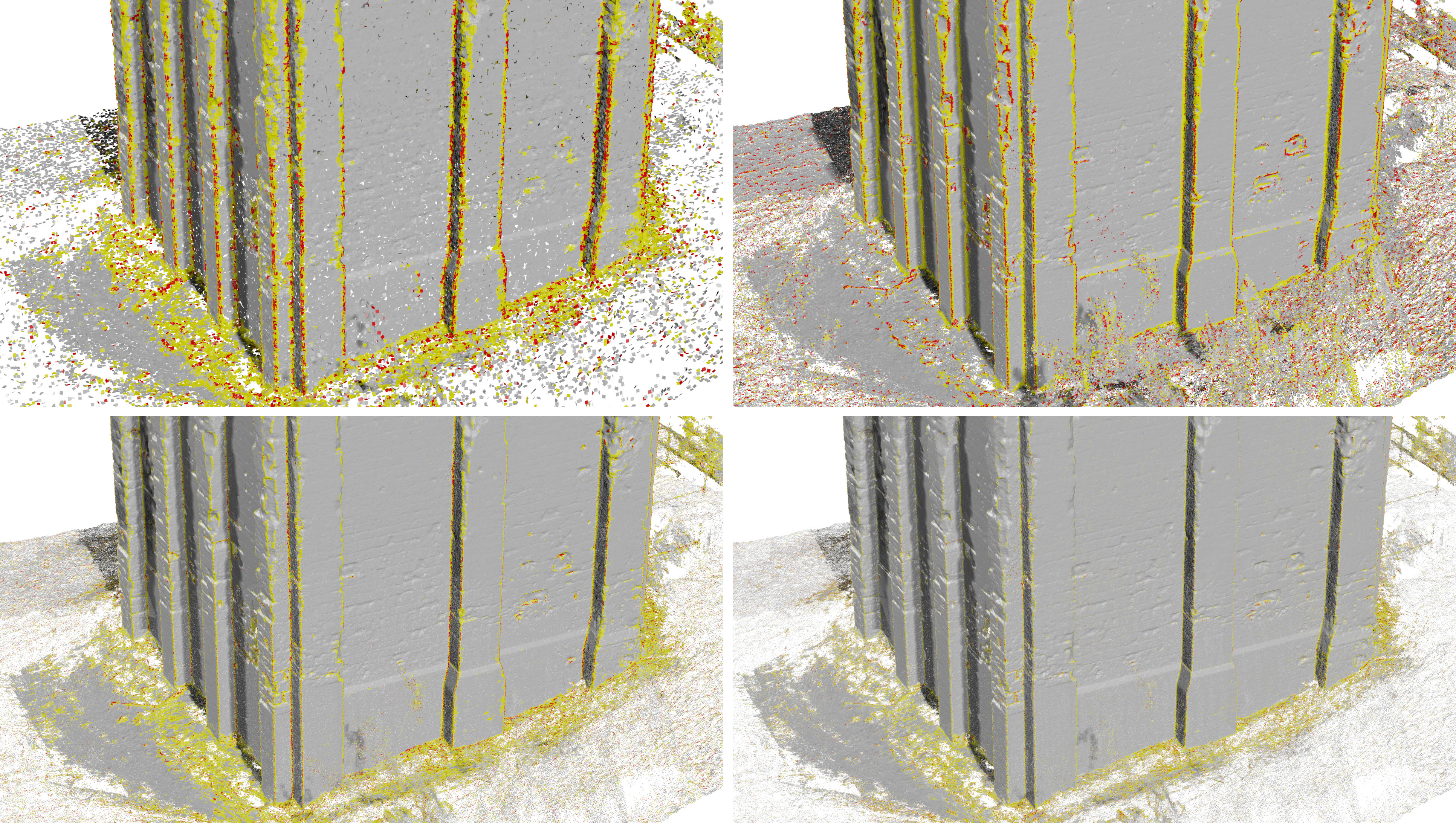}
    \caption{\dataLoudun: \pced classification with different densities: 1, 5, 15 and 35 million points from top left to bottom right.}
    \label{fig:loudoncomp}
\end{figure}


\begin{table*}[]
    \footnotesize
    \nico{
\begin{tabular}{|c|c|c|c|c|c|c|c|}
\hline
\bf{Training}      & \pcp(\dtsDFLT)    & \ecnet(\dtsEC) & \pienet (\dtsABC) & GLS   & \cnn(average)                          & \fc(average)  & \pced(average)     \\ \hline
 Time $t_{\text{K}}$ & 8.34*   & 7.32* & 25.77*  & 0.05  & 14.68 (\Sum{14.68}{0.05})     & 3.97 (\Sum{3.97}{0.05})   & \textbf{0.43 (\Sum{0.43}{0.05})} \\ \hline
                   Energy $E_{\text{K}}$ & 2167.42* & 1831.05* & 6443.78* & 9.16
                   & 1321.43 (\Sum{1321.43}{9.16}) & 357.60 (\Sum{357.60}{9.16}) & \textbf{38.60 (\Sum{38.60}{9.16})}  \\ \hline
\end{tabular}
}
\caption{
\nico{Times $t_{\text{K}}$ ($2^{nd}$ row) and processing unit energy consumption $E_{\text{K}}$ ($3^{rd}$ row) required for processing 1K points when training the different networks ($1^{st}$ row) denoted as \emph{name(training dataset)}. \emph{(average)} represents the average of the times obtained when training on the different datasets \dtsDFLT, \dtsABC and \dtsSHREC. \pcp is trained on an NVIDIA TITAN Quadro RTX 6000 GPU, and \ecnet and \pienet are trained on an NVIDIA TITAN X GPU. The times and energy consumption for \ecnet and \pienet are computed respectively from the statistics provided in~\cite{ecnet2018} and~\cite{PieNet2020}.}
\label{tab:energy_training}
}
\end{table*}


\begin{table*}[]
    \footnotesize
    \nico{
\begin{tabular}{|c|c|c|c|c|c|c|c|c|c|}
 \hline
  \bf{Classification}    &\ca   & \fee & \pcp   & \ecnet   & \pienet (8K pts) & GLS   & \cnn                        & \fc  & \pced     \\ \hline
 Time $t_{\text{K}}$ & \textbf{0.015} & 0.16 & 2.28*    & 1.32*  & 0.062*   & 0.023 & 0.043 (\Sum{0.043}{0.023})  & 0.0024 (\Sum{0.0024}{0.023}) & \textbf{0.0026} (\Sum{0.0026}{0.023})       \\ \hline
                   Energy $E_{\text{K}}$   & \textbf{1.36}   & 14.79 & 592.87*  & 345.77* & 15.63* & 4.24
                   & 3.87 (\Sum{3.87}{4.24})   & 0.22  (\Sum{0.22}{4.24})   & \textbf{0.23}  (\Sum{0.23}{4.24})  \\ \hline
\end{tabular}
}
\caption{
\nico{Times $t_{\text{K}}$ ($2^{nd}$ row) and processing unit energy consumption $E_{\text{K}}$ ($3^{rd}$ row) required for classifying 1K points with the different methods ($1^{st}$ row). \pcp and \ecnet are run on an NVIDIA TITAN Quadro RTX 6000 GPU, and \pienet is run on an NVIDIA TITAN X GPU. The times and energy consumption for \pienet are computed from the statistics provided in~\cite{PieNet2020}.}
\label{tab:energy_classif}
}
\end{table*}

\nico{
\subsection{Energy efficiency}
\label{subsec:power}
Energy consumption is a critical aspect of processing methods when considering their application in embedded systems or their environmental impact on 
global warming.
In addition to evaluating the results in terms of classification and timings, we present an empirical study of the energy consumption of the different approaches for training and classification, based on their hardware requirement and processing time.

\paragraph{Methodology}
Ideally, energy consumption should be measured during processing in order to account for the modulation of processor charge. 
Since we do not have such measuring device yet, we propose to 
roughly estimate the energy consumption based on the Thermal Design Power (TDP) of the processing unit (CPU or GPU), which measures the energy consumption under high workload.
This thus provides over-estimations of the energy consumption that remain useful for the comparison of 
approaches with significant differences
(in our case, up to several orders of magnitude, as can be seen in Tables~\ref{tab:energy_training} and~\ref{tab:energy_classif}).
The TDP is expressed in Watts and it is provided by manufacturers. 
We denote $E_{\text{K}}$ the energy, expressed in Joule (1J = 1Ws), used for processing 1K points, and we define it as follows:
    $$E_{\text{K}} = TDP . t_{\text{K}}\qquad\text{and}\qquad
    t_{\text{K}} = \frac{Pt . 1000}{N}~,$$
where $t_{\text{K}}$ is the time required for processing 1K points, $Pt$ is the processing time and $N$ is the number of processed points in the time $Pt$.
The TDP of our 10 cores Intel Xeon E5-2640 v4 @2.4 GHz is 90W, the one of an NVIDIA TITAN X GPU is 250W and for an NVIDIA Quadro RTX 6000 it is 260W. 
Based on Tables~\ref{tab:training_stats},~\ref{table:trainingtimeours},~\ref{table:trainingtime},~\ref{tab:visualTime}, and the statistics provided by~\citet{PieNet2020} and~\citet{ecnet2018} for training and classification times and number of points, we show in Tables~\ref{tab:energy_training} and \ref{tab:energy_classif} the mean processing unit energy consumption $E_{\text{K}}$ of each approach for respectively training and classifying 1K points.

When training (see Table~\ref{tab:energy_training}), our approach requires significantly less energy than our baselines (one order of magnitude) and than very deep neural networks running on the GPU, e.g. \pcp, \ecnet and \pienet (up to two orders of magnitude).
The low efficiency of \cnn can be explained by its higher number of weights (about 4 times higher than \pced) and by its implementation based on Tensorflow. 

Similar trends can be observed during classification (see Table~\ref{tab:energy_classif}).
We report the performance of \pienet provided by~\citet{PieNet2020} on models composed of around 8 thousand points. 
Other approaches are evaluated on larger models, up to several millions of points. 
Complementary experiments would be required to better evaluate \pienet classification efficiency on larger point clouds.
\ca requires a very little energy while \pced and its baselines remains more efficient than \fee.
}


\subsection{Complementary experiments}
\label{sec:compexp}

\paragraph{Variation of sampling (minimum scale)}
The features computed in the SSM are influenced by the minimum scale $s_\text{min}$, estimated from the density of the point cloud.
We show in Figure~\ref{fig:loudoncomp} how our approach behaves when changing the point cloud density by subsampling, while keeping the automatic estimation of $s_\text{min}$.
This is illustrated on the Loudun tower, initially composed of 35 millions points, and subsampled down to 15, 5 and 1 million points.
Unsurprisingly, decreasing the resolution reduces the quantity of details and leads to thicker edges.
However, our classification remains stable and \pced still finds the position of edges, even for low densities, and without requiring the user to set any parameter value.

\begin{figure}[t]
    \centering
{
    \footnotesize
    \def\svgwidth{\linewidth}
\begingroup%
  \makeatletter%
  \providecommand\color[2][]{%
    \errmessage{(Inkscape) Color is used for the text in Inkscape, but the package 'color.sty' is not loaded}%
    \renewcommand\color[2][]{}%
  }%
  \providecommand\transparent[1]{%
    \errmessage{(Inkscape) Transparency is used (non-zero) for the text in Inkscape, but the package 'transparent.sty' is not loaded}%
    \renewcommand\transparent[1]{}%
  }%
  \providecommand\rotatebox[2]{#2}%
  \newcommand*\fsize{\dimexpr\f@size pt\relax}%
  \newcommand*\lineheight[1]{\fontsize{\fsize}{#1\fsize}\selectfont}%
  \ifx\svgwidth\undefined%
    \setlength{\unitlength}{5760bp}%
    \ifx\svgscale\undefined%
      \relax%
    \else%
      \setlength{\unitlength}{\unitlength * \real{\svgscale}}%
    \fi%
  \else%
    \setlength{\unitlength}{\svgwidth}%
  \fi%
  \global\let\svgwidth\undefined%
  \global\let\svgscale\undefined%
  \makeatother%
  \begin{picture}(1,0.27604167)%
    \lineheight{1}%
    \setlength\tabcolsep{0pt}%
    \put(0,0){\includegraphics[width=\unitlength,page=1]{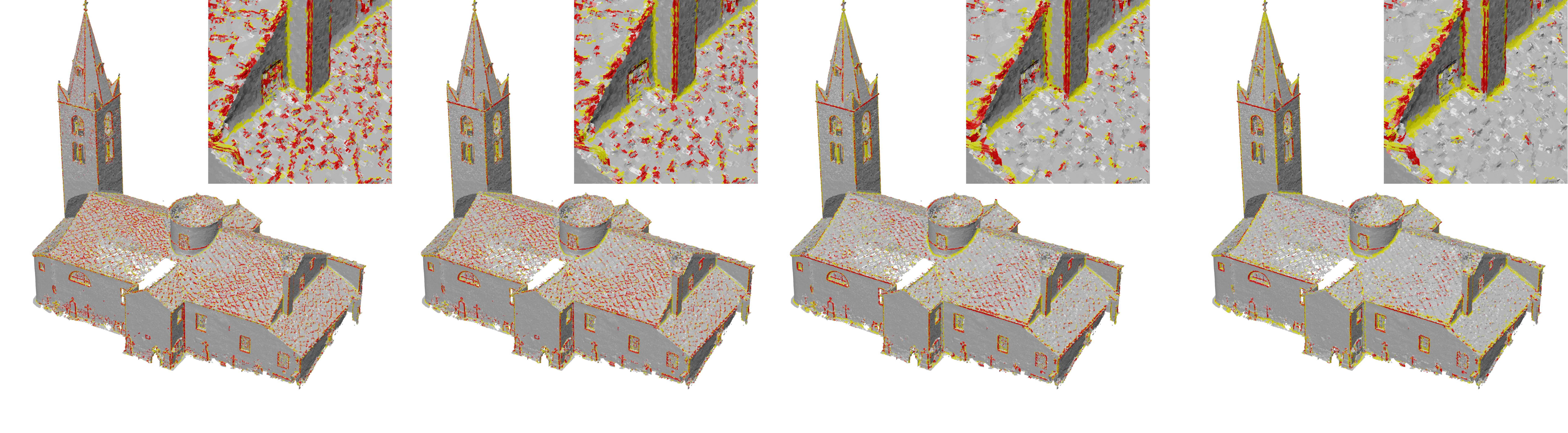}}%
    \put(0.07991319,0.00248836){\makebox(0,0)[lt]{\lineheight{1.25}\smash{\begin{tabular}[t]{l}$s_{\max} = 0.02$\end{tabular}}}}%
    \put(0.32837674,0.00211986){\makebox(0,0)[lt]{\lineheight{1.25}\smash{\begin{tabular}[t]{l}$s_{\max} = 0.05$\end{tabular}}}}%
    \put(0.56328068,0.00220674){\makebox(0,0)[lt]{\lineheight{1.25}\smash{\begin{tabular}[t]{l}$s_{\max} = 0.20$\end{tabular}}}}%
    \put(0.83202298,0.00197671){\makebox(0,0)[lt]{\lineheight{1.25}\smash{\begin{tabular}[t]{l}$s_{\max} = 0.50$\end{tabular}}}}%
  \end{picture}%
\endgroup%

}
    \caption{Classification results when varying the maximum scale when computing the GLS for the analyzed object. 
    }
    \label{fig:maxscale}
\end{figure}
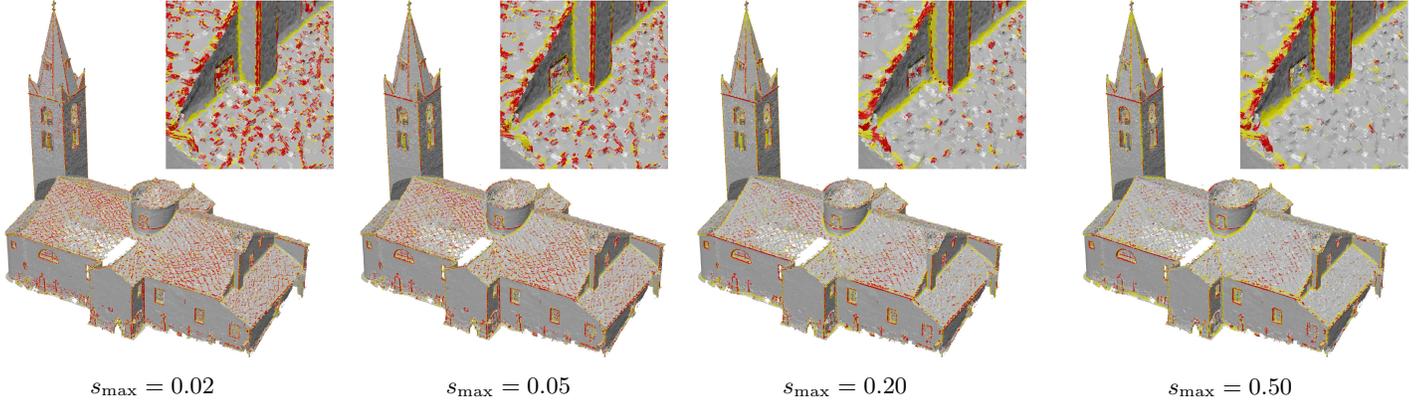

\paragraph{Variation of maximum scale}
We also show in Figure~\ref{fig:maxscale} how our classification behaves when changing the maximum scale.
Thanks to the logarithmic scale sampling and the scale invariance property of the SSM entries, our approach provides stable results even though the maximum scale is divided by 10 or multiplied by 2.5.

\begin{figure}[htb]
    \centering
    
   
   \includegraphics[width=\linewidth]{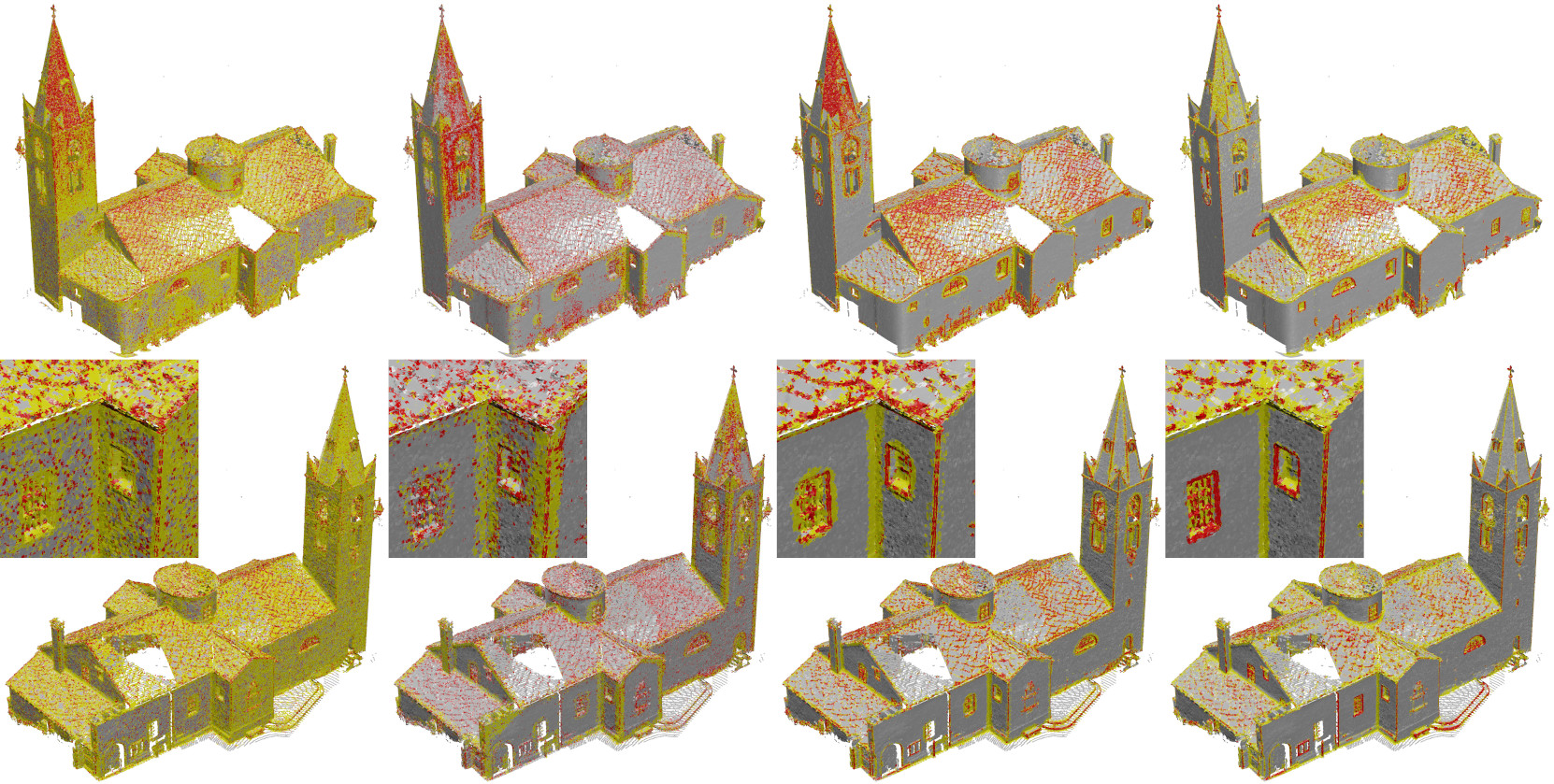}
    \caption{Impact of the surface reconstruction algorithm on the classification. From left to right: covariance plane fitting (used by ~\citet{Bazazian15}), point set surfaces~\cite{Alexa:2001}, algebraic sphere fitting, algebraic point set surfaces~\cite{Guennebaud:2007:APS} (used in this work).}
    \label{fig:spm}
\end{figure}

\paragraph{Surface reconstruction algorithm}
Our network is parameterized using the Algebraic Point Set Surfaces (APSS)~\cite{Guennebaud:2007:APS}, which are known to be stable and reliable even at large scales.
We illustrate in Figure~\ref{fig:spm} the performance of our classifier when computing the parameterization with different approaches: covariance plane fitting (also used in~\citet{Bazazian15}), plane-based point set surfaces~\cite{Alexa:2001} and algebraic sphere fitting (same fitting as APSS but without the Moving Least Squares -MLS- projection).
For each variant we compute similar values as our GLS descriptor, with derivatives estimated using finite differences.
We clearly observe that the use of sphere fitting rather than plane fitting improves the classification, and the best results are always obtained using MLS projection.
Eventually, recent more stable derivative evaluations of the APSS may also be experimented~\cite{lejemble2021}. 

\section{Discussion}

{\paragraph{SSM and patch-based architectures.} Our network relies on the SSM, a point cloud parameterization based on a set of GLS descriptors. 
An open question is whether the SSM could be efficiently used to provide additional parameters to the points used as input in patch-based point cloud deep processing architectures such as PointNet.
This may be investigated, but we point out two issues that should be considered. 
The first is the significant memory overhead that will be generated on architectures that are already resource-demanding. 
The second is more conceptual: \nico{the} goal of using the SSM is to capture local multi-scale shape descriptors at each point without having to explicitly handle their neighborhood in the network. 
This makes a lot of sense for a point cloud processed by point individually in the network, as we do for edges.
This is less clear for a processing relying on point patches (e.g. for style or large features recognition).
The first layers of these networks learn features from the spatial organisation of neighbor points and it is not easy to predict which additional parameter would be redundant (i.e. the information it brings is already captured by the network first layers) and which parameter would increase the network efficiency.}

{\paragraph{Need of local surface reconstruction.} The need for locally reconstructing an approximating surface at different scales may be seen as a drawback of our method. 
In fact, this reconstruction does not allow to explicitly reconstruct or detect edges, and once the surface reconstructed, the problem remains unsolved. 
However, it enables the computation of the SSM shape descriptors that, used as input in our network, allow us to efficiently label edge points.} 

\section{Limitations and conclusion}

We introduced a new parameterization together with its dedicated neural network architecture (\pced) specially designed for the classification of edges in point clouds. \pced outperforms both state of \thib{the} art methods such as \ca, \fee, \ecnet, 
\pcp{} \nico{and \pienet}, and baselines as \cnn and \fc. 
\pced is also remarkably compact by being only composed of about 2100 weights. 
Given this small size, it is faster than the other approaches tested in this work. 
Both \cnn and \pced achieve similar very good results on synthetic point clouds, which shows that our parameterization based on GLS descriptors is very efficient to encode the point clouds features required for edge classification. 
Coupled with \pced, it provides a very compact multi-scale representation that captures local geometric properties with reduced sensitivity with respect to noise, as can be seen on the various tests performed on real point cloud scans.
The training and classification of our approach is also sufficiently fast to enable interactive training and classification from direct user inputs.
\paragraph{Limitations.} APSS requires oriented normals, which may be tedious to compute accurately in some cases. 
For all our experiments with point clouds without normals, we 
\lo{obtained} good classification results by estimating consistent normal vectors using Meshlab~\cite{Meshlab}, without requiring human intervention.
An interesting future work would be to consider alternative fitting techniques that do not require oriented normals~\cite{Chen2013}.
Regarding performance, SSM precomputation is currently the bottleneck of the approach, however we believe that a GPU implementation would enable near real-time classification, the most computationally intensive task being the neighborhood queries. 
Theoretically, SSM is by nature limited to surfaces only, alternative representations~\cite{Digne:2018} might be considered in future work to handle lines and volumes.

\paragraph{Perspectives.} Other experiments may be conducted to improve the network layout, for instance n by n scales concatenation. 
An interesting direction for future work could be the study of the extension of this architecture to other geometrical labelling tasks, but also to semantic analysis.


\section{Acknowledgements} 

We would like to thank Archeovision for the courtesy of the \dataLoudun dataset. This work has been partially funded by the ANR CaLiTrOp project (ANR-11-BS02-0006). Point clouds processing and visualization have been implemented in the Radium Engine~\cite{radium2021}.
\revision{\dataRueMadame is part of the \emph{Paris-rue-Madame database: MINES ParisTech 3D mobile laser scanner dataset from Madame street in Paris}.
MINES ParisTech created this special set of 3D MLS data for the purpose of detection-segmentation-classification research activities, but does not endorse the way they are used in this project or the conclusions put forward.}

\bibliographystyle{ACM-Reference-Format}
\bibliography{egbib}

\appendix
\section*{Appendices}

\section{GLS Descriptors and Derivatives}
\label{app:gls}

The APSS~\cite{Guennebaud:2007:APS} defines a scalar field $S(\mathbf{p})_\mathbf{u} = [ 1  \; \; \mathbf{p}^T \; \; \mathbf{p}^T\mathbf{p} ] \cdot \mathbf{u}$, where $\mathbf{u} = [ u_c \; \; \mathbf{u}_\ell \; \; u_q ]$ is the vector of field parameters. 
The parameters of the algebraic sphere are obtained by normalizing these field parameters: $\mathbf{\hat{u}} = \mathbf{u} / \sqrt{|| \mathbf{u}_\ell ||^2-4u_c u_q}$.
In the GLS~\cite{Mellado:2012:GLS}, the algebraic sphere parameters are reparameterized to compute its geometric parameters: $\tau=S(\mathbf{p})_\mathbf{\hat{u}}(\mathbf{p})$ the local relief, $\mathbf{\eta} = \frac{\nabla S(\mathbf{p})_\mathbf{\hat{u}}(\mathbf{p})}{|| \nabla S(\mathbf{p})_\mathbf{\hat{u}}(\mathbf{p}) ||}$ the normal vector and $\kappa=2\hat{u}_q$ the mean curvature.

The Scale-Space Jacobian of the GLS parameters if defined as a $5\times{}4$ matrix:
$$
\begin{bmatrix}
\frac{\delta{}\tau}{\delta\mathbf{x}} & \frac{\delta\mathbf{\eta}_x}{\delta\mathbf{x}} & \frac{\delta\mathbf{\eta}_y}{\delta\mathbf{x}} &
\frac{\delta\mathbf{\eta}_z}{\delta\mathbf{x}} &
\frac{\delta{}\kappa}{\delta\mathbf{x}} \\
\frac{\delta{}\tau}{\delta{}t} & \frac{\delta\mathbf{\eta}_x}{\delta{}t} & \frac{\delta\mathbf{\eta}_y}{\delta{}t} &
\frac{\delta\mathbf{\eta}_z}{\delta{}t} &
\frac{\delta{}\kappa}{\delta{}t} 
\end{bmatrix},
$$
where $\delta\mathbf{x}$ and $\delta{}t$ are the derivatives in scale and space respectively.
$k1$ is computed by projecting $\frac{\delta\mathbf{\eta}}{\delta\mathbf{x}}$ on the surface tangent plane, which provides an estimate of the second fundamental form.

\section{Scores used for quantitative comparison}
\label{app:scores}
The scores used in Section~\ref{sec_results} are defined as follows:

Precision (also denoted positive predictive value -- PPV) measures the proportion of positive identifications that are actually correct (the higher, the better). It is defined as:
$$\displaystyle {\text{precision}} = \frac{TP}{TP + FP}.$$

Recall (also denoted sensitivity, hit rate, or true positive rate  -- TPR) measures the proportion of actual positives that are correctly identified (the higher, the better). It is defined as:
$$\displaystyle {\text{recall}} = \frac{TP}{TP + FN}.$$

The Matthews Correlation Coefficient (MCC) is a correlation coefficient between the observed and predicted binary classifications; it returns a value in [-1 : 1]. A coefficient of 1 represents a perfect prediction, 0 no better than random prediction and -1 indicates a total disagreement between prediction and observation. It is defined as:
$$\displaystyle {\text{MCC}}={\frac {TP\times TN-FP\times FN}{\sqrt {(TP+FP)(TP+FN)(TN+FP)(TN+FN)}}}.$$

F1 score is a measure of a test accuracy. For binary classification, F1 is defined as follows:
$$\displaystyle {\text{F1}} = 2 . \frac{\displaystyle {\text{precision}} \times \displaystyle {\text{recall}}}{\displaystyle {\text{precision}} + \displaystyle {\text{recall}}} = \frac{2 TP}{2 TP + FP + FN} .$$

Accuracy measures is the fraction of predictions our model got right. For binary classification, accuracy is defined as:
$$\displaystyle {\text{accuracy}} = \frac{TP+TN}{TP + TN + FP + FN}.$$

\lo{The Intersection over Union score (IoU) is a value between $0$ and $1$ specifying the overlap between the prediction and the observation. A value of $1$ means that the union of the predicted and the reference sets is the same as their overlap (intersection) while a value of $0$ means that there is no overlap between the predicted and the reference sets. It is defined as follows:
$$\text{IoU} = \frac{TP}{TP+FP+FN}.$$

}

\end{document}